\documentclass[11pt,hidelinks]{article}
\usepackage[body={17.5cm, 22cm},right=2cm]{geometry}
\usepackage{color}
\usepackage{amssymb,graphicx}
\usepackage{amsmath}
\usepackage{epsfig}
\usepackage[bookmarks=true,pdfborder={0 0 0}]{hyperref} 
\usepackage{cite}
\pdfoutput=1
\usepackage{subcaption}

\usepackage{pifont}

\usepackage[utf8]{inputenc}
\usepackage{verbatim}   
\usepackage{amsfonts}
\usepackage{amssymb}
\usepackage{graphicx}
\usepackage{slashed}
\usepackage[normalem]{ulem}
\usepackage[multiple]{footmisc}
\usepackage{enumitem}

\newcommand{\km}{{\, {\rm km}}}

\newcommand{\kpc}{{\, {\rm kpc}}}

\newcommand{\cm}{{\, {\rm cm}}}

\newcommand{\eV}{{\, {\rm eV}}}

\newcommand{\MeV}{{\, {\rm MeV}}}
\newcommand{\GeV}{{\, {\rm GeV}}}

\def\beq{\begin{equation}}
\def\eeq{\end{equation}}
\def\bea{\begin{eqnarray}}
\def\eea{\end{eqnarray}}
\def\bitem{\begin{itemize}}
	\def\eitem{\end{itemize}}
\newcommand{\bec}{\begin{center}}
	\newcommand{\eec}{\end{center}}
\newcommand{\ba}{\begin{array}}
	\newcommand{\ea}{\end{array}}
\def\bar#1{\overline{#1}}

\def\abs#1{\left| #1\right|}

\def\inv{^{\raise.15ex\hbox{${\scriptscriptstyle -}$}\kern-.05em 1}}
\def\lbar{{\lower.35ex\hbox{$\mathchar'26$}\mkern-10mu\lambda}} %lambda bar

\def\to{\rightarrow}

\let\<=\langle
\let\>=\rangle

\let\+=\uparrow

\begin{document}

\begin{titlepage}
	~\vspace{1cm}
	\begin{center}

		{\LARGE \bf Dark Photon Stars:
			\\ \vspace{0.1mm}
			 {\Large Formation and Role as Dark Matter Substructure}
	\\
	\vspace{3.1mm}
		}

		\vspace{0.7cm}

		{\large
			Marco~Gorghetto$^a$,
			Edward~Hardy$^b$,
			John March-Russell$^c$,\\
			Ningqiang Song$^{b}$,
			and Stephen M. West$^d$

			}
		
		\vspace{.6cm}
		{\normalsize { \sl $^{a}$ 
				Department of Particle Physics and Astrophysics, Weizmann Institute of Science,\\
				Herzl St 234, Rehovot 761001, Israel }}
		
		\vspace{.2cm}
		{\normalsize { \sl $^{b}$ Department of Mathematical Sciences, University of Liverpool, \\ Liverpool, L69 7ZL, United Kingdom}}

	    \vspace{.2cm}
		{\normalsize { \sl $^{c}$ Rudolf Peierls Centre for Theoretical Physics, University of Oxford, \\ Oxford OX1 3PU, United Kingdom}}
		
	    \vspace{.2cm}
		{\normalsize { \sl $^{d}$ Department of Physics, Royal Holloway, University of London, \\ Egham, Surrey, TW20 0EX, United Kingdom}}
	
	\end{center}
	\vspace{1cm}
	\begin{abstract}

Any new vector boson with non-zero mass (a `dark photon' or `Proca boson') that is present during inflation is automatically produced at this time from vacuum fluctuations and can comprise all or a substantial fraction of the observed dark matter density, as shown by Graham, Mardon, and Rajendran. We demonstrate, utilising both analytic and numerical studies, that such a scenario implies an extremely rich dark matter substructure arising purely from the interplay of gravitational interactions and quantum effects. Due to a remarkable parametric coincidence between the size of the primordial density perturbations and the scale at which quantum pressure is relevant, a substantial fraction of the dark matter inevitably collapses into gravitationally bound solitons, which are fully quantum coherent objects. The central densities of these `dark photon star', or `Proca star', solitons are typically a factor $10^6$ larger than the local background dark matter density, and they have characteristic masses of $10^{-16} M_\odot (10^{-5}{\rm eV}/m)^{3/2}$, where $m$ is the mass of the vector. 
During and post soliton production a comparable fraction of the energy density is initially stored in, and subsequently radiated from, long-lived quasi-normal modes. Furthermore,
the solitons are surrounded by characteristic `fuzzy' dark matter halos in which quantum wave-like properties are also enhanced relative to the usual virialized dark matter expectations. 
Lower density compact halos, with masses a factor of $\sim 10^5$ greater than the solitons, form at much larger scales. We argue that, at minimum, the solitons are likely to survive to the present day without being tidally disrupted. This rich substructure, which we anticipate also arises from other dark photon dark matter production mechanisms, opens up a wide range of new direct and indirect detection possibilities, as we discuss in a companion paper.

	\end{abstract}

\end{titlepage}

	\thispagestyle{empty}

{\fontsize{11.5}{10.5}
\tableofcontents
}

%%%%%%%%%%%%%%%%%%%%%%%%%%%%%%%%%%%%%%%%%%%%%%%%%

\newpage

\section{Introduction and Summary}

Despite the overwhelming evidence for the existence of dark matter (DM), from scales spanning from astrophysical to cosmological, its specific nature remains unknown. Candidates range from primordial black holes or massive super-Planckian composite states, through sub-Planckian particles with mass $\gtrsim 1\eV$, to particles with mass $\lesssim 1\eV$ which are best described as semi-classical `wave' dark matter in galaxies such as the Milky Way.
So far, in the particle case, we have no information about the DM spin, and only extremely limited information about its mass and possible non-gravitational interactions with the Standard Model (SM). 
Among the numerous candidates, a minimal possibility is a new bosonic particle of spin-0 or spin-1. The presence of such particles is expected from string theory compactifications \cite{Svrcek:2006yi,Arvanitaki:2009fg,Arvanitaki:2009hb,Goodsell:2009xc,Cicoli:2011yh,Arias:2012az}. 
 
 Irrespective of their couplings to the SM, elementary spin-0 particles that exist as states at high scales are automatically produced in the early Universe via the so-called misalignment mechanism \cite{Preskill:1982cy,Abbott:1982af,Dine:1982ah}, and, if stable, form a component or all of the dark matter. There are a variety of other production mechanisms that could lead to a cosmologically interesting relic density of spin-0 particles, though misalignment production is attractive because of its minimality.

For vector bosons, the abundance from misalignment is generically suppressed \cite{Arias:2012az}. However, if a massive vector state is present during inflation, its longitudinal component is automatically produced by inflationary fluctuations \cite{Graham_2016}. 
 The resulting relic abundance $\Omega_A$ is
 \begin{equation}\label{eq:OmegaDM}
 	\frac{\Omega_A}{\Omega_{\rm DM}} \simeq \sqrt{\frac{m}{6\cdot 10^{-6}\eV}} \left(\frac{H_I}{10^{14}\GeV} \right)^2 ~,
 \end{equation}
 where $m$ is the mass of the vector, $H_I$ is the Hubble scale during inflation and $\Omega_{\rm DM}$ is the observed DM abundance \cite{Graham_2016}. 
This expression assumes that the vector mass does not change
during evolution of the Universe from the inflationary epoch until today, and that there are no charged or Higgs-like states close in mass or lighter than the vector boson.\footnote{Such a situation is possible in, e.g. the string theory context.}
Given the current upper bound on $H_I$ from non-observation of gravitational waves and the requirement that $m\ll H_I$, such production can provide the observed DM density for $10^{-5} \eV\lesssim m\lesssim 10^8 \GeV$.

The vector is produced because the longitudinal component acts as a scalar during inflation by the Goldstone equivalence theorem and -- in the same way as any other scalar present at this time -- obtains an approximately scale-invariant spectrum of energy density perturbations.\footnote{The transverse components are not produced since they are equivalent to a massless vector.} Crucially, after inflation, the perturbations in the vector at large scales redshift faster than they would for a scalar. Consequently, the primordial power spectrum reaches a form that, in addition to the standard adiabatic perturbations, has an isocurvature component peaked at cosmologically tiny scales $\simeq 10^{11} {\rm km} \, (10^{-5}{\rm eV}/m)^{1/2}$, corresponding to the size $H^{-1}$ of the Hubble horizon when $H=m$. The suppression of the perturbations at large scales makes the spectrum automatically consistent with bounds on isocurvature perturbations, since the primordial power spectrum is essentially unconstrained at those small distances. Production by inflationary fluctuations is therefore both robust and, at least in the absence of further interactions, unavoidable.\footnote{Very massive scalars are automatically produced from inflationary fluctuations by a different process \cite{Chung:1998zb}.}

\begin{figure}[t]
	\begin{center}
\includegraphics[width=1.\textwidth]{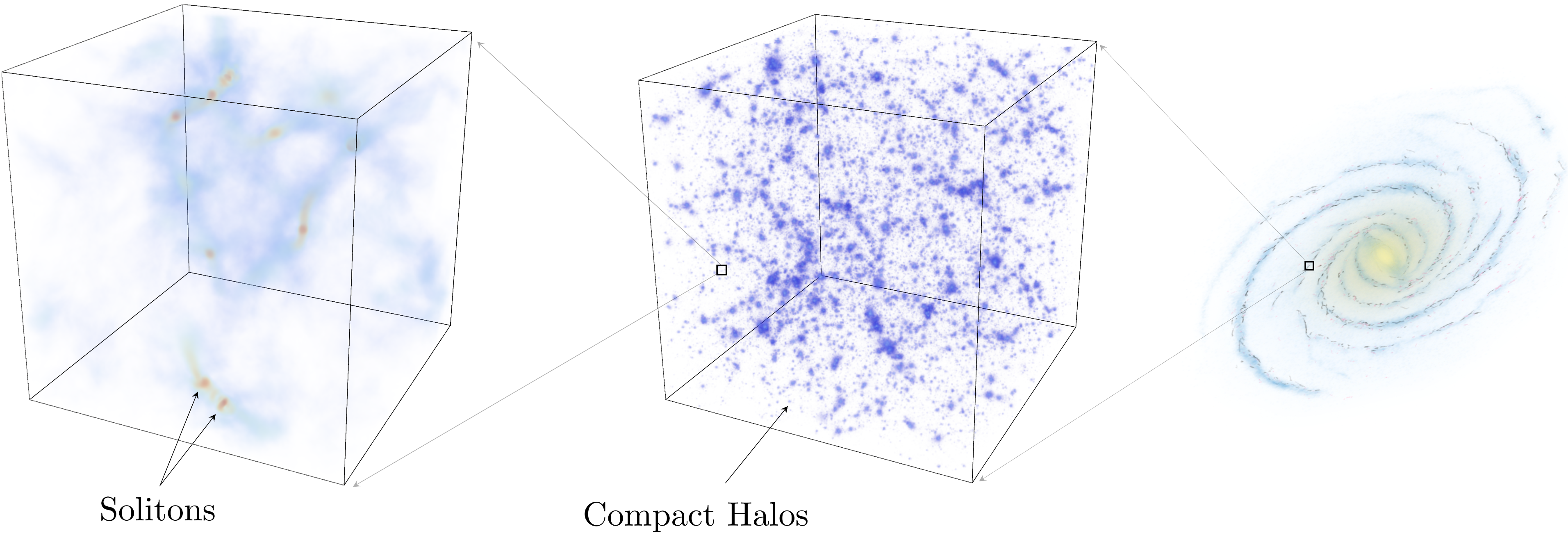}
	\end{center}
	\caption{A schematic representation of the dark photon dark matter distribution today. At subgalactic scales, roughly parsec and smaller, the dark matter is bound in compact halos that arise from the collapse of small-scale primordial inhomogeneities (centre panel). The left panel shows structure at smaller scales: solitons, i.e. dark photon stars, their surrounding fuzzy halos, and dark matter filaments connecting them.} \label{fig:DM_sub_pic}
\end{figure}

The purpose of this paper is to investigate the \emph{model-independent} dynamical process of structure formation from these small-scale primordial inhomogeneities. We will show that they lead to a strikingly rich structure of gravitationally bound objects, depicted in Figure~\ref{fig:DM_sub_pic}, which are normally absent in conventional cold dark matter structure formation.\footnote{We note that other types of interesting exotic compact objects can occur in more complex dark sectors involving dark photons \cite{Chang:2018bgx}.} 
As we will see in Section~\ref{section:SPsection}, the typical length scale of the inhomogeneities is so tiny that the quantum pressure of the bosons is relevant in the dynamics. Indeed, although the small-scale inhomogeneities are already non-linear before matter-radiation equality (MRE), quantum pressure prevents their collapse until after MRE.
 At that point, the overdensities collapse into gravitationally bound objects fully supported by quantum pressure, with typical mass of order $10^{-16} M_\odot(10^{-5} {\rm eV}/m)^{3/2}$. The number of bound particles is ${\cal N}\simeq 10^{55} (10^{-5} {\rm eV}/m)^{5/2}$, and for all vector masses $m\lesssim {\rm few}\times 10^7\GeV$ the occupation numbers of the quantum states are so large that this bound object is well-described as a quantum soliton of the semi-classical vector field.
 
 We find that these \emph{dark photon star} or \emph{Proca star} solitons make up an order one fraction of the DM, a remarkable possibility given that the solitons and both macroscopic and intrinsically quantum.
 Moreover they attract surrounding DM, which therefore gets bound around the solitons in a `fuzzy' halo, see Figure~\ref{fig:DM_sub_pic} (left). Additionally, the solitons are produced with excited quasi-normal modes, which are long lived and decay via the emission of dark photon (spherical) waves.
 
 The evolution during inflation also leaves the vector field with overdensities at larger length scales, with a $k^3$ power spectrum. Overdensities on increasingly large scale become non-linear after MRE, and when they collapse they induce a `primordial' structure formation, producing small halos, which we dub \emph{compact halos}. These have mass of order $10^{-11} M_\odot(10^{-5} {\rm eV}/m)^{3/2}$ and will contain some of the  solitons, see Figure~\ref{fig:DM_sub_pic} (centre). The compact halos will become subhalos of the larger galactic halos that are produced when the standard adiabatic almost-scale invariant fluctuations sourced by the inflaton become non-linear. 

We will see that the solitons, fuzzy halos and the densest compact halos are likely to remain undisrupted during the formation and evolution of the Milky Way, and therefore are likely to persist to the present day. The dark matter substructure in Figure~\ref{fig:DM_sub_pic} is an unavoidable property of vector boson DM produced by inflation fluctuations. It could therefore lead to smoking-gun signatures for dark photon dark matter, singling out its particle nature and production mechanism.

This is possible even in the most challenging scenario in which the dark photon has no non-gravitational interactions with the SM. The dark matter substructure can be investigated with gravitational-only probes. Depending on the mass of the substructures, the possibilities include, at minimum, pulsar timing arrays \cite{Baghram:2011is,Ramani:2020hdo,Lee:2020wfn,Lee:2021zqw}, microlensing, \cite{Paczynski:1985jf,Griest:1990vu,Zurek:2006sy,Fairbairn:2017dmf,Fairbairn:2017sil,Croon:2020ouk,Croon:2020wpr,Fujikura:2021omw}  photometric microlensing \cite{Dai:2019lud,Blinov:2019jqc,Arvanitaki:2019rax,Blinov:2021axd}, and extra-galactic strong gravitational lensing \cite{Arvanitaki:2019rax}. In particular, we will see that the size and mass of the solitons and compact halos are in one-to-one correspondence with the dark photon mass. As a result, experimental evidence of such DM substructure will allow us to infer the mass of the dark photon. This would lead to a prediction of the Hubble scale during inflation, which -- if confirmed e.g. by observation of tensor modes in the cosmic microwave background -- would be compelling evidence for this type of dark matter.

In the presence of direct interactions of the massive vector with the SM, e.g. kinetic mixing with hypercharge and thus the SM $Z$-boson and photon \cite{Holdom:1985ag,Babu:1996vt,Dienes:1996zr,Babu:1997st,Abel:2008ai}, the substructure leads to a plethora of signatures in both direct and indirect detection experiments. Solitons (and the fuzzy halos around them) have typical energy density of the order of the Universe's density at MRE, many orders of magnitude larger than the local DM density in the vicinity of the Earth today. Similarly, compact halos have average density of a few orders of magnitude larger than the local density. We will see that solitons can encounter the Earth and other astrophysical objects frequently. These will lead to significant changes to direct detection prospects and new astrophysical and cosmological signals. In a companion paper we discuss the resulting detection and observational signals in detail.\footnote{Although not the focus of our present work, we note wave-like dark matter can also have other interesting structure  \cite{Hui:2020hbq,Lisanti:2021vij}.}

Additionally, the primordial perturbations of a vector produced from inflationary fluctuations have surprisingly similar qualitative features to those in other new physics scenarios. For example, axions in the post-inflationary scenario have qualitatively similar primordial perturbations, although the dynamics that produces them is totally different. 
There are also other model-dependent mechanisms that could lead to a relic abundance of vector dark matter \cite{Agrawal:2018vin,Co:2018lka,Dror:2018pdh,Bastero-Gil:2018uel,Machado:2018nqk,Long:2019lwl,Bastero-Gil:2021wsf} (in the same way as there are other production mechanisms besides misalignment for scalars). These might also lead to similar initial conditions, e.g. if the vector is produced via parametric resonance or topological defects. Therefore, although in this paper we focus on a particularly minimal and predictive theory, we expect that our approach and results may be at least qualitatively applicable to a wide range of scenarios.

The paper is structured as follows: In Section~\ref{s:initial} we describe the production of a vector boson during inflation and calculate the power spectrum of primordial density inhomogeneities during radiation domination, following \cite{Graham_2016} (see also \cite{Ema:2019yrd,Ahmed:2020fhc,kolb2021completely} for related work). A reader familiar with \cite{Graham_2016} or solely interested in the later development of small-scale structure and solitons may safely skim rapidly through this Section, the primary results being eqs.~\eqref{eq:PAt1} and \eqref{eq:lambdastar} along with the precise numerically-derived spectrum shown in Figure~\ref{fig:Pspectra} (left panel).  In Section~\ref{section:SPsection} we first discuss the post-inflation dynamics of the fluctuations and the importance of the `quantum pressure' term arising from the Heisenberg uncertainty principle. We then highlight aspects of the physics of the vector solitons that are particularly important for phenomenology, before discussing in detail the non-perturbative evolution of inhomogeneities around the time of MRE and the formation of solitons from their collapse. In Section~\ref{s:compact_halos} we describe the `primordial' structure formation at larger scales and the compact halos. In Section~\ref{s:late_time} we discuss the survival of the vector dark matter substructure until today and the collision rate of solitons with the Earth. In Section~\ref{sec:Conclusion} we summarise our results, describe improvements and extensions, and propose future directions, some of which will be covered in our companion paper which focuses on the potential observational and experimental implications of the solitons and their fuzzy halos. Appendices provide details on the initial conditions from inflation, analytic analysis of the evolution of overdensities, our approach to numerically solving the Schr\"odinger-Poisson equations, analytic treatment of the soliton and compact halo mass distributions, and finally an extensive discussion of the survival of solitons, fuzzy halos and compact halos to the present day.

\section{Initial Conditions from Inflation} \label{s:initial}

Consider a vector boson $A^\mu$ (we will also  use the name `dark photon' interchangeably) with field strength $F_{\mu\nu}%=\partial_\mu A_\nu-\partial_\nu A_\mu
$, described by the action
\begin{equation}\label{eq:lagrangianA}
	S=\int dt d^3x \sqrt{-g}\left[-\frac14g^{\mu\rho}g^{\nu\sigma}F_{\mu\nu}F_{\rho\sigma}+\frac12m^2g^{\mu\nu}A_\mu A_\nu\right] , \
\end{equation}
with metric $ds^2\equiv g_{\mu\nu}dx^\mu dx^\nu$. We remain agnostic about the dynamics giving rise to the mass as long 
as $m$ remains constant during and after the inflationary epoch,
and that no other light fields significantly coupling to $A^\mu$ are present. All that matters in the following is that the action in eq.~\eqref{eq:lagrangianA} describes the vector field during inflation and in the subsequent evolution of the Universe.\footnote{Swampland conditions might require $m\gtrsim 0.3~\eV$ for such an effective theory to be embedded in a UV completion \cite{Reece:2018zvv}, however there are constructions that claim to evade these limits \cite{Craig:2018yld}. Moreover, we will see that the case $m\gtrsim \eV$ leads to interesting phenomenology and experimental signals.} 
As we will see, efficient inflationary production requires $m\ll H_I$.\footnote{If the mass $m$ is produced by a dark Higgs mechanism, our analysis applies if the mass of the dark Higgs is much larger than the inflationary Hubble scale, so it can be neglected during inflation.} For the purpose of determining the relic abundance it is sufficient to consider the homogeneous background FRW metric $ds^2=-dt^2+a^2(t)d\vec{x}^2$. 

The three propagating degrees of freedom of the vector field are $\vec{A}\equiv A_i$, while the component $A_0$ does not have a kinetic term in eq.~\eqref{eq:lagrangianA} and corresponds to an auxiliary field. We can eliminate $A_0$ % from $\vec{A}$
by writing eq.~\eqref{eq:lagrangianA} in terms of the Fourier modes $\tilde{A}^\mu =\int %\frac{d^3k}{(2\pi)^3}
d^3x\exp(-i\vec{k}\cdot\vec{x})A^\mu$, where $k$ is the comoving momentum (we will drop the tilde in the following). The equations of motion of $A^0$ become algebraic, can be solved explicitly, and their solution plugged back into eq.~\eqref{eq:lagrangianA} to get rid of $A^0$. This leads to $S=S_T+S_L$ with
%\begin{equation}
\begin{align}\label{eq:lagrangianLT1}
	S_T = &\int  \frac{a^3 d^3k \, dt}{(2\pi)^3}  \frac{1}{2a^2} \left[|\partial_t\vec{A}_T|^2-\left(\frac{k^2}{a^2}+m^2\right)|\vec{A}_T|^2\right], \\\label{eq:lagrangianLT2}
	S_L = &\int  \frac{a^3 d^3k \, dt}{(2\pi)^3}  \frac{1}{2a^2} \left[\frac{a^2m^2}{k^2+a^2m^2}|\partial_tA_L|^2-m^2|A_L|^2\right],
\end{align}
where we have decomposed $\vec{A}$ in terms of the longitudinal and transverse modes $A_L$ and $\vec{A}_T$, defined by $\vec{k}\cdot\vec{A}=k A_L$ and $\vec{k}\cdot\vec{A}_T=0$.

The actions $S_T$ and $S_L$ describe transverse and longitudinal modes, which are decoupled from each other given the FRW form of the metric. Eqs.~\eqref{eq:lagrangianLT1} and \eqref{eq:lagrangianLT2} hold both during inflation -- when the Hubble parameter $H\equiv \dot{a}/a=H_I$ is approximately constant -- and subsequently in the early Universe (although we will see that the transverse and longitudinal modes become coupled together around MRE).

We consider the evolution of a generic mode, starting from when it is subhorizon during inflation ($k/a>H_I$) in the vacuum state to when it becomes nonrelativistic ($k/a<m$) and subhorizon, in the radiation dominated era. 
First note that all the modes that start subhorizon during inflation ($k/a>H_I$) are relativistic during inflation, since $k/a>H_I\gg m$, and remain relativistic at horizon exit (i.e. when $a=k/H_I$). 
Using conformal time $d\eta= d t/a$, the action for the transverse modes reads 
$S_T= (2\pi)^{-3}\int d^3kd\eta \frac{1}{2} \left
(|\partial_\eta\vec{A}_T|^2-(k^2+a^2 m^2)|\vec{A}_T|^2\right
)$. Since $m$ is negligible for relativistic modes, $S_T$ is time-translation invariant (in fact, conformally invariant). The vacuum state of the transverse modes therefore does not change, and they are not produced. We therefore set $\vec{A}_T=0$ in the reminder of this Section.\footnote{This is a standard result for a massless vector, and in the relativistic limit the transverse components act like a massless vector by the Goldstone Equivalence theorem.}

On the other hand, in the relativistic limit the action for the longitudinal modes reduces  to that of a free real scalar $\varphi\equiv (m/k)A_L$, i.e. $S_L=\int a^3 d^3x dt \frac12[(\partial_t\varphi)^2-|\nabla\varphi|^2/a^2]$, where we Fourier transformed back to coordinate space. It is well known that the vacuum of this theory is time-dependent \cite{osti_4802789,Bunch:1978yq}. Modes that are in the vacuum at early times get populated -- after they exit the horizon -- with a Gaussian power spectrum $\mathcal{P}_\varphi=(H_I/2\pi)^2$, where we defined the power spectrum $\mathcal{P}_X$ of a generic field~$X$ as %$\langle \phi(k)\phi(k')\rangle=$
\begin{equation}\label{eq:PX}
	\langle X^*(t,\vec{k})X(t,\vec{k}')\rangle\equiv (2\pi)^3\delta^3(\vec{k}-\vec{k}')\frac{2\pi^2}{k^3}\mathcal{P}_X(t,k) \ .
\end{equation}
This implies that 
$\mathcal{P}_{A_L}=(k H_I/2\pi m)^2$ at horizon exit, $a=k/H_I$. The expectation value of the energy density $\rho\equiv  T^{00}$ reads
\begin{equation}\label{eq:rho}
	\langle\rho\rangle=\int d\log k\, \frac{1}{2a^2}\left[\frac{a^2m^2}{k^2+a^2m^2}\mathcal{P}_{\partial_t A_L}+m^2\mathcal{P}_{A_L}\right] ~,
\end{equation}
and the energy density spectrum at horizon exit is scale invariant,  $\partial \rho/\partial \log k\simeq H_I^4/(2\pi)^2$. This is the well known result for a scalar field, and indeed the longitudinal component of the vector in the relativistic limit reproduces a massless scalar by the Goldstone Equivalence theorem.

After inflation the modes evolve classically, following the equations of motion of eq.~\eqref{eq:lagrangianLT2},
\begin{equation}\label{eq:eomAL}
	\left[\partial_t^2+\frac{3k^2+a^2m^2}{k^2+a^2m^2}H\partial_t+\frac{k^2}{a^2}+m^2\right]A_L=0 \ .
\end{equation}

We can calculate the evolution of each mode with initial condition $A_L=A_{L,0}$ and $\dot{A}_L\simeq 0$ at $a=k/H_I$, until it reenters the horizon and becomes nonrelativistic. Since eq.~\eqref{eq:eomAL} is linear, $A_L$ will still have a Gaussian distribution during such %evolution
evolution, with a power spectrum at a generic time given by
\begin{equation}\label{eq:PAt}
	\mathcal{P}_{A_L}(t,k)=\left(\frac{k H_I}{2\pi m}\right)^2\left(\frac{A_L(t,k)}{A_{L,0}}\right)^2 \, ,
\end{equation}
where we used eq.~\eqref{eq:PX}.\footnote{Note that the actual value of $A_{L,0}$ is not needed for $\mathcal{P}_A$, so practically one can solve eq.~\eqref{eq:PAt} with $A_{L,0}=1$.} Unfortunately, the solution for $A_L$ cannot be evaluated analytically for a generic $k$. However, it is possible to understand the behaviour of $A_L(t,k)/A_{L,0}$ analytically. Here we summarise the main results, leaving complete derivations to  Appendix~\ref{app:initial_conditions}.

All the modes of interest will eventually  become subhorizon and nonrelativistic, and it is convenient to classify them into two classes: (1) Those that become nonrelativistic while still superhorizon (`low frequency') and (2) Those that reenter the horizon while relativistic (`high frequency'), and become nonrelativistic only afterwards. We define the mode that becomes nonrelativistic exactly when it reenters the horizon $k_\star/a_\star\equiv m =H(a_\star) $, so the low (high) frequency modes satisfy $k<k_\star$ ($k>k_\star$) respectively.

While relativistic and superhorizon $\rho\propto a^{-2}$ for all modes. Modes with $k<k_\star$ are suppressed because they become nonrelativistic while superhorizon and subsequently their energy density decreases as $\rho\simeq m^2A_L^2/a^2\propto a^{-2}$ until they enter the horizon. This is the crucial difference compared to a scalar field, for which $\rho$ is frozen for nonrelativistic superhorizon modes. The difference is due to the form of the mass term,  which controls the energy density in such modes: 
for a scalar, $\frac12m^2\varphi^2\propto{\rm const}$, while $\frac12m^2g^{ij}A_iA_j\propto a^{-2}$ for a vector. Meanwhile, the modes with $k>k_\star$ are suppressed because they enter the horizon while still relativistic. They  have $\rho\propto a^{-4}$ after they enter the horizon but before they become nonrelativistic and  $\rho \propto a^{-3}$ subsequently. 

\begin{figure}[t]
	\begin{center}
		\includegraphics[width=0.51\textwidth]{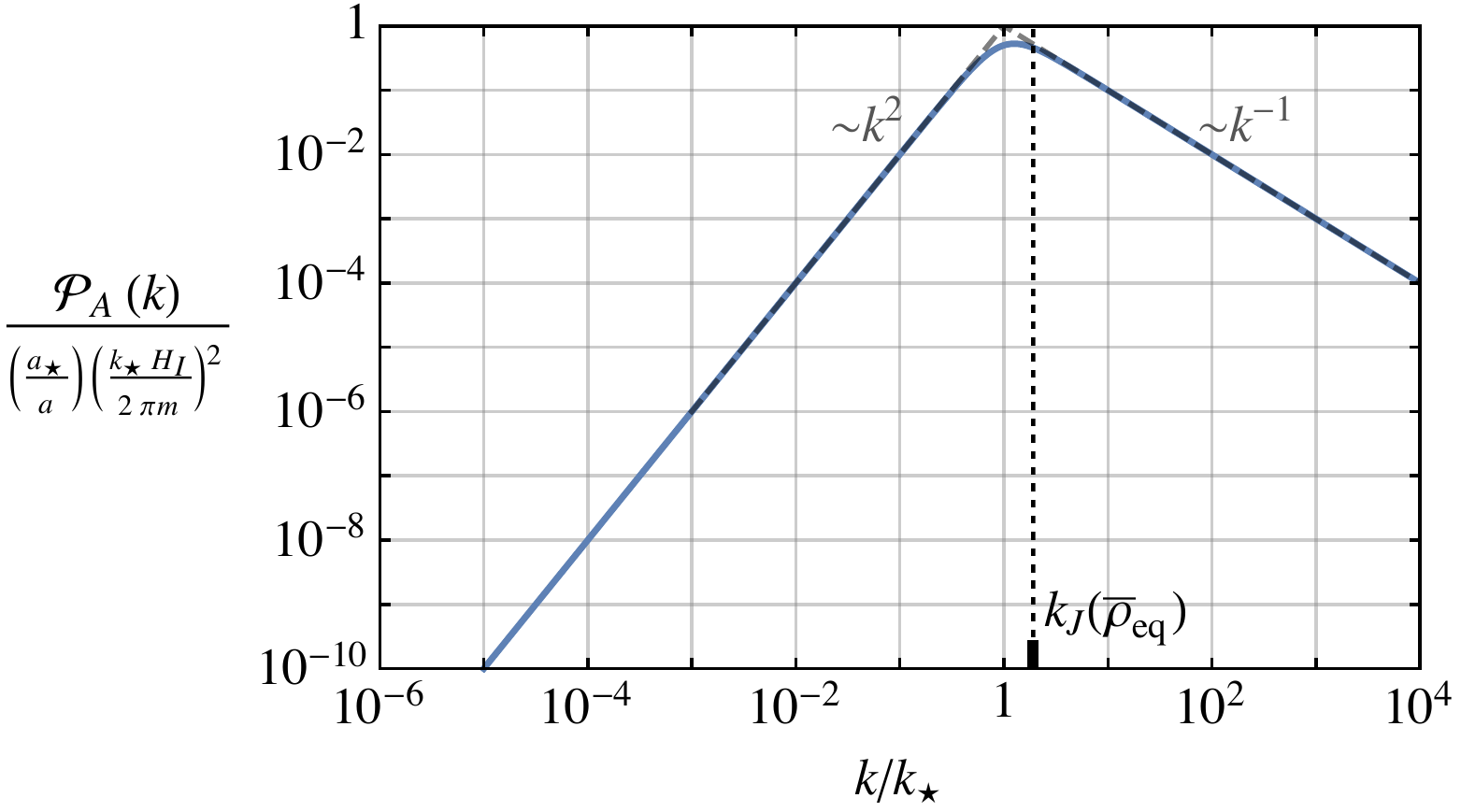}
		\includegraphics[width=0.46\textwidth]{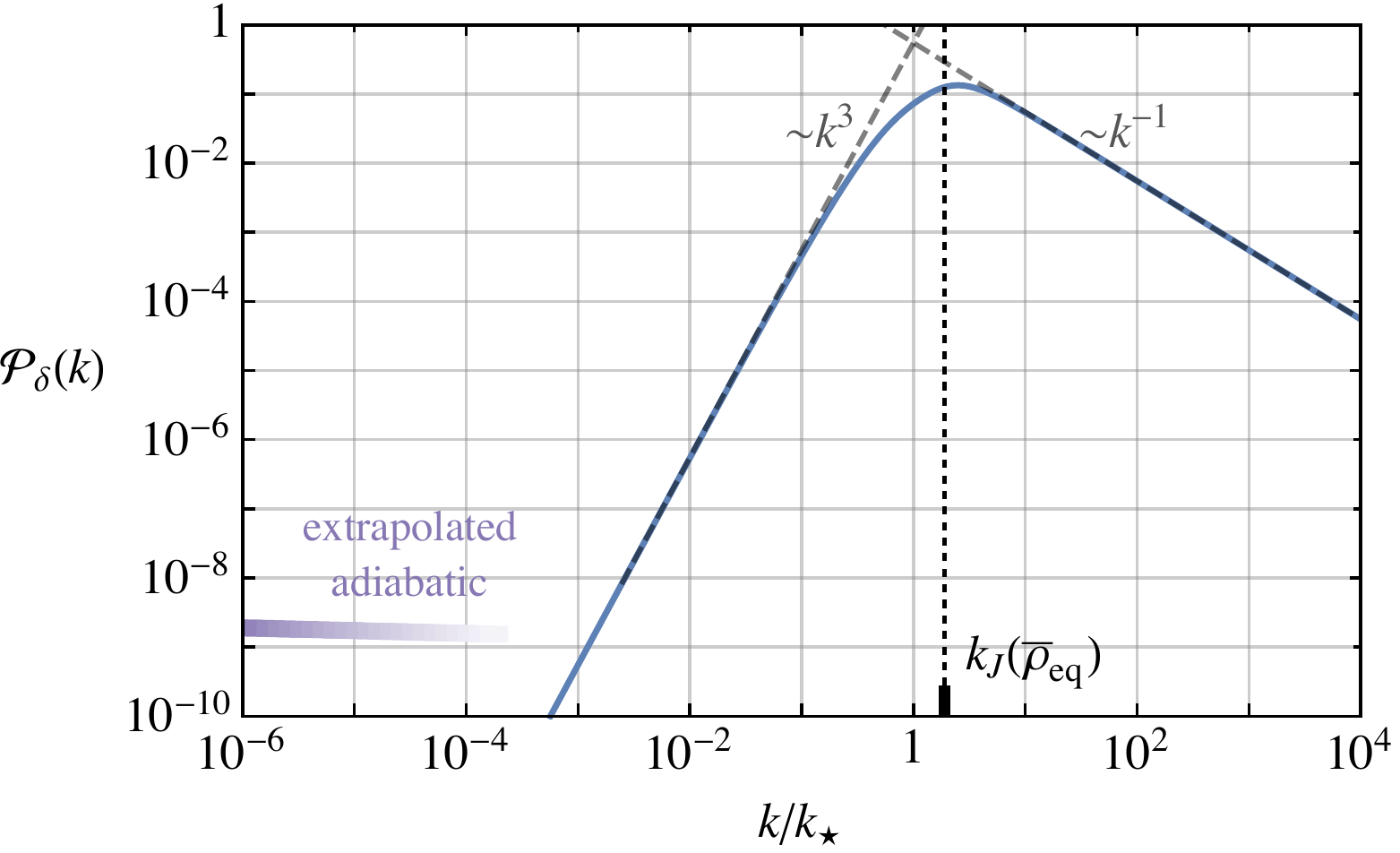}
	\end{center}
	\caption{{\bf \emph{Left:}}  The  power spectrum of the longitudinal vector field component as is automatically produced by inflationary fluctuations, during radiation domination and after the modes have become subhorizon and nonrelativistic. The spectrum is peaked at $k_\star$, corresponding to the momentum equal to the Hubble parameter $H$ when $H=m$. 
		{\bf \emph{Right:}} The  power spectrum $\mathcal{P}_\delta$ of the (non-Gaussian) overdensity field. Importantly for later soliton formation, at matter radiation equality the quantum Jeans momentum $k_J(\bar{\rho})$ associated with the mean DM density (defined in Section~\ref{ss:SP_analytic}) coincides with $k_\star$ up to an order one factor, regardless of the dark photon mass.
		\label{fig:Pspectra}} 
\end{figure}

The result of this is that the spectrum is peaked at momentum $k_\star$. Modes with $k\ll k_\star$ are suppressed since they stayed in the superhorizon nonrelativistic regime the longest, and those with $k\gg k_\star$ are suppressed because they underwent subhorizon relativistic redshift before becoming nonrelativistic, with larger $k$ suppressed more since they were subhorizon and relativistic for longer. The power spectrum of $A_L$ is, to a very good approximation, given by  
\begin{equation}\label{eq:PAt1}
	\mathcal{P}_{A_L}(t,k)\simeq \left(\frac{ k_\star H_I}{2\pi m}\right)^2\left(\frac{a_\star}{a}\right) \frac{(k/k_\star)^2}{1+(k/k_\star)^3}\, .
\end{equation}
The exact form of $\mathcal{P}_{A_L}(t,k)$, plotted in Figure~\ref{fig:Pspectra} (left), can be extracted by solving eq.~\eqref{eq:eomAL} numerically. 
The least suppressed mode, $k_\star$, corresponds to subgalactic scales today: defining $\lambda_\star \equiv 2\pi/k_\star$ 
\begin{equation}\label{eq:lambdastar}
a_0 \lambda_\star = \frac{2\pi a_0}{m a_\star} \simeq 10^{11} \text{km}\left(\frac{10^{-5}\text{eV}}{m}\right)^{1/2}\, ,
\end{equation} where $a_0$ is the FRW scale factor today.  Moreover, the misalignment mechanism is related to the energy density in the zero mode, which gets a huge suppression, and is therefore ineffective.

The energy density of the vector behaves as matter at late times, and forms a component of the DM abundance. Given the peaked form of $\mathcal{P}_{A_L}$, the DM abundance is approximately given by the energy density in modes of momentum $k_\star$, redshifted from horizon exit at $a=k_\star/H_I$ to $a=a_\star$  and then to today %(with $\rho\propto a^{-3}$)
 when the scale factor is $a\equiv a_0$. Therefore,  $\rho(a_0)\simeq H_I^4/(2\pi)^2((k_\star/H_I)/a_\star)^2(a_\star/a_0)^3 \propto H_I^2 m^{1/2}$. A full calculation leads to 
 the relic abundance given in eq.~\eqref{eq:OmegaDM}. 
The Hubble scale during inflation is bounded by the non-observation of tensor modes in the cosmic background,  assuming single field slow roll inflation. The latest data combination from Planck and BICEP2 \cite{Akrami:2018odb}
bounds  $H_I< 6\cdot 10^{13} \GeV$.\footnote{In particular, the bound is on the  tensor-to-scalar ratio $r<0.056$, which is related to the Hubble scale during inflation by $H_I= 8\cdot10^{13} \sqrt{r/0.1} \GeV$.} As a result, the vector can account for the full dark matter abundance if $m\gtrsim 10^{-5}$ eV.

\begin{figure}[t]
	\begin{center}
		\includegraphics[width=0.52\textwidth]{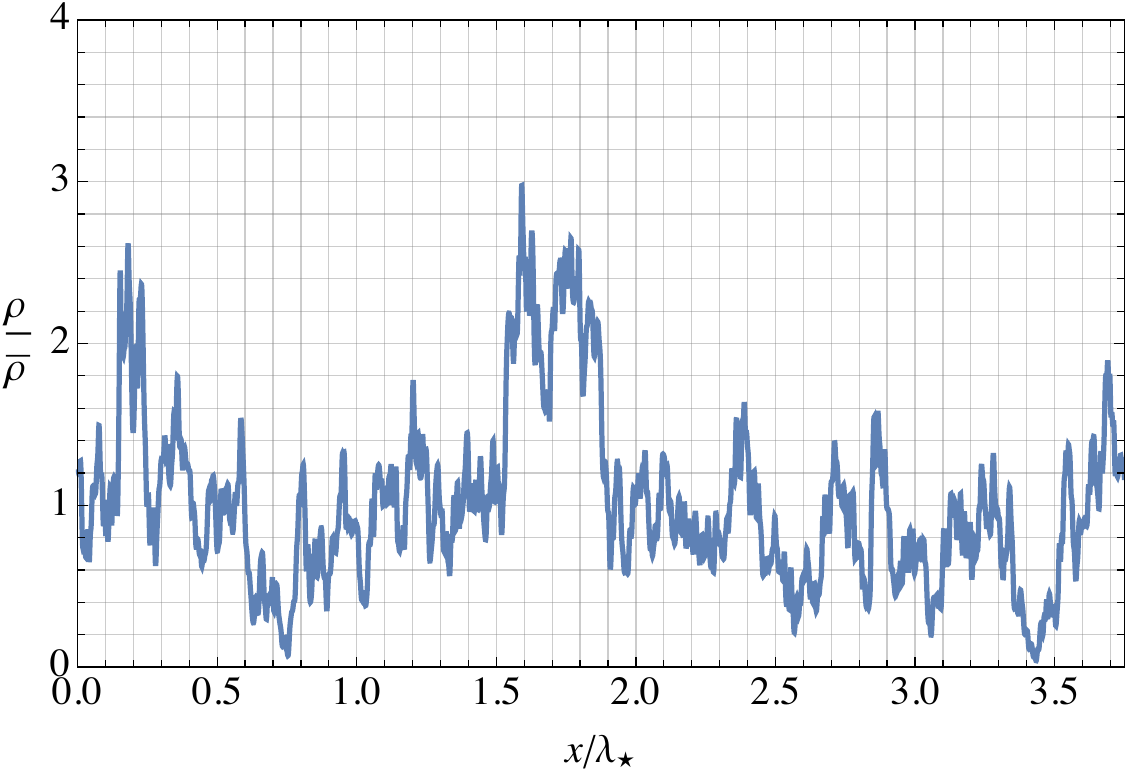}
	\end{center}
	\caption{A one-dimensional slice of the vector field energy density $\rho$, during radiation domination, relative to its mean density $\bar{\rho}$. The typical fluctuations are on scales $\lambda_\star=2\pi/k_{\star}$, eq.~\eqref{eq:lambdastar}, %a_0 \simeq10^{11} \text{km}(10^{-5}\text{eV}/m)^{1/2}$,
	corresponding to the $\mathcal{P}_{\delta}$ peak in Figure~\ref{fig:Pspectra}. There are also small overdensities at scales larger than $\lambda_\star$ (not distinguishable in the plot), which correspond to the $(k/k_\star)^3$ part of $\mathcal{P}_\delta$, and considerable fluctuations on smaller scales because of the slow $(k_\star/k)$ fall-off
	of the power spectrum for $k\gg k_\star$.
		\label{fig:1dslice}} 
\end{figure}

The  structure in the power spectrum of $A_L$ leads to small-scale overdensities in the energy density field $\rho=(\dot{A}_L^2+m^2A_L^2)/2$. During radiation domination and once all the relevant modes have become non-relativistic, the properties of typical fluctuations remain constant. In Figure~\ref{fig:1dslice} we plot a section of the vector's energy density over a line at this stage, normalised to its average energy density $\bar{\rho}\equiv \langle\rho\rangle$. There are obvious ${\cal O}(1)$ fluctuations in the energy density (the peaks have $\rho/\bar{\rho}\simeq 2\div4$), with spatial size of order $\lambda_\star$.

It is convenient to introduce the overdensity field 
\begin{equation}\label{eq:deltax}
	\delta(x)\equiv\frac{\rho(x)-\bar{\rho}}{\bar{\rho}} \, .
\end{equation}
The distribution of inhomogeneties is encoded in the power spectrum of overdensities, $\mathcal{P}_{\delta}$, defined from $\delta(x)$ as in eq.~\eqref{eq:PX}.  
Once all the relevant modes are non-relativistic, $\mathcal{P}_\delta$ is constant during radiation domination. Using the fact that $\partial_t A_L$ and $A_L$ are independent Gaussian fields, one finds 
\begin{equation}\label{eq:Pdelta}
	\mathcal{P}_{\delta}(t,k) \simeq \frac{\sqrt{3} (k/k_\star)^3}{\pi  \left((k/k_\star)^{3/2}+1\right)^{8/3}} \ ,
\end{equation}
which is a useful analytic approximation that captures  the asymptotic limits $k/k_\star\to\{0,\infty\}$ exactly (see Appendix~\ref{app:initial_conditions} and \cite{Graham_2016} for full expressions). In Figure~\ref{fig:Pspectra} (right) we plot $\mathcal{P}_{\delta}$, which as expected is peaked at $k/k_\star\simeq1$, and decreases as $(k/k_\star)^3$ and $(k_\star/k)$ at small and large $k$, respectively.  Since $\delta(x)$ does not have a Gaussian distribution (indeed, it is asymmetric around $\delta=0$) it is not fully described by $\mathcal{P}_{\delta}$.\footnote{Note that  $\rho(x)\sim(\partial A(x))^2+A(x)^2$ has local (quadratic) non-Gaussianities, since $A$ and $\partial A$ are Gaussian variables. Thinking of $\langle\rho(x)\rangle$ as a constant, appropriate in the large volume limit, $\delta(x)$ has the same property.} The power spectrum however still provides useful information about the variance of the field and the magnitude of the overdensities. Note that at length scales much larger than $\lambda_\star = 2\pi/k_\star$, $\delta(x)$ is Gaussian and can be fully reconstructed from $\mathcal{P}_\delta$ alone.

The fluctuations in the vector's energy density are isocurvature perturbations, since they are induced only in the vector during inflation. As mentioned in the Introduction, these fluctuations are only allowed to be ${\cal O}(1)$ because perturbations at much larger scales, which are strongly constrained by observations, are automatically suppressed thanks to the $k^3$ behaviour.  This is in contrast to a scalar field, for which the power spectrum from inflationary fluctuations is flat at $k<k_\star$, and order one fluctuations are completely excluded unless the scalar is a tiny fraction of the total DM. The smallest scales at which the power spectrum has been observed are roughly $k_{\rm obs}=7 {\rm Mpc}^{-1}$ from Lyman-alpha \cite{Croft:2000hs}, so $k_{\rm obs}/k_\star\simeq 10^{-11} \left(\eV/m \right)^{1/2}$, and the observed modes are far off the left of the plot in Figure~\ref{fig:Pspectra} (right) for all relevant dark photon masses. 

Inflation also sources perturbations of the inflaton, which are metric perturbations with a change of gauge, i.e. $ds^2=-(1-2\Phi)dt^2+(1+2\Phi)a^2d\vec{x}^2$. The gravitational potential $\Phi$ has small differences in different patches after inflation. These lead to the same relative perturbations in all form of energy, i.e. adiabatic perturbations, including in the  vector overdensity field $\delta(x)$. Its power spectrum $\mathcal{P}_\delta$ therefore automatically acquires also the almost-scale-invariant contribution, as shown by the purple line in Figure~\ref{fig:Pspectra} (right), as is necessary to be consistent with observations.

We have considered a vector with only the action of eq.~\eqref{eq:lagrangianA}, in which case the relic density discussed in this Section is an unavoidable contribution to the dark matter abundance. The situation is more complicated if the vector has self-interactions, couples to other particles or has a non-minimal coupling to gravity~\cite{Arias:2012az}. Such possibilities do not necessarily affect the production during inflation, and indeed it would be surprising if fluctuations of order $H_I$ were prevented. However, the subsequent evolution will be affected in some theories, which could alter the relic abundance. For instance this is the case if there are `dark electrons' \cite{Arvanitaki:2021qlj}.\footnote{If an interaction leads to an operator of the form $F^4/\Lambda^4$ in the effective theory at $H_\star$, when the most important modes re-enter the horizon, the subsequent evolution is unaffected provided $\Lambda \gg (m H_I)^{1/2} \simeq (m/\eV)^{3/2} \MeV$ where the last equality holds if $\Omega_A = \Omega_{\rm DM}$. We leave the effects on the evolution of the superhorizon modes for future work.} A key direction for future work is to systematically study the impact of different possible interactions.

\section{Collapse of Inhomogeneities: Vector Solitons} \label{section:SPsection}

 During radiation domination the vector field evolves freely, i.e. as in the absence of a gravitational potential.\footnote{This is the reason we neglected $\Phi$ in the previous Section.} The gravitational potential becomes relevant at around MRE. If the vector constitutes a sizeable fraction of the dark matter abundance, the large overdensities corresponding to the peak of $\mathcal{P}_\delta$ at $k_\star$, start evolving nonlinearly under the effect of the gravitational interactions at around this time (more precisely, at $a\simeq a_{\rm eq}/\delta$ where $a_{\rm eq}$ is the scale factor at MRE, see Appendix~\ref{aa:SPMRE} for further details). In what follows we will assume that the vector makes up all the DM.%

 Since they are already ${\cal O}(1)$, as soon as the gravitational potential becomes important the overdensities cannot be studied using the standard perturbative treatment of the density field. They are expected to clump into bound objects, and -- as we will see next -- this happens in the regime where the fuzzy dark matter properties of the boson, i.e.  quantum pressure, are important.

\subsection{Dynamics of fluctuations and quantum pressure}\label{ss:SP_analytic}

The modes relevant for the DM abundance are nonrelativistic at around MRE. It is therefore convenient to rewrite the equations of motion (EoM) of the Lagrangian in eq.~\eqref{eq:lagrangianA} and the component~00 of the Einstein equations in their nonrelativistic form. To do so, we work in terms of $\psi_i$ defined by
\begin{equation}\label{eq:psidef}
    A_i\equiv\frac{1}{\sqrt{2 m^2 a^3}}(\psi_i e^{-imt}+{\rm c.c.})
\end{equation}
in the limit where $\psi$ is slowly evolving so $\dot{\psi}_i\ll m\psi_i$ and $\ddot{\psi}_i\ll m^2\psi_i$. In terms of $\psi_i$ the EoM become the Schr\"odinger--Poisson (SP) system \cite{Adshead:2021kvl}
\begin{align}\label{eq:sp1a}
	\left(i\partial_t+\frac{\nabla^2}{2m}-m\Phi\right)\psi_i&=0 ~,\\
	\nabla^2\Phi=\frac{4\pi G}{a}\sum_{i}&\left(|\psi_i|^2-\langle|\psi_i|^2\rangle\right) ~.\label{eq:sp1b}
\end{align}

Apart from involving the three propagating components of $\vec{A}$, these equations have exactly the same form as those for a nonrelativistic scalar field. (The fourth vector field component, $A^0$, is non-dynamical and may be recovered from the Lorenz-Proca constraint $\nabla_\mu A^\mu=0$ which follows as a consistency relation from the EoM.)  At around MRE when $\Phi$ is non-negligible, the EoM are nonlinear and couple together all the components of the vector through the gravitational potential. In particular the longitudinal and transverse modes are coupled and no longer evolve independently. Consequently, although initially vanishing, the transverse mode is sourced by the longitudinal mode via $\Phi$.\footnote{Note that the energy overdensity in baryonic matter $\rho_{b}-\langle\rho_{b}\rangle$ should also appear on the right hand side of eq.~\eqref{eq:sp1b}, as it contributes to the total stress energy tensor. 
However, before recombination (in particular at MRE) baryons are strongly coupled to photons, and their evolution is dominated by interactions with photons, instead of the gradient of $\Phi$. As a result, at very subhorizon scales baryons are practically homogeneous, and $\rho_{b}-\langle\rho_{b}\rangle\simeq 0$. Therefore, effectively $\Phi$ is only sourced by DM, and baryons act as a background that only drives the Universe's expansion.}

For a full treatment, eqs.~\eqref{eq:sp1a} and~\eqref{eq:sp1b} need to be solved numerically with initial conditions in eq.~\eqref{eq:PAt1}. However we can qualitatively understand the evolution by doing the Magdelung transformation, i.e. writing eqs.~\eqref{eq:sp1a} and~\eqref{eq:sp1b} in terms of density and velocity fields, $\rho_i$ and $\vec{v}_i$, defined by $\psi_i=\sqrt{a^3\rho_i}e^{i\theta_i}$ with $\vec{v}_i\equiv (am)^{-1}\nabla\theta_i$. The imaginary and real parts of the Schr\"odinger equation become the continuity and Euler equations of a three-component perfect fluid with local density $\rho_i=|\psi_i|^2/a^3$ and velocity $\vec{v}_i$, i.e.
\begin{align}
	\partial_t\rho_i +3H\rho_i+a^{-1}\nabla\cdot(\rho_i \vec{v}_i)=& \ 0  \label{eq:cont}\\
	\partial_t\vec{v}_i+H\vec{v}_i+a^{-1}(\vec{v}_i\cdot\nabla)\vec{v}_i=&-a^{-1}(\nabla\Phi+\nabla\Phi_{Q i}) \label{eq:Euler}\\
	\label{eq:Poisson}\nabla^2\Phi=&\ 4\pi G a^2(\rho-%\langle \rho\rangle)
	\bar{\rho}) \ ,
\end{align}
where $\rho=\sum_i\rho_i$ and $\bar\rho\equiv\langle\rho\rangle$ is the average energy density of the vector field, and we defined
\begin{equation}
	\Phi_{Qi}\equiv-\frac{\hbar^2}{2a^2m^2}\frac{\nabla^2\sqrt{\rho_i}}{\sqrt{\rho_i}} \ ,
\end{equation}
where for clarity we have restored the factors of $\hbar$ in this one expression. Their appearance is due to the fact that the mass $m$ of a particle is independent of $\hbar$ while, due to the $\hbar$ that appears in the Planck-Einstein-de~Broglie relation, the mass parameter that appears in the field action, eq.~\eqref{eq:lagrangianA}, is really $m/\hbar$. From now on we return to $\hbar=c=1$ units.
From eq.~\eqref{eq:Euler}, the fluid is subject to the gravitational potential $\Phi$ and to the `quantum pressure' potential $\Phi_Q$. The gravitational potential tends to increase the overdensities, while the quantum pressure tends to make overdensities fluctuate. This can be seen from the fact that the quantum pressure has the opposite sign as the gravitational potential, or more rigorously in the perturbative treatment of small overdensities, which we summarise in Appendix~\ref{app:over}. 

The importance of $\Phi_Q$ with respect to $\Phi$ can be understood by comparing the last two terms of eq.~\eqref{eq:Euler}, i.e. if $\nabla\Phi\ll\nabla\Phi_Q$ the quantum pressure dominates. Taking the divergence of this relation and using eq.~\eqref{eq:Poisson}
 we get $-8\pi G\rho m^2a^4\ll \nabla^2(\nabla^2\sqrt{\rho}/\sqrt{\rho})$.   
Going to Fourier space, for comoving momenta $k$ much larger than the comoving `quantum' Jeans momentum associated with physical density $\rho$
\begin{equation}\label{eq:quantumJeans}
	k_J(\rho)\equiv a(16\pi G\rho m^2 )^{1/4} \, ,
\end{equation}
$\Phi_Q$ is much more important than $\Phi$, and in the opposite limit it is irrelevant. This implies that overdensities at comoving length scales smaller than $\lambda_J\equiv 2\pi/k_J$ are prevented from collapsing, while those at length scales much larger than $\lambda_J$ evolve as in the absence of quantum pressure. Note that this is not the conventional Jeans scale, which is proportional to the inverse of the sound speed %$c_s^2=\delta P/\delta \rho$
and in this context is infinity.
 
\subsection{The remarkable coincidence} \label{ss:kstarkJ}

Crucially, if  the vector boson makes up an ${\cal O}(1)$ fraction of the dark matter,  at the time of MRE the quantum Jeans scale $k_J(\bar{\rho})$ (corresponding to the average density) is parametrically close to the typical scale of spatial fluctuations $k_\star$, independently of $m$. 
In particular, the ratio between $k_J(\bar{\rho})$ and $k_\star$ at MRE is
\begin{equation}\label{eq:kJoks_2}
	\left.\frac{k_J(\bar{\rho})}{k_\star}\right|_{a=a_{\rm eq}}=
		\frac{(16\pi G\bar{\rho}_{\rm eq} m^2)^{1/4}}{m (a_\star/a_{\rm eq})}=g_R\left(12\frac{\bar{\rho}_{\rm eq}}{\rho^{\rm tot}_{\rm eq}}\right)^{1/4}=g_R\left(6 \frac{\Omega_{A}}{\Omega_{\rm M}} \right)^{1/4}, \, %\simeq 1.9 \left( \frac{\Omega_{A}}{\Omega_{\rm DM}} \right)^{1/4}\ ,
\end{equation}
where $\Omega_{\rm M}$ is the matter energy density normalised to the critical density, and $\rho^{\rm tot}_{\rm eq}$ is the total matter energy density at MRE. 
In eq.~\eqref{eq:kJoks_2} we defined $g_R\equiv (g_{{\rm eq},\epsilon}/g_{\star,\epsilon})^{1/4} (g_{\star,s}/g_{{\rm eq},s})^{1/3}$, where $g_{s}$, $g_{\epsilon}$ denote the effective number of relativistic degrees of freedom for entropy and energy, and as usual ${\rm eq}$ and $\star$ denote quantities at MRE and when $H=m$. This factor accounts for the change in the number of degrees of freedom between $H_\star$ and MRE, which affects the value of $k_\star$. 
The cancellation of $m$, $G$ and $\rho_{\rm eq}$ in eq.~\eqref{eq:kJoks_2} occurs because, up to numerical factors, $a_\star/a_{\rm eq} \simeq T_{\rm eq}/T_\star \simeq \left(G\rho_{\rm eq}^{\rm tot}/m^2\right)^{1/4}$ (making the excellent approximation that only radiation contributes to the energy density at $T_\star$).\footnote{More explicitly, $a_\star/a_{\rm eq}=g_R^{-1} (4 \pi G \rho^{\rm tot}_{\rm eq}/3 m^2 )^{1/4}$. We also note for future use that $a_\star/a_{\rm eq}$ can be rewritten as $a_\star/a_{\rm eq}= g_R^{-1} (H_{\rm eq}/(\sqrt{2}m))^{1/2}$, using $H_{\rm eq}^2=8\pi G\rho^{\rm tot}_{\rm eq}/3$.}

For dark photon masses of interest, $m\gtrsim 10^{-5}$ eV,  $T_\star \gtrsim 200$ GeV, so we set $g_{\star,s}$ and $g_{\star,\epsilon}$ to their Standard Model high temperature values, which gives $g_R \simeq 1.27$.\footnote{This is a reasonable assumption provided there are not $\gg 10^2$ new degrees of freedom close to the TeV scale.} 
Therefore, if the vector comprises the full dark matter abundance,
\begin{equation}\label{eq:kJoks}
	\left.\frac{k_J(\bar{\rho})}{k_\star}\right|_{a=a_{\rm eq}} \simeq 1.9\ .
\end{equation}
Eq.~\eqref{eq:kJoks} means that the quantum pressure term (i.e. the wave-like properties of the boson) cannot be neglected and will affect the evolution and collapse of the ${\cal O}(1)$ overdensities in Figure~\ref{fig:1dslice}, regardless of the value of $m$, and even of $G$. In Figure~\ref{fig:Pspectra} we show the value $k_J^{\rm eq}$ of $k_{J}(\bar{\rho})%|_{a=a_{\rm eq}}
$ at MRE. In fact, the numerical value is close to the peak of $P_\delta$, which is at about $2k_\star$.

\subsection{Dynamics around matter radiation equality}

Eq.~\eqref{eq:kJoks} leads us to conclude that at $a=a_{\rm eq}$\,:
\vspace{-2mm}
\begin{itemize}[leftmargin=0.2in] \setlength\itemsep{0.15em}
	\item
For modes with $k\lesssim k_J^{\rm eq}
\simeq k_\star$ the dynamics is as in the absence of quantum pressure. In particular, over cosmological scales -- much larger in length than $\lambda_\star$ -- the quantum pressure  is irrelevant and the field behaves as conventional DM.  This applies to the adiabatic modes and those in the $k^3$ IR tail of $\mathcal{P}_\delta$, see Figure~\ref{fig:Pspectra}.  Since the fluctuations in the $k^3$ tail are small at MRE, at least initially the field at these distances is in the perturbative regime and its evolution can be calculated analytically. Deep in radiation domination such isocurvature fluctuations (i.e. the $k^3$ modes) are frozen, and once in radiation domination they grow linearly. To a good approximation $\delta\propto1+\frac32\frac{a}{a_{\rm eq}}$ and $\mathcal{P}_\delta\propto(1+\frac32\frac{a}{a_{\rm eq}})^2$ (we give further details in Appendix~\ref{app:over}).

\item The dynamics of modes with $k\gtrsim k_J^{\rm eq}$ is dominated by the quantum pressure. As mentioned, some of these modes are already nonlinear, but are prevented from collapsing into bound objects, and instead just oscillate. 

\item Modes $k\simeq k_J^{\rm eq}$ are on the boundary at which quantum pressure is relevant at MRE. Owing to the coincidence between $k\simeq k_J^{\rm eq}$ and $k_\star$, these are at the peak of $\mathcal{P}_\delta$ and there are order one fluctuations on these scales. Such modes oscillate because of quantum pressure before MRE, and -- as we will see next -- collapse around MRE.%, when the two independent factors preventing their collapse (quantum pressure and radiation domination) both pass.\footnote{Similar dynamics can occur for axion-like-particles in the large misalignment regime \cite{Arvanitaki:2019rax}.}

\end{itemize}
\vspace{-1mm}

The fact that the modes with $k\lesssim k_\star$ are linear allows us to (at least initially)  treat the evolution of the %
 modes with $k\gtrsim k_\star$ independently. As time increases the comoving quantum Jeans momentum increases as $k_J(a)=k_J^{\rm eq}(a/a_{\rm eq})^{1/4}$, see eq.~\eqref{eq:quantumJeans}, i.e. the dashed line in Figure~\ref{fig:Pspectra} moves to the right (here and in the following, whenever $k_J$ takes a scale factor $a$ as an argument we mean $k_J(\bar{\rho}(a))$). 
  As a result, the modes at larger comoving momentum, previously prevented from collapsing, will be able to clump and are expected to form compact bound objects. In other words, collapse of the overdensities with $k\gtrsim k_\star$ is prevented until the comoving quantum Jeans length is smaller than the comoving size of the overdensity, when they start being dominated by the gravitational potential instead of the quantum pressure. 
 In particular, given the coincidence in eq.~\eqref{eq:kJoks}, the modes at $k\simeq k_\star$ (where $\mathcal{P}_\delta$ is peaked) collapse into bound objects at around $a\simeq a_{\rm eq}$, when the two independent factors preventing their collapse (quantum pressure and radiation domination) both pass.\footnote{Similar dynamics can occur for axion-like-particles in the large misalignment regime \cite{Arvanitaki:2019rax}.}

Once formed the objects rapidly decouple from the Hubble flow and will be described by a stationary solution of eqs.~\eqref{eq:sp1a} and~\eqref{eq:sp1b} in the absence of expansion.\footnote{Therefore in comoving coordinates the objects' size decreases.} Since the collapse happens at a scale where the quantum pressure is still relevant in the dynamics, we expect that for such solutions the last term in eq.~\eqref{eq:Euler} is of the same order as the others, in particular of $\nabla\Phi$. In fact, as we will see shortly, these bound objects are \emph{solitons}. For these, the gravitational potential term is fully balanced by the quantum pressure term $(\Phi_Q=-\Phi)$, instead of being balanced by the velocity terms as in a conventional halo. Moreover, we expect that the mass of the objects is parametrically set by the dark matter mass inside a region of the size of the collapsing perturbation, which is given by the quantum Jeans scale at the time of collapse, so a soliton formed at time $a$ has mass

\begin{equation}\label{eq:MJ}
M(a)=c_M M_{J}(a), \qquad \text{with} \qquad M_J(a)\equiv \frac{4\pi}{3} \bar{\rho} a^3\lambda_J^3(a)\propto a^{-3/4} \, ,
\end{equation}
and $c_M$ a dimensionless time-independent ${\cal O}(1)$ coefficient. As time increases, $M_J(a)$ decreases, and solitons with smaller and smaller masses are produced, from the collapse of the smaller scale fluctuations. When integrated over $a>a_{\rm eq}$, this leads to a relic abundance of solitons with a nontrivial mass distribution.

As we will discuss in detail in Section~\ref{s:compact_halos}, as time progresses also the modes with $k\lesssim k_\star$ become nonperturbative and collapse to form halos, which unlike solitons are supported by the velocity term in eq.~\eqref{eq:Euler}. We dub these \emph{compact halos}. 
These are expected to include some of the solitons previously produced. Inside the compact halos the modes with $k\gtrsim k_\star$ can no longer be treated separately, as they are no longer decoupled. However, outside the compact halos the field at $k\ll k_\star$ is still perturbative and solitons are still expected to form following eq.~\eqref{eq:MJ}. Once the majority of the DM is bound in compact halos at around $a/a_{\rm eq}\simeq 30$, $z\simeq 100$ (see Figure~\ref{fig:bound_frac} in Section~\ref{s:compact_halos}), the production of solitons is expected to decrease, eventually approaching zero.

Note that, crucially, it is thanks to the coincidence in eq.~\eqref{eq:kJoks} that (dense) solitons form. If $\mathcal{P}_\delta$ were peaked at a momentum $k_\star$ much smaller than $k_J^{\rm eq}$, the overdensities would have collapsed into halos, as quantum pressure would have been negligible. If instead $k_J^{\rm eq}\gg k_\star$, the collapse would have been prevented until late times (if at all) and resulted in much less dense solitons (since their density is set by the DM density at the time of formation, as discussed below).

\subsection{Vector Solitons (Quantum Dark Photon Stars)}

A vector soliton is the self-gravitating stationary solution of eqs.~\eqref{eq:sp1a} and~\eqref{eq:sp1b} in (asymptotically) flat spacetime $a=1$ that minimises the total energy $E$ at fixed, finite, value of the vector-boson particle number
\begin{equation}\label{eq:particlenumber}
{\cal N} \equiv \frac{1}{m}\int d^3x\sum_i|\psi_i|^2~.   
\end{equation}
This particle number is conserved as the fundamental action for the massive vector boson, eq.~\eqref{eq:lagrangianA}, contains no number-changing self-interactions. (The implied interactions with gravitons, $h_{\mu\nu}$, following from the action are ineffective as they do not lead to number-changing processes such as $A_i A_j\rightarrow h, 2h,\ldots$ for any of the solutions we consider.)\footnote{Note that a free theory possesses an infinite number of conserved quantities as the particle and antiparticle numbers for every single wavenumber and spin direction are individually conserved. In the presence of classical gravitational interactions a only a finite number of these quantities, including ${\cal N}$, remain conserved due to non-linear mode mixing.  In the presence of additional interactions with, e.g., Standard Model fields, the effective conservation of particle number over the cosmological timescales of interest must be reconsidered. We emphasise that contrary to statements in the literature the existence of a dark photon, or equivalently Proca star does not require $A_\mu$
to be promoted to a complex field.
This conservation law is made explicit in the non-relativistic reduction, eqs.~\eqref{eq:psidef}, \eqref{eq:sp1a} and \eqref{eq:sp1b}, as the
fields $\psi_i$ are complex (despite the fact that $A_i$ are real) and the effective action for $\psi_i$ associated to these SP equations has a global U(1) invariance leading to a conserved particle number. Thus vector solitons are an
example of a non-topological Q-ball soliton \cite{Coleman:1985ki} stabilised against decay due to their being the lowest energy state at fixed ${\cal N}$, in particular lower than the energy $E={\cal N} m$ of a collection of ${\cal N}$ far-separated vector bosons.}

In any case, the validity of a semi-classical analysis in terms of fields requires ${\cal N}\gg 1$. Since we will always be in a regime where the gravitational binding energy is small compared to the total bare rest-mass $M\equiv m {\cal N}$ it is convenient for later purposes to use $M$ as an effective conserved mass parameter of the soliton. In the limit in which we work, and in terms of $M$, the total soliton energy is
\begin{equation}\label{eq:totalenergy}
E=M+\int d^3x \sum_i\left( \frac{1}{2m^2}|\nabla\psi_i|^2+\frac12\Phi|\psi_i|^2 \right)~.
\end{equation}

Although not mathematically proven, it has been conjectured that ground state vector solitons have the form $\psi_i\propto u_i$ where $u_i$ is a generic complex vector with unit norm ($u_i^*u_i=1$), and have spherically symmetric energy density \cite{Adshead:2021kvl,Jain:2021pnk}.\footnote{Ref.~\cite{Brito:2015pxa} considers a full relativistic treatment of the closely related case of a \emph{complex} massive vector field $A^\mu$ coupled to gravity, showing that there exist gravitationally bound `Proca-star' solutions. However the lowest energy solution identified in \cite{Brito:2015pxa} is in fact an excited state (see also the discussion in~\cite{Adshead:2021kvl}).  Moreover, as we have emphasised, a complex $A^\mu$ with a manifest U(1) symmetry is not necessary for the existence of an effectively conserved particle number and an associated stable vector soliton.% In addition the important quantum nature of the soliton solutions is not emphasised.
}
In this case, the solution space of ground state solitons as a function
of their mass can be constructed by a rescaling of the basic \emph{ansatz}
\begin{equation}\label{eq:basicsol}
\psi_i=\frac{m}{\sqrt{4\pi G}}e^{-i\gamma mt}\chi_1(m r)u_i  \ ,  \qquad \Phi=\Phi_1(m r) ~,
\end{equation}
where $r\equiv\abs{\vec{x}}$ is the distance from the soliton's centre and $\gamma$ a fixed real constant. Here $\chi_1(x)$ and $\Phi_1(x)$ are monotonic functions that satisfy $\chi_1(x)\to(1,0)$ and $\Phi_1(x)\to({\rm const},0)$ for $x\to(0,\infty)$. 

Numerical solution of the equations arising from the use of this \emph{ansatz} shows that for the ground state soliton,  $\gamma\simeq-0.65$, while an excellent analytic approximation for $\chi_1(x)$ is $\chi_1(x)\simeq(1+a^2x^2)^{-b}$ with $(a,b)\simeq(0.228,4.071)$. Similarly, $\Phi_1(x)$ is well approximated by $\Phi_1(x)\simeq c(1+a^2x^2)^{-b}$, with $(a,b,c)\simeq (0.465, 0.676,- 1.31 )$.  Given the form of $\chi_1$, the soliton's energy is localised at its centre. All other ground state soliton solutions of eq.~\eqref{eq:sp1a} and~\eqref{eq:sp1b} have the form in eq.~\eqref{eq:basicsol} with the rescalings $\chi_1(x)\to\alpha^2\chi_1(\alpha x)$, $\Phi_1(x)\to\alpha^2\Phi_1(\alpha x)$ and $\gamma\to \alpha^2\gamma$, for any $\alpha>0$.

Given eqs.~\eqref{eq:totalenergy} and \eqref{eq:basicsol} and $A_i\propto \psi_i e^{-imt}$, the quantity $-\gamma\alpha^2$ can be interpreted as the binding energy per unit mass.
Our non-relativistic, weak-gravitational field, approximation thus requires $\alpha \ll 1$ for consistency. \footnote{As $\alpha\rightarrow {\cal O}(1)$ the ground state soliton enters the strong gravitational field regime, potentially becoming unstable to gravitational collapse to a black hole or disruption.} The corresponding rescaled soliton solution has mass and half-mass radius 
\begin{equation}\label{eq:alphasol_mass_radius}
M\simeq \frac{2\alpha}{Gm}~,\qquad
R\simeq \frac{1.9}{\alpha m}~,
\end{equation}
respectively. All the soliton solutions that we consider in this work have $\alpha\ll 1$ and are stable.
 
It will be useful to us that the combination $MR$ is independent of the soliton mass and is given by
\begin{equation}\label{eq:MR}
MR \simeq\frac{3.9}{Gm^2} \, .
\end{equation}
The corresponding central density of the solitons is 
\begin{equation}\label{eq:rhos}
\rho_{s} =\frac{\alpha^4 m^2}{4\pi G} \simeq \frac{1}{G m^2 R^4} \simeq \frac{G^3m^6M^4}{64\pi} \, .
\end{equation}
The quantum Jeans length associated to this density is $\lambda_J(\rho_{s})\simeq 2.3 R$, and is therefore of order of the size of the object. When studying the properties of solitons here and subsequently $\lambda_J(\rho_s)= 2\pi/(16\pi G \rho_s m^2)^{1/4}$ refers to the physical rather than comoving quantum Jeans length. As discussed, this means that the quantum pressure term is relevant for this solution. In fact, since $\vec{v}=0$ from eq.~\eqref{eq:basicsol}, solitons are configurations fully supported by quantum pressure, in the sense that $\Phi_Q=-\Phi$ (this can be used as an alternative equivalent definition of solitons). In Figure~\ref{fig:massa} (left) we plot the soliton density profile  in terms of $\lambda_J(\rho_{s})$.

Solitons can be thought of as bound configurations of bosons with gravitational energy balanced by their intrinsic kinetic energy. This comes from their `irreducible' velocity, of order $1/(mR)$, that stems from the uncertainty principle associated to the Schr\"odinger eq.~\eqref{eq:sp1a} and the fact that the particles are localized in a finite region of space. Indeed, the de Broglie wavelength of the particles is approximately equal to the size of the soliton. Note that given eq.~\eqref{eq:MR}, this intrinsic velocity is the order of the virial velocity $%v_{\rm vir}\equiv
   (GM/R)^{1/2}$ that the particles would have in a gravitationally bound configuration, consistent with the particle interpretation.

Given our expectation that the masses of cosmologically produced solitons are determined by the quantum Jeans length at the time of creation, eq.~\eqref{eq:MJ}, the density of the solitons is parametrically set by the average DM energy density when they are created, independently of $m$. Indeed, plugging the expected value of the solitons masses in eq.~\eqref{eq:MJ} into eq.~\eqref{eq:rhos}, the density of the solitons produced at time $a$ is $\rho_s= 4.5\cdot10^{4} c_M \bar{\rho}(a)$.

It is noteworthy that ground state vector solitons with a fixed mass form an infinite set parameterized by $u_i$, i.e. a generic direction in coordinate space, which indicates the direction of the vector field. This moduli space of ground state solutions is due to the $U(3)$ symmetry of  eqs.~\eqref{eq:sp1a} and~\eqref{eq:sp1b}. For a fixed direction, the soliton profile resembles the well known scalar soliton, see e.g. \cite{Hui:2016ltb}.

We also observe that a soliton is only an exact solution in the idealised case where the total mass $\int d^3x\sum_i|\psi_i|^2$ is finite. When the mass is infinite, for instance in the presence of a constant density background $\bar{\rho}$, solitons of central density $\rho_s\gg \bar{\rho}$ are still good approximations of stationary solutions near their cores. However, away from their centres the solitons' profiles will be deformed, and, since they gravitationally attract the background matter, they will be surrounded by a halo.

Importantly, as argued in detail in Ref.~\cite{Jain:2021pnk}, depending on the choice of $u_i$, and in the absence of other interactions, the ground state solitons can either carry zero total angular momentum $J$, or possess macroscopically large $J$ without energy cost or alteration of the radial profile $\chi_1(r)$, at least at leading order in an expansion in $G$.  Since all known
physically-realised astrophysical objects carry angular momentum, it is interesting to discuss in more detail the physics of these $J\neq 0$ vector solitons.

The total angular momentum $\vec{J}$ is composed of two parts, an orbital angular momentum contribution, $\vec{L}$, and an intrinsic spin density which integrates to $\vec{S}$. In terms of $\psi_i$, and in the non-relativistic weak-gravity limit in which we are working, they are given by
\begin{align}\label{eq:angmomdef}
    L_p & =   \frac{i}{2m}\epsilon_{pqr} \int d^3x~ (\psi_m^* \partial_q \psi_m x^r - {\rm c.c.}) \\
    S_p & = \frac{i}{m}\epsilon_{pqr} \int d^3 x ~ \psi_q \psi_r^*~.
\end{align}
On the ground state soliton solutions, eq.~\eqref{eq:basicsol}, the first of these expressions  gives $\vec{L}=0$.  On the other hand, the intrinsic angular momentum of these solitons is $\vec{S}=i(\vec{u}\times\vec{u}^*) \mathcal{N}\hbar$, where once again we have temporarily restored the factors of $\hbar$ for clarity. Since $|\vec{u}\times\vec{u}^*|\leq1$, $S$ can acquire any value in $[0,{\cal N}\hbar]$ and can even vanish depending on $u_i$.

This can be made explicit by employing an orthonormal basis, ${\bf \epsilon}^{(\lambda)}_{\hat n}$, for the
$\lambda = \pm 1, 0$ spin quantization states along a fixed axis ${\hat n}$. Choosing, without loss of generality, ${\hat n}= {\hat z}$ these polarisation states are
\begin{align}
    {\bf \epsilon}^{(\pm 1)}_{\hat z} = \frac{1}{\sqrt{2}}  \begin{pmatrix} 
    1 \\ \pm i \\ 0  
    \end{pmatrix} \, , \qquad {\bf \epsilon}^{(0)}_{\hat z} = \begin{pmatrix}
    0 \\ 0 \\ 1  
    \end{pmatrix} ~.
\end{align}
Then one may expand $\psi_i$ and thus $A_i$ in this basis; specifically for the $\vec{L}=0$ solutions
\begin{equation}\label{eq:basicsolpol}
\psi_i=\frac{m}{\sqrt{4\pi G}}e^{-i\gamma mt}\sum_\lambda\chi^{(\lambda)}(\vec{x})
({\bf \epsilon}^{(\lambda)}_{\hat z})_i~,
\end{equation}
where in principle the functions $\chi^{(\lambda)}(\vec{x})$
could differ. However in the non-relativistic weak-gravity limit of interest to us, $\chi^{(\lambda)}(\vec{x}) = a_{(\lambda)} \chi_1(mr)$
with $a_{(\lambda)}$ complex coefficients satisfying $\sum_{\lambda}|a_{(\lambda)}|^2=1$, and with $\chi_1(mr)$
satisfying exactly the same equation as before.\footnote{At higher order in the weak gravity $G$-expansion the functions $\chi^{(\lambda)}(\vec{x})$ will depend non-trivially on $\lambda$ due to the metric dependence on $\vec{J}$ which is (asymptotically) given by $g_{t\phi} \simeq 2GJ \sin^2\theta /r$ as $r\rightarrow \infty$ for $\vec{J}$ oriented along the ${\hat z}$ axis.}
For instance, for $|a_{(\pm 1)}|=1$ the resulting total
spin angular momentum parallel to the ${\hat z}$ axis is maximum and reads $J_z=\pm {\cal N}\hbar$.

Linearly or partially polarised ground state solitons are possible if non-trivial complex linear combinations of the angular momentum eigenstates result from the non-linear dynamics of structure and vector soliton formation (in the absence of additional interactions the decoherence time can be parametrically long). In fact, we will see in the next Section that, in the minimal theory we study, the solitons are generically formed with a nontrivial polarisation. In particular, they have intrinsic angular momentum uniformly distributed in the range $S\in [0,{\cal N}\hbar]$, and, as expected, negligible orbital angular momentum $\vec{L}$. This is not surprising given that, as mentioned, there is no energy difference between solutions with different intrinsic angular momentum. We note that, if present, interactions of the dark photons with the Standard Model, themselves or other new light states could lead to solitons with a particular angular momentum being energetically favoured.

\subsection{Comparison with numerical simulations} \label{ss:SP_simulations}

To confirm our analytic expectations, measure the unfixed numerical coefficient $c_M$ in eq.~\eqref{eq:MJ}, and determine the soliton mass distribution, we study the dynamics of the system numerically. 
It is straightforward to show that the equations of motion~\eqref{eq:sp1a} and~\eqref{eq:sp1b}, the initial condition in eq.~\eqref{eq:PAt1} and $a(t)$ only depend on the number of relativistic degrees of freedom at $T_\star$ and $\Omega_A/ \Omega_{\rm DM}$ (see Appendix~\ref{aa:SPMRE}). The evolution is therefore independent of the values of $m$ and $G$, which is why the ratio between $k_\star$ and $k_J^{\rm eq}$ in eq.~\eqref{eq:kJoks} does not depend on $m$ explicitly. We continue to fix the number of relativistic degrees of freedom at $T_\star$ to the SM high temperature value and we assume that the dark photon makes up all the DM.

We solve eqs.~\eqref{eq:sp1a} and~\eqref{eq:sp1b} numerically on a discrete lattice in a periodic box of constant comoving size $(3.75\lambda_\star)^3$, starting deep in radiation domination at $a/a_{\rm eq}\ll0.01$ (this choice is not important as long as $a/a_{\rm eq}\ll1$), from a realisation of the initial conditions with the Gaussian power spectrum in eq.~\eqref{eq:PAt1}. The final simulation time is limited by loss of resolution of the soliton cores and by the growth of density perturbations on the scale of the box%\mcol{, of order $4\lambda_\star$}
, which once non-perturbative lead to finite volume systematic uncertainties. With our available numerical resources we can run to $a/a_{\rm eq}\simeq 7$. 
Our main results are obtained averaging over  approximately 100 individual simulations. Further details of the evolution and tests of the systematic uncertainties are given in Appendix~\ref{app:SP}.

\begin{figure}[t!]
	\begin{center}
			\begin{subfigure}[c]{0.4\textwidth}
	\includegraphics[width=\linewidth]{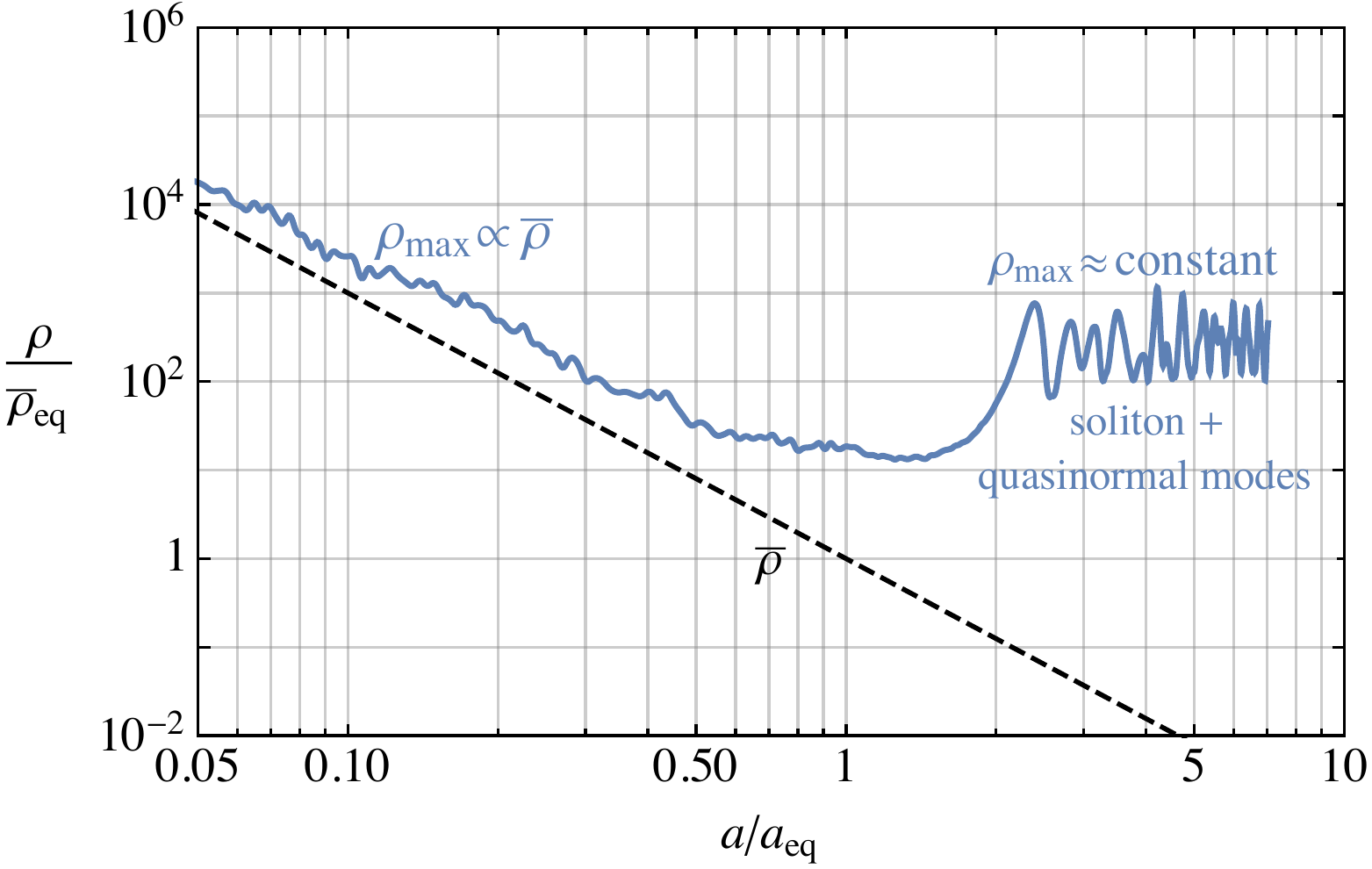}
\end{subfigure}		\ \ \ \ \ \ 
		\begin{subfigure}[c]{0.42\textwidth}
			\includegraphics[width=\linewidth]{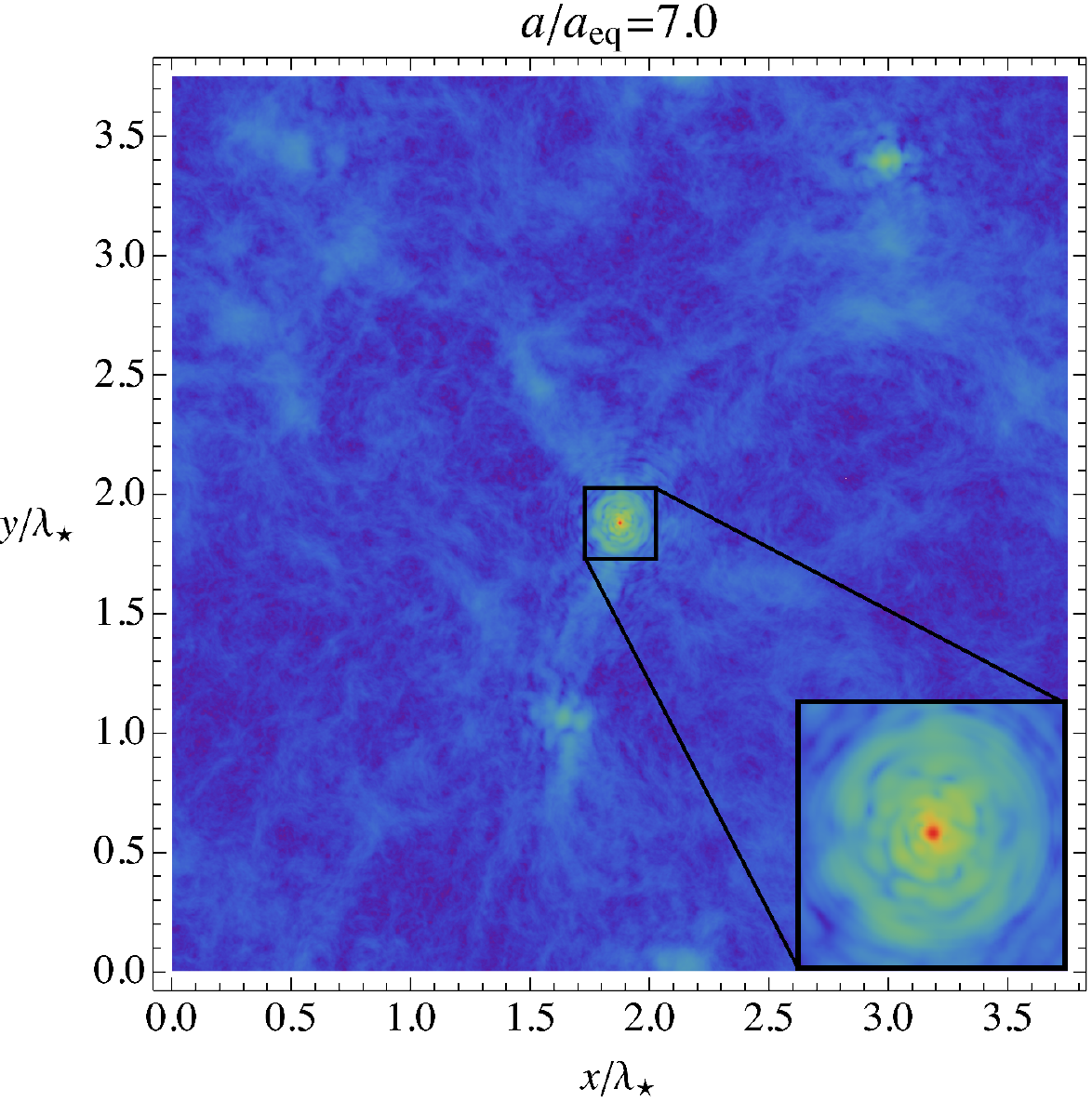}
		\end{subfigure}		
		\begin{subfigure}[c]{0.1\textwidth}
			\includegraphics[width=\linewidth]{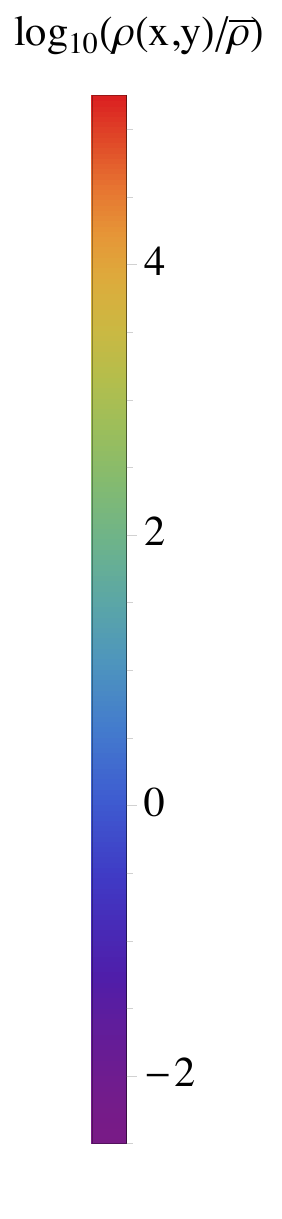}
		\end{subfigure}		
	\end{center}
	\caption{
	{\bf \emph{Left:}} 
	The evolution of the maximum value $\rho_{\rm max}$ of the energy density field of the vector through MRE, in a simulation with volume $(3.75\lambda_\star)^3$. Time is parameterised by the scale factor $a$ relative to that at MRE, $a_{\rm eq}$. The mean vector energy density $\bar{\rho}$ is also plotted. At early times $\rho_{\rm max}$ follows $\bar{\rho}$,
	  with small fluctuations due to the oscillation of modes with $k\gtrsim k_J$, driven by quantum pressure. The collapse of overdensities with $\delta \gtrsim 1$, which in the absence of quantum pressure would occur at $a/a_{\rm eq}\simeq 1/\delta$, is hindered until after MRE. Once $k_J/k_\star \propto a^{1/4}$ has grown sufficiently, overdensities collapse. After the collapse, the maximum density is at a point inside a soliton. The soliton is produced with excited quasinormal modes, so the maximum density subsequently oscillates. 
 {\bf \emph{Right:}}
		A slice of the energy density at $a/a_{\rm eq}=7$, in the same simulation as is plotted in the left panel. The slice passes through the point that has the largest density at this time, which is at the centre of a soliton. The soliton (red region in inset) is surrounded by a 
		 spherical `fuzzy' halo (yellow/green region) and there are cosmic filaments connecting it to other solitons. Spherical waves can be seen around the soliton, which are due to the emission of energy from quasinormal modes. A video showing the evolution can be found at \cite{video}. \label{fig:denslice}} 
\end{figure}

To verify that the broad features of the dynamics of Section~\ref{ss:SP_analytic} do indeed occur, in Figure~\ref{fig:denslice} (left) we show the time evolution of the maximum density $\rho_{\rm max}$ in a single, typical, simulation run.  As expected, prior to MRE $\rho_{\rm max}/\bar{\rho}$ is, on average, constant. Due to quantum pressure, density fluctuations on small scales, $k\gtrsim k_J(a)$, oscillate even during radiation domination, which leads to small oscillations in $\rho_{\rm max}/\bar{\rho}$ during this time. This would not happen in the limit $k_J/k_\star \to \infty$, in which case $\rho(\vec{x})$ would be almost completely frozen for a nonrelativistic field.\footnote{Such oscillations can be seen for perturbative density fluctuations, which we analyse in Appendix~\ref{app:over}, and we have confirmed using numerical simulations that this remains the case for large fluctuations.}

In the absence of quantum pressure, the largest overdensities in the initial conditions $\delta \simeq 3$ would collapse at $a/a_{\rm eq} \simeq 1/\delta\simeq 1/3$. As expected, collapse is actually delayed to a later time, $a\gtrsim a_{\rm eq}$, when $k_J(a)$ has exceeded $k_\star$. 
  At $a>a_{\rm eq}$, the maximum density increases fast, until it reaches an approximately constant value, while the mean dark matter density continues to decrease. This indicates that a bound object, in our case a soliton, has formed and decoupled from the Hubble flow. Additionally, the maximum density has clear oscillations, which are due to the soliton forming with excited quasi-normal modes. We study the growth of density perturbations and the evolution of the density power spectrum in more detail in Appendix~\ref{app:moreSP}.

In Figure~\ref{fig:denslice} (right) we plot the density field $\rho$ through the slice of the same simulation that contains the point with the largest density, at $a/a_{\rm eq}=7$. 
There is a central soliton (red region). 
The soliton is surrounded by a spherical fuzzy halo (yellow/green region) extending far from its core, the maximum density of which is about two orders of magnitude smaller than the soliton core density. Finally, the early stages of a cosmic web connecting different solitons have formed (see also Figure~\ref{fig:DM_sub_pic} left, where we show a 3D version of the same energy density). Spherical waves can be seen beyond the halo. These are due to energy released by the decay of the soliton's quasi-normal modes. 

\begin{figure}[t]
	\begin{center}
		\includegraphics[width=0.46\textwidth]{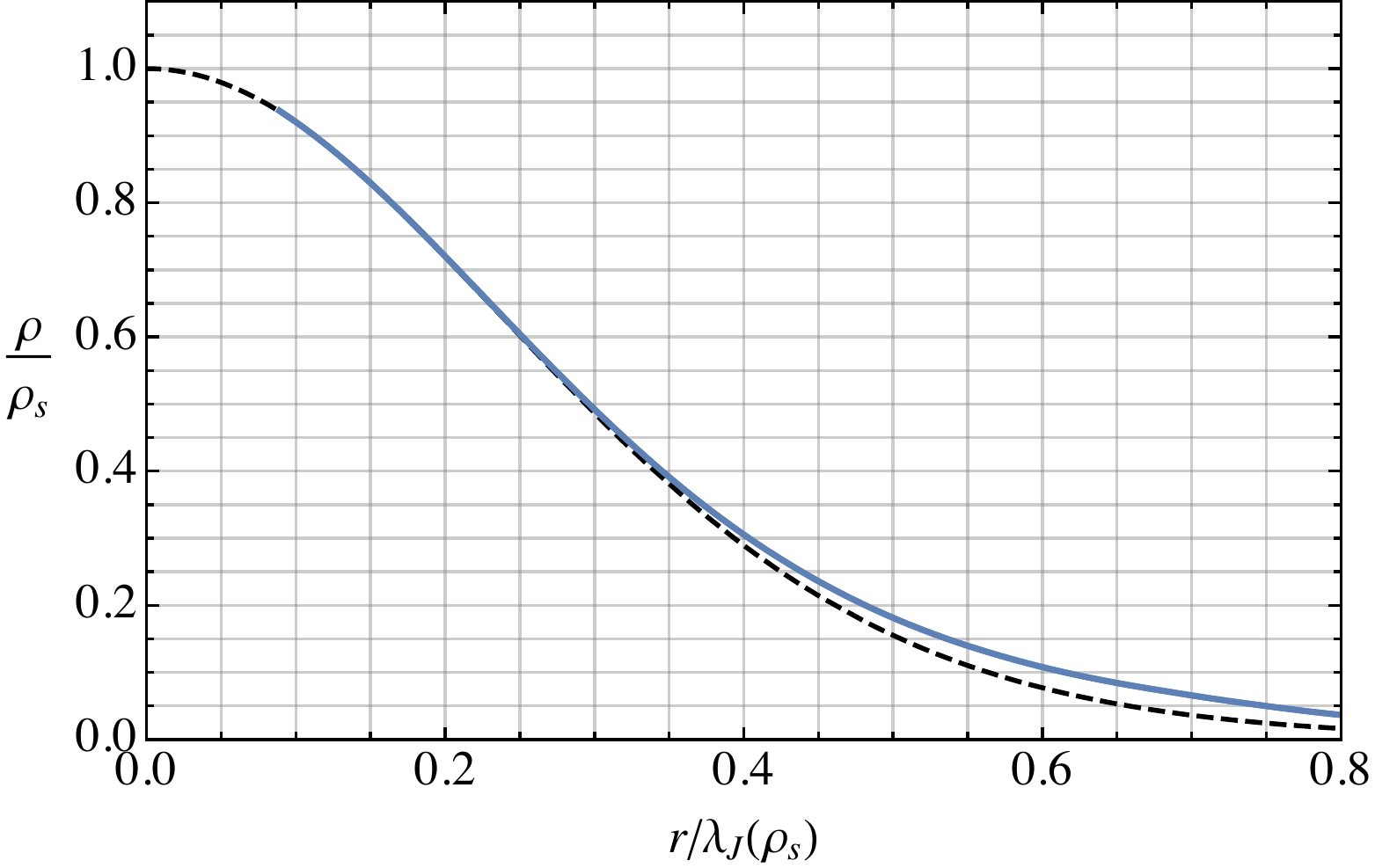} \ \ \  \ 
		\includegraphics[width=0.5\textwidth]{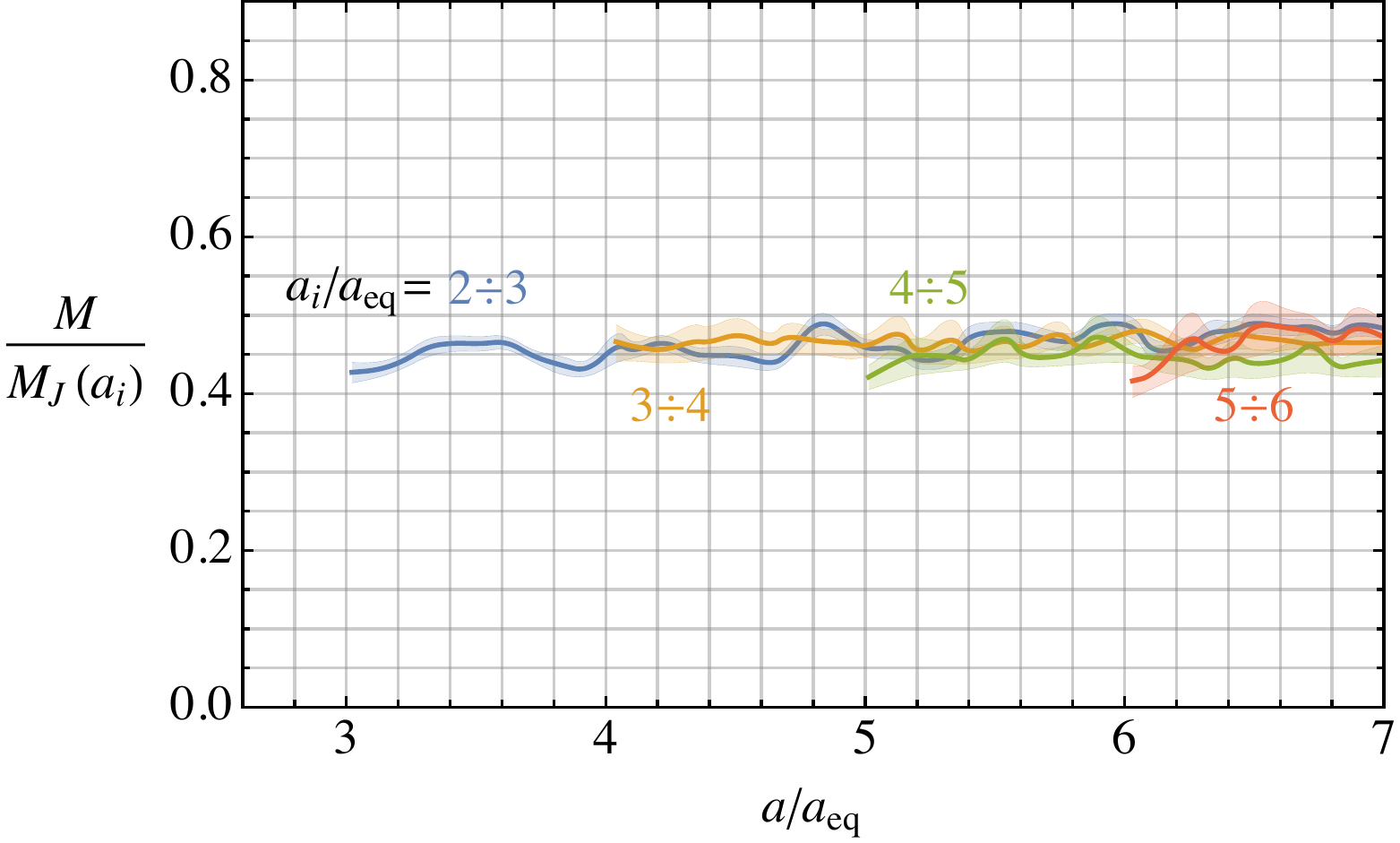}
	\end{center}
	\caption{
				{\bf \emph{Left:}} The spherically averaged density profile of collapsed objects as a function of the distance from their centre (blue line). The density is plotted relative to that at the object's centre $\rho_{s}$, and the distance is normalised to the (physical) quantum Jeans length $\lambda_J(\rho_{s})=2\pi/k_J(\rho_{s})$. We average over all objects, with barely visible statistical uncertainties. The profile in the inner region closely matches that of a soliton (dashed black line, see eq.~\eqref{eq:basicsol}). 
				{\bf \emph{Right:}}		
		The time evolution of the masses $M$ of the solitons, normalised to the parametric expectation $M_J(a_i)=(4\pi/3)  a_i^3 \bar{\rho} \lambda_J^3(a_i) \propto a_i^{-3/4}$, where $a_i$ is the time when the soliton forms and $\lambda_J(a_i)=2\pi/ k_J(\bar{\rho}(a_i))$. The solitons are binned based on $a_i$, with statistical error bars. Once produced the soliton masses are approximately constant. Moreover, on average, $M(a_i)=c_M M_J(a_i)$ with $c_M\simeq 0.45$ independent of $a_i$. \label{fig:massa}} 
\end{figure}

To understand the nature of the collapsed objects, in Figure~\ref{fig:massa} (left) we plot the spherically averaged density profile around the centre of the objects at $a/a_{\rm eq}=5$, averaged over all the objects in our full set of simulations. 
To enable the profiles of objects with different mass to be combined, for each object the density profile is normalised to its central density $\rho_{s}$ and the distance from its centre to the quantum Jeans length $\lambda_J(\rho_{s})$ corresponding to its central density $\rho_{s}$. As it is clear from Section~\ref{ss:SP_analytic}, in terms of these variables the soliton density profile is $\chi_1^2(x/\lambda_J(\rho_{s}))$ and is independent of the soliton mass.

 Evidently the collapsed objects have a profile that is remarkably close to the soliton, out to a distance $\simeq \lambda_J(\rho_{s})/2\simeq 1.5 R$. This confirms our expectation that the objects formed around MRE are supported by quantum pressure. Spatial angular momentum initially carried by the dark matter that later becomes a soliton could be lost during formation or transferred to the fuzzy halo that surrounds the solitons.  $80\%$ of the total mass in the pure soliton solution  is within $ \lambda_J(\rho_{s})/2$, so to a good approximation we can identify the mass in the soliton-like part of the collapsed objects as being the same as the total mass of the vacuum soliton.  That gravitational interactions result in the dark photon no longer being purely longitudinal and indeed the vector field in the soliton is not longitudinal. 
 
As a further check of the nature of the solitons, we have evaluated the variance of $\psi_i(\vec{x})/|\vec{\psi}|^2$ over the soliton cores (defined as the region in which the spherically averaged density exceeds $\rho_s/2$). This would represent the spatial variation of the (would-be) constant unit vector $u_i$ in eq.~\eqref{eq:basicsol}. As expected, the variance is tiny ($\lesssim 0.02$) for almost all of the objects that form, confirming that the solutions are indeed close to the pure soliton solution. We also calculated the intrinsic angular momentum of the solitons, defined in eq.~\eqref{eq:angmomdef}, in units of $\mathcal{N}$. We find that all the components of $\vec{S}/\mathcal{N}$ in the soliton core have a flat probability distribution between $-1$ and $1$.

We note that the initial perturbative linear growth of the $k\lesssim k_\star$ modes  matches the analytic expectation (we analyse this in Appendix~\ref{app:moreSP}). However, due to the limited simulation time available only few such modes have become nonperturbative by the final time. The compact halos resulting from their collapse are still almost non-existent and contain only a very negligible fraction of DM (see also Figure~\ref{fig:collapsef} in the next Section), so we do not attempt to study them in these small scale simulations.

Finally, as mentioned, initially the solitons are produced with quasinormal modes. Although the quasinormal modes are expected to have disappeared by today, they are long-lived with effective Q-factor $>10^3$ \cite{Guzman:2004wj} and could have interesting, possibly observable, consequences, which would be worth exploring in the future. In particular, a substantial fraction of the DM is in spherical waves emitted due to the quasinormal modes (since quasinormal modes initially store an order one fraction of the soliton energy density). Their wavelength is of the physical size of the solitons when they are first emitted.

\subsection{The soliton mass distribution} \label{ss:solitonmassf}

To check that the mass of the solitons produced at a given time is set, on average, by the quantum Jeans scale at that time, $M(a)=c_M M_J(a)\propto a^{-3/4}$, and to fix the unknown coefficient $c_M$, in Figure~\ref{fig:massa} (right) we plot the time evolution of the average value of the masses $M$ of the solitons, grouping the solitons based on the scale factor $a_i$ when they form, and normalising their masses to $M_J(a_i)$.  We calculate the soliton masses starting from their central densities $\rho_{s}$ using the relation in eq.~\eqref{eq:rhos}.\footnote{In Appendix~\ref{app:moreSP} we show that measuring the soliton masses from the density profile leads to values that are consistent, to the precision we require.}  We see that once formed the soliton masses are, on average, approximately constant. Moreover the anticipated proportionality $M(a) = c_M M_J(a)$ is reproduced remarkably well, with a universal constant coefficient $c_M \simeq 0.45$. This works equally well for the solitons produced e.g. at $a/a_{\rm eq}=2\div3$ and $a/a_{\rm eq}=6\div7$. On other other hand, despite $c_M$ being constant on average, at any time solitons with a range of masses are produced, approximately within $0.3 \lesssim c_M \lesssim 0.7$. For reference, in physical units the quantum Jeans mass and the mass of the solitons are
\begin{equation} \label{eq:MJeq}
M_J^{\rm eq}\equiv M_J(a_{\rm eq}) =5.2\cdot 10^{-23} M_{\odot}  \left( \frac{\eV}{m}\right)^{3/2} ~,
\end{equation}
and
\begin{equation}\label{eq:Ma}
M(a)  = 2.3 \cdot 10^{-23} M_\odot \left(\frac{c_M}{0.45} \right) \left(\frac{a_{\rm eq}}{a} \right)^{3/4}  \left( \frac{\eV}{m}\right)^{3/2} ~.
\end{equation}
As we will see in more detail shortly, the solitons produced through the evolution have masses approximately in the range $M \simeq (0.05 \div 0.5) M_J(a_{\rm eq}) = (2.6\cdot 10^{-24} \div 2.6\cdot 10^{-23}) M_\odot (\eV/ m)^{3/2}$. Ultimately, the solitons have masses inversely proportional to $m^{3/2}$, because the total mass initially contained in a region of volume $k_\star^{-3}$ has this dependence.

Classically the soliton ground state is stable. Quantum mechanically there is the possibility of decay to gravitons, but this process is exponentially suppressed \cite{Eby:2015hyx}.
Therefore, unless destroyed by e.g. tidal disruption (studied in Section~\ref{ss:destroy}), they constitute an irreducible component of the DM abundance. One of the most interesting  quantities is the soliton mass distribution, 
\begin{equation}\label{eq:dfDMdlog}
\frac{d f_{s}(a,M)}{d \log M}=\frac{M}{\bar{\rho}}\frac{d n(a,M)}{d \log M} \, ,
\end{equation}
 which counts the fraction of dark matter in solitons per unit log mass (in plots we use the base 10 logarithm so that $f_s$ can easily be estimated). In eq.~\eqref{eq:dfDMdlog}, $f_{s}(a,M)$ is the fraction of dark matter in solitons with mass less than $M$, and can be related to the number density of solitons $n(a,M)$ with mass less than $M$ at time $a$.\footnote{Consequently, the fraction of dark matter in solitons with mass such that $a\leq \log_{10} M \leq b$ is $\int_a^b df/d\log_{10} M ~d\log_{10} M$.}

\begin{figure}[t]
	\begin{center}
		\includegraphics[width=0.7\textwidth]{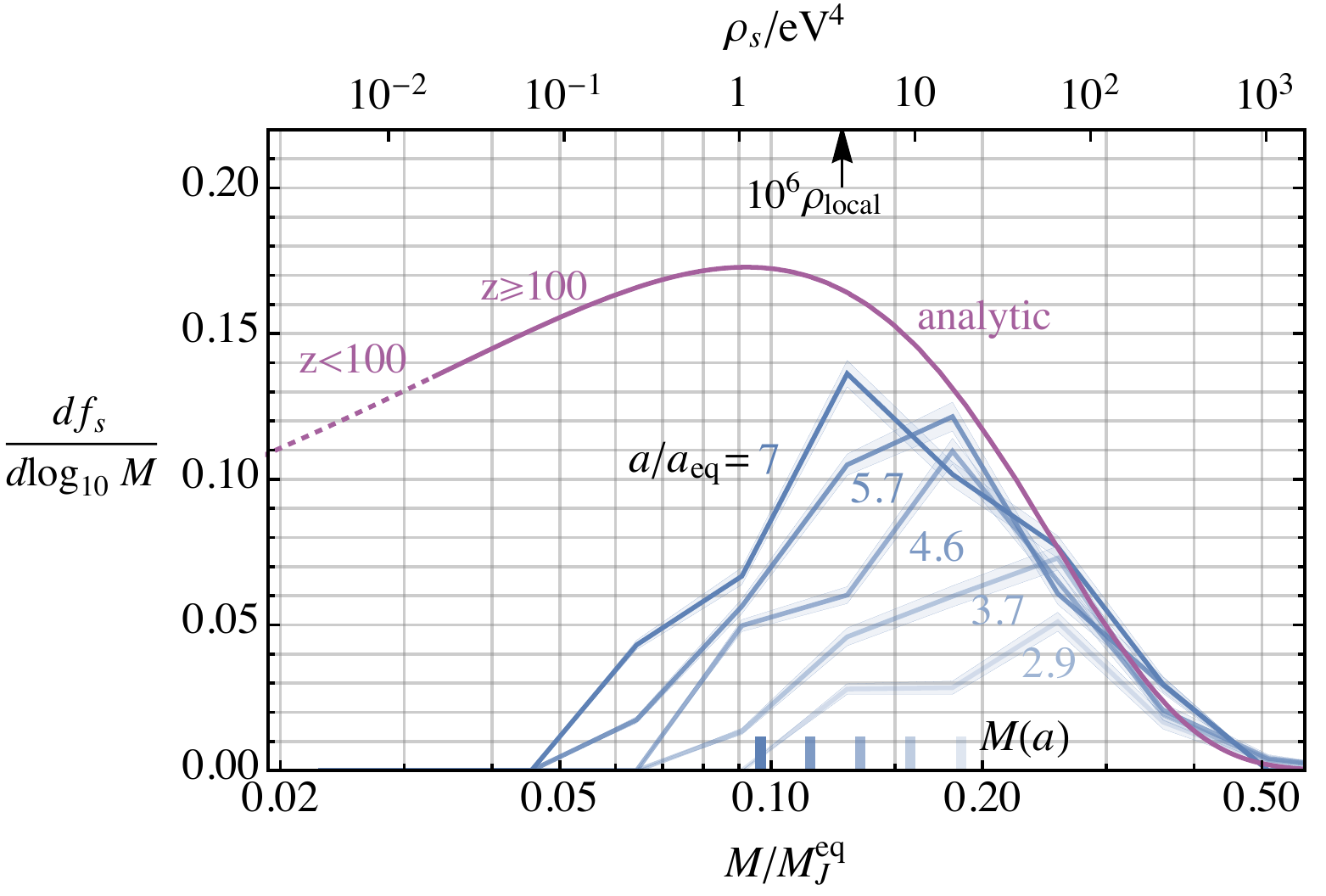}
	\end{center}
	\caption{ The time evolution of soliton mass distribution $df_s/d\log_{10}M$ as a function of the soliton mass $M$ relative to $M_J(a_{\rm eq}) = 5.2\cdot 10^{-23} M_\odot (m/\eV)^{-3/2}$, with statistical error bars.  This counts the percentage of DM in solitons per unit mass. The low mass cutoff in the mass distribution tracks the average mass of solitons being produced at a given time, $M(a)=c_M M_J(a)\propto a^{-3/4}$, shown by coloured ticks above the lower axis. By the end of the simulations at $a/a_{\rm eq}=7$ the distribution has reached an approximately constant form for solitons of mass $M \gtrsim 0.15 M_J(a_{\rm eq})$. The analytic expectation described in the main text is also plotted, extrapolating to $z=100$, when most of the dark matter becomes bound in compact halos, see Section~\ref{s:compact_halos}. 
		 The central densities of the solitons $\rho_{s}$ are indicated on the upper axis. These densities are orders of magnitude larger than the local dark matter density in the vicinity of the Earth, taken to be $\rho_{\rm local}= 0.5\GeV/ \cm^3$.  \label{fig:HMF_soliton}} 
\end{figure}

The soliton mass distribution in eq.~\eqref{eq:dfDMdlog} can be evaluated in numerical simulations,  at different times. The result is shown in Figure~\ref{fig:HMF_soliton}.  As expected, the distribution is time-dependent, 
and has a break at low masses that tracks  $M=M(a) \equiv c_M M_J(a)$ (indicated with ticks above the lower axis), because only a few solitons with $M\lesssim M(a)$ are produced before $a$. 
The shape at masses $M\gg M(a) \equiv c_M M_J(a)$ has an approximately time-independent form, since no new solitons of these masses are being produced. 
On the upper axis of the same Figure we show the density of the solitons $\rho_{s}$ at their core, which is related to their mass via eq.~\eqref{eq:rhos}. 
$\rho_{s}$ is independent of $m$ and  parametrically coincides with the average DM density at the time of formation. In particular, it corresponds to the density of the Universe at MRE for the most massive solitons, and is many orders of magnitude larger than the local DM energy density $\rho_{\rm local}$ today.

Unfortunately the limited time range of simulations means we cannot capture the solitons that are produced at $a/a_{\rm eq}\gtrsim7$ (evidently, the soliton mass distribution is still evolving at small $M$ in Figure~\ref{fig:HMF_soliton}). Given that no additional heavy solitons will be subsequently produced, Figure~\ref{fig:HMF_soliton} gives a lower bound on $f_{s}$, and additional production will only strengthen e.g. direct detection signals. We will also see in  Section~\ref{ss:destroy} that the densest solitons, for which we do have a reliable prediction, are most likely to survive to the present day in the Milky Way.  

Interestingly, despite the complicated dynamics, the shape of the soliton mass distribution can be understood via a simple analytic argument based on the initial power spectrum $\mathcal{P}_\delta$ of Figure~\ref{fig:Pspectra}. This gives theoretical control of the soliton mass distribution, and also allows us to estimate the extrapolation of the numerical results in Figure~\ref{fig:HMF_soliton} to smaller soliton masses. Estimating the mass distribution requires two inputs: 1) the mass and 2) the number density of the solitons produced, as a function of time $a$. To fix 1) we crudely approximate that the solitons produced at every time have a unique mass $M=M(a)$ given by eq.~\eqref{eq:MJ}, with $c_M\simeq 0.45$. For 2), we assume that as soon as the comoving quantum Jeans scale drops below the size of an order one fluctuation, this will collapse into a soliton. Therefore, the number  
of solitons with masses $M \div M+dM$ that are produced is expected to be proportional to the `frequency' with which there are corresponding fluctuations in the initial conditions that are larger than some critical value $\delta_c$ of order one. To estimate this frequency, we make the crude assumption that the dark photon density field is Gaussian, with power spectrum of Figure~\ref{fig:Pspectra} (right), although in reality this is not the case.  In  Appendix~\ref{app:analytic_prediction} we derive the resulting soliton mass function.

Despite involving rough approximations, in Figure~\ref{fig:HMF_soliton} we see that our analytic argument reproduces the data at large masses, where this has already converged to its late-time value, remarkably well.\footnote{In this we have set the unfixed parameter in our analysis $\delta_c =0.22$ to reproduce the soliton production rate measured in simulations.}   The analytic prediction does not account for the decrease in the soliton production rate due to DM becoming bound in compact halos at larger scales, which becomes important around $a/a_{\rm eq}\simeq 30$, corresponding to $z\simeq 100$. We therefore indicate on the plot the solitons that are produced before this time, for which the prediction applies. 

We note that the argument we used is very similar to that usually employed to estimate the abundance of primordial black holes from small scale curvature perturbations that could be produced during inflation (see, e.g. \cite{Biagetti:2021eep}), with the quantum Jeans scale functioning as an effective `horizon', in the sense that both prevent the collapse of the order one fluctuations, until they cross the size of that perturbation.

Although we have focused on solitons produced from initially large density fluctuations, 
 we  note that additional solitons might form later in compact halos. This could happen directly when the compact halo forms, although in Section~\ref{s:compact_halos} we will see that only a small fraction of the mass in the halos will end up in a soliton this way.\footnote{During collapse, the density in the compact halo increases, usually by a factor $\simeq 200$. If the fluctuation corresponds to spatial scales only slightly larger than $k_\star^{-1}$ this could be enough for quantum pressure to become relevant and most of the mass might end up in a soliton with a relatively large mass might.} Alternatively it could occur later, on longer timescales, by gravitational relaxation, see \cite{Levkov:2018kau,Eggemeier:2019jsu,Chen:2020cef}. Additionally, the solitons already present might increase their mass by accreting the background DM via gravitational relaxation, or solitons might merge together if they become bound into compact halos. This lead to an uncertainty on their mass. We do not try to study these potentially important issues in our present paper.

\subsection{Fuzzy halos around solitons}\label{ss:fuzzy_halo}

\begin{figure}[t]
	\begin{center}
		\includegraphics[width=0.6\textwidth]{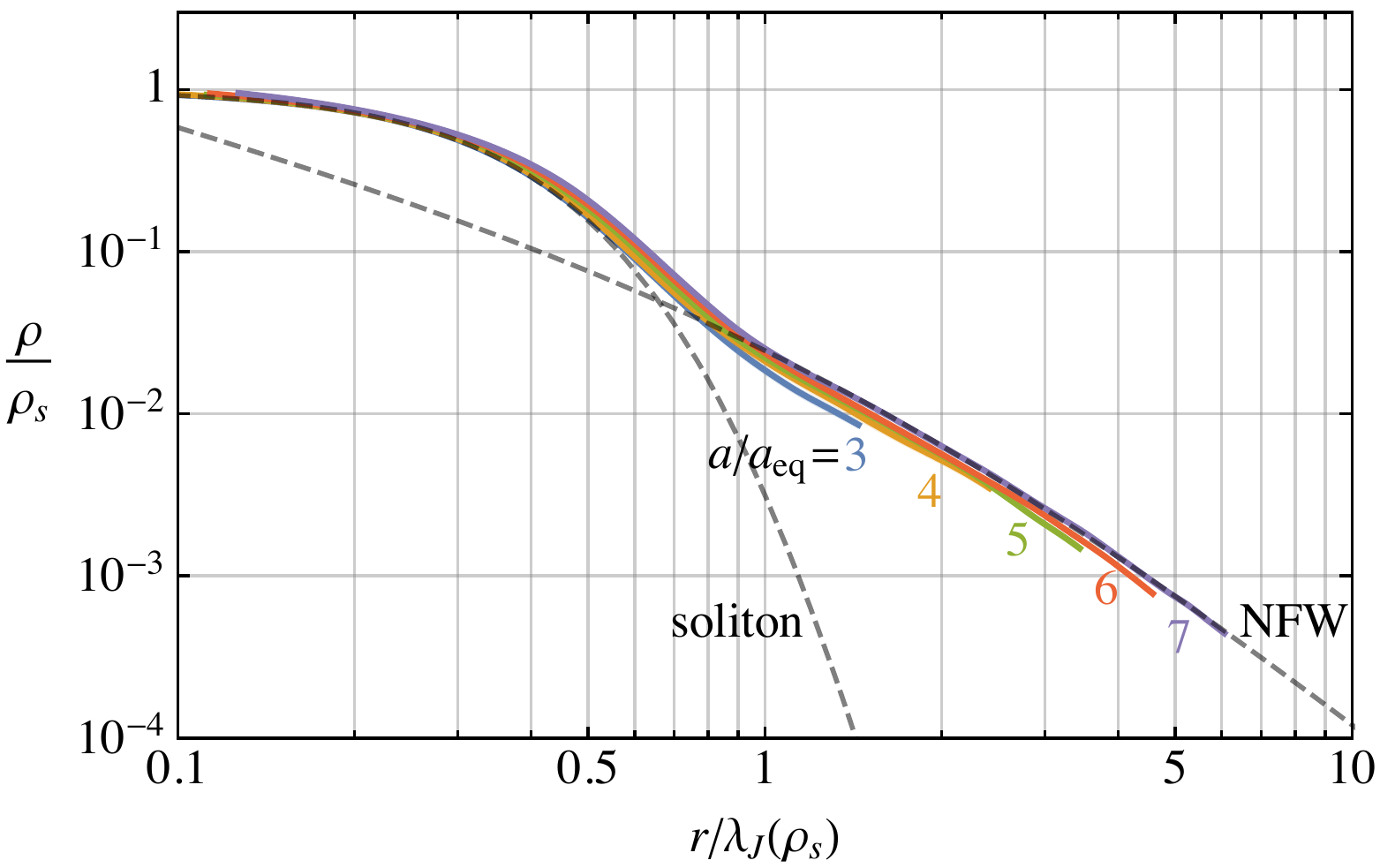}
	\end{center}
	\caption{
	The (spherically averaged) density profile around solitons relative to the density at their core $\rho_{s}$, as a function of the distance from the centre in units of the quantum Jeans scale at the core $\lambda_J(\rho_{s})$. Results are shown averaging over all heavy solitons with $M>0.3 M_J(a_{\rm eq})$, which form relatively early. The density profile matches that of the pure soliton solution for $r/\lambda_J(\rho_{s})\lesssim 1/2$, and the `fuzzy' halo outside is well fit by an NFW profile, see eq.~\eqref{eq:NFWfuzzy}. The profiles are cut off at a radius where the dark matter density drops below $20$ times the mean dark matter density.
		\label{fig:profile_log}} 
\end{figure}

Although the bound objects seen in simulations resemble the pure soliton solution of eq.~\eqref{eq:basicsol} at small distances, as anticipated in Section~\ref{ss:SP_analytic} their profile deviates at distances larger than the soliton half-mass radius. Indeed, soliton cores are surrounded by a halo. 
We dub this a \emph{`fuzzy' halo},
because the quantum Jeans length associated with its typical density is only marginally smaller than the size of the halo and the wave-like properties of the vector are still relevant. 
Indeed, the fuzzy halo is a (time-dependent) solution of eq.~\eqref{eq:Euler} where the gravitational potential is locally balanced by a combination of both the quantum pressure and velocity term, in particular $\vec{v}_i\neq0$. 
Such a halo can be seen surrounding the soliton in Figure~\ref{fig:denslice}. 
Fluctuations on distances of order the de Broglie wavelength can be seen.

In Figure~\ref{fig:profile_log} we plot the density profile averaged over relatively heavy solitons, with mass $M \gtrsim 0.3M_J(a_{\rm eq})$. As in Figure~\ref{fig:massa}, we combine different objects' profiles by plotting their density relative their core density $\rho_{s}$ and measuring the radius in units of $\lambda_J(\rho_{s})$. The profile at a given time is cut off at the radius where the density drops below $20$ times the mean DM density. 

The fuzzy halo is evident as a deviation from the soliton profile for $r\simeq \lambda_J(\rho_{s})/2$. 
 The scaling property of the soliton solution appears to also apply to their halos, and  the profiles from different objects take a universal form (this suggests the existence of a soliton-fuzzy halo relation). 
At later times the halo extends further from the soliton core, with the region closer to the soliton remaining with a fixed density. The inner part of the halo might form when the overdensity collapses. However, at $a/a_{\rm eq}\simeq 7$ the fuzzy halo extends almost down to the mean DM density at $a/a_{\rm eq}\simeq 3$, so the external parts of the halo are most likely related to the accretion,  as the background DM is attracted to the soliton.

The fuzzy halo turn out to be well described by an NFW profile
\begin{equation}\label{eq:NFWfuzzy}
\frac{\rho(r)}{\rho_{s}}=  \frac{\rho_0}{\frac{r /\lambda_J(\rho_{s})}{r_0} \left(1 + \frac{r /\lambda_J(\rho_{s})}{r_0} \right)^2  }~,
\end{equation}
where $\rho_0$ and $r_0$ are dimensionless parameters that are universal for all the halos. Fitting the density profile at $a/a_{\rm eq}=7$ in the interval $1\leq r/\lambda_J(\rho_{s}) \leq 7$ (where the lower limit is due to the profile transitioning to soliton form) we obtain that $\rho_0 \simeq 0.042$ and $r_0\simeq 1.56$ accurately reproduce the data.\footnote{In Appendix~\ref{app:moreSP} we show that lighter solitons also have a surrounding fuzzy halo, still well fitted by an NFW profile with the same $\rho_0$ and $r_0$. Since they form later, for such solitons the fuzzy halo had less time to grow outwards and at the final simulation time it extends less far from the core than in Figure~\ref{fig:profile_log}, but is still growing.}  
We do not know how far out the fuzzy halos will grow at times beyond the reach of simulations. 
In Section~\ref{ss:destroy} we will see that the outer parts of the fuzzy halos are destroyed in the late Universe and the parts that are most likely to survive are mostly already formed at $a/a_{\rm eq}\simeq 7$, the final simulated time. 

Eq.~\eqref{eq:NFWfuzzy}, together with the soliton mass distribution and an input about how far the fuzzy halos extend, allows the distribution of fuzzy halos to be calculated. 
The fuzzy halos surrounding solitons  contain much more DM mass than the solitons themselves. As an indication, the part of a fuzzy halo within $8 \lambda_J(\rho_{s})$ (which is a typical distance out to which a fuzzy halo is likely to survive disruption in the late Universe) has mass $\simeq  6 M$ where $M$ is the mass of the central soliton. 
 The corresponding mass and size in physical units can be read off from eqs.~\eqref{eq:Ma} and \eqref{eq:lambdaJkm}.

Finally, we note that Figure~\ref{fig:denslice} shows that there are overdense filaments connecting solitons analogous to the standard cosmic web, which forms much later at much larger scales. However, these are much less dense than the fuzzy halos and are probably destroyed in the subsequent evolution.

\section{Compact Halos and Primordial Structure Formation} \label{s:compact_halos}

In this Section we focus on scales larger than $\lambda_\star=2\pi/k_\star$ and reconstruct the evolution of the modes in the $k^3$ part of the spectrum in Figure~\ref{fig:Pspectra} (right).  
As we will see, as they become nonperturbative, they induce the formation of a chain of heavier and heavier compact halos: a `primordial' structure formation. This happens before (and is normally not present in) canonical structure formation, because of the additional small-scale inhomogeneities. 

These dynamics are similar to the formation of compact halos in the case of post-inflationary axions, which also has a density power spectrum with a $k^3$ dependence in the IR.\footnote{With a peak of roughly $\mathcal{P}_\delta\simeq 1$ at a scale approximately set by $k_\star$ \cite{Kolb:1993zz,Kolb:1994fi,Kolb:1995bu}, although there are large uncertainties related to the decay of the string-domain wall system.} The process of compact halo formation in this case has been studied extensively,  both analytically \cite{Zurek:2006sy,Fairbairn:2017dmf,Enander:2017ogx,Fairbairn:2017sil,Dai:2019lud,Kavanagh:2020gcy,Ellis:2020gtq,Blinov:2021axd} and numerically  \cite{Zurek:2006sy,Eggemeier:2019khm,Xiao:2021nkb}. Additionally \cite{Lee:2020wfn,Blinov:2021axd} studied compact halos in vector dark matter produced by inflationary fluctuations using the Press-Schechter approach, similar to our analytic analysis.\footnote{In some theories fluctuations in the inflaton can lead to similar dynamics~\cite{Niemeyer:2019gab,Eggemeier:2020zeg}.}

\subsection{Dynamics of the IR modes
}

As shown in Section~\ref{ss:SP_analytic}, after MRE, modes with $k<k_\star$ (both the $k^3$ and adiabatic modes) are not affected by quantum pressure and grow linearly. Once a $k^3$ mode %with $k<k_\star$ 
becomes of order one, the linear approximation to eqs.~\eqref{eq:sp1a} and~\eqref{eq:sp1b} is no longer applicable. 
Qualitatively, we expect that at this point the DM density contained in that perturbation collapses into a gravitationally bound object -- a `compact' halo -- incorporating already bound objects inside the region, including solitons.  Since perturbations on larger and larger scales become nonlinear at later and later times (given that $\mathcal{P}_\delta(k)\propto k^3$), heavier and heavier compact halos will  progressively form. When the adiabatic modes become nonperturbative, at around $z\simeq 15$, they trigger canonical structure formation. We expect that after this time few new compact halos form, and most of those already present eventually become small subhalos of much larger galactic halos, as in Figure~\ref{fig:DM_sub_pic}.\footnote{Adiabatic modes on small spatial scales collapse slightly before those on larger scales because adiabatic fluctuations grow logarithmically during radiation domination once they have re-entered the horizon.} Standard structure formation occurs as in cold dark matter cosmology, with the only difference that some of the DM is bound in compact halos.

The small-scale substructure of the compact halos 
(e.g. any solitons they contain) 
is clearly affected by quantum pressure and the same may be true very close to the centre of the halos. 
However, given that they are generated from scales larger than the quantum Jeans scale,  if we focus on their properties 
at scales sufficiently larger than $\lambda_\star$ (effectively smoothing out small scales), the effect of quantum pressure is expected to be mostly irrelevant.  
Moreover, at large enough scales the (effective) initial velocity of the field is negligible (because the IR modes do not oscillate, being unaffected by quantum pressure, and the field is highly nonrelativistic). 
This means that at these scales the equations of motion in eqs.~\eqref{eq:cont},~\eqref{eq:Euler},~\eqref{eq:Poisson}  reduce to those of a single component perfect fluid (i.e. with $\Phi_Q=0$)  
subject only to 
 gravitational interactions, with density $\rho=\rho_1+\rho_2+\rho_3$, and $\vec{v}\simeq\vec{v}_i$ ($\simeq0$ at $a<a_{\rm eq}$), i.e. 
\begin{align}
	\partial_t\rho +3H\rho+a^{-1}\nabla\cdot(\rho \vec{v})=& \ 0  \label{eq:cont1c}\\
	\partial_t\vec{v}+H\vec{v}+a^{-1}(\vec{v}\cdot\nabla)\vec{v}=&-a^{-1}\nabla\Phi \label{eq:Euler1c}\\
	\label{eq:Poisson1c}\nabla^2\Phi=&\ 4\pi G a^2(\rho-
	\bar{\rho}) \ .
\end{align}
Despite still being nonlinear, the dynamics of the system at these large scales is simpler and can be analysed by combining standard analytic and numerical approaches. On the analytical side, the so-called Press--Schechter (PS) method  handles the evolution of eqs.~\eqref{eq:cont1c},~\eqref{eq:Euler1c},~\eqref{eq:Poisson1c} by determining the number of halos present at every time, based on the power spectrum in the initial conditions.  Although this is a model rather than a first principles calculation, it has been shown to capture the main qualitative features, and provide a reasonable quantitative prediction of halo mass functions in many settings~\cite{Lacey:1994su}.  The same equations can be also investigated via N-body simulations. 
By evolving a system of discrete particles interacting only gravitationally, these reproduce the dynamics of a perfect fluid. Owing to the discretisation into particles, N-body simulations do not lose resolution of collapsed objects in the way that direct SP simulations do, and are therefore better suited to studying the successive chain of structure formation.

\subsection{Formation of compact halos} \label{ss:form_compact}

In the following we treat the $k^3$ modes separately from the adiabatic modes, which will be accurate until the adiabatic fluctuations start becoming non-linear, and determine the abundance of compact halos. 
A halo is a set of gravitationally bound matter. 
In the remainder of this Section we will not consider the substructure of compact halos in terms of subhalos or solitons. As we will discuss in Section~\ref{ss:destroy} we expect that many of the solitons survive intact inside compact halos, although they could also be destroyed, merge with each other or increase in mass by accretion. 
Compact halos will subsequently be bound inside adiabatic halos.

We can predict the distribution of compact halo masses using the Press-Schechter approach. 
At distances larger than $\lambda_\star$, while the linear approximation is valid the overdensity field grows as $\delta(t,\vec{x})=\delta(t\ll t_{\rm eq},\vec{x})D[a]$ with $D[a]=1+\frac32\frac{a}{a_{\rm eq}}$. We expect that at every time regions of space where the overdensity has exceeded an order one critical value $\delta_c$ have collapsed into a halo. 
To assign a mass to these regions, we consider the field smoothed over a distance $R$, and the mass contained in each of them is $M=(4\pi/3)R^3\bar{\rho}$. Since these regions are expected to collapse into halos of mass up to $M$, the Press-Schechter anzats~\cite{Press:1973iz} is that the fraction of DM in halos with mass $>M$ equals the probability that the field $\delta_s$ smoothed over $R=(3M/4\pi\bar{\rho})^{1/3}$ is larger than $\delta_c$. 
This probability is fully determined by the power spectrum $\mathcal{P}_\delta$ since  $\delta$ is initially Gaussian at scales larger than $\lambda_\star$, and so only by the variance $\sigma _s^2\equiv\langle \delta_s^2(t,\vec{x})\rangle$, with $\mathcal{P}_\delta$ at $t\ll t_{\rm eq}$ as in eq.~\eqref{eq:Pdelta}.

In Appendix~\ref{app:compact_halos}, we show that the resulting fraction of DM bound in compact halos $f_h$ then satisfies
\begin{equation}\label{eq:dfhdlogM}
	\frac{d f_{\rm h}(a,M)}{d\log M}=%\sqrt{\frac{2}{\pi}}\frac{\delta_c}{\sigma_s(M)}e^{-\frac{\delta^2_c}{2\sigma^2_s(M)}}\left|\frac{d\log \sigma_s(M)}{d\log M}\right|\simeq
	\sqrt{\frac{2}{\pi}}\nu e^{-\nu^2/2}\left|\frac{d\log \nu}{d\log M}\right|\simeq\sqrt{\frac{8M}{M_\star}}\frac{\pi\delta_c}{3^{3/4} D^2[a]}\exp\left[{-\frac{  \pi^3 \delta_c^2}{3\sqrt{3} D^2[a]}\frac{8M}{M_\star}}\right]~,
\end{equation}
where $\nu(M)\equiv \delta_c/\sigma_s(M)$ and we introduced an extra factor of $2$ to address the well known cloud-in-cloud problem, as originally done in~\cite{Press:1973iz}. In the last equality we defined $M_\star\equiv (4\pi/3)\bar{\rho}a^3(2\pi/k_\star)^3\simeq 6.9 M_J^{\rm eq}$ (see eqs.~\eqref{eq:kJoks} and~\eqref{eq:MJ}) and approximated the spectrum with a single power law $k^3$, considering only modes with $k<k_\star$ since modes with $k\gtrsim k_\star$ are affected by quantum pressure, and form solitons.

As evident from eq.~\eqref{eq:dfhdlogM}, the halo mass distribution is peaked at
\begin{equation}\label{eq:Mhalopeak}
M\simeq 9.5M_\star \left[\frac{1.7}{\delta_c}\right]^2\left[\frac{100}{z+1}\right]^2\simeq 65M_J^{\rm eq} \left[\frac{1.7}{\delta_c}\right]^2\left[\frac{100}{z+1}\right]^2\, , % M_J^{\rm eq} 
\end{equation}
with an exponential cutoff at higher masses and a power law suppression ($\propto M^{1/2}$) at lower masses. Therefore, as anticipated, as time increases the most frequent halos are increasingly heavy. Also as expected, the compact halos are much heavier than the solitons, see eq.~\eqref{eq:Ma}, and in fact will contain some of the solitons. 
Eqs.~\eqref{eq:dfhdlogM} and~\eqref{eq:Mhalopeak} are reliable only at large enough masses, which come from the largest modes that are least affected by quantum pressure.

\begin{figure}%[h!]
	\begin{center}
		\includegraphics[width=0.65\textwidth]{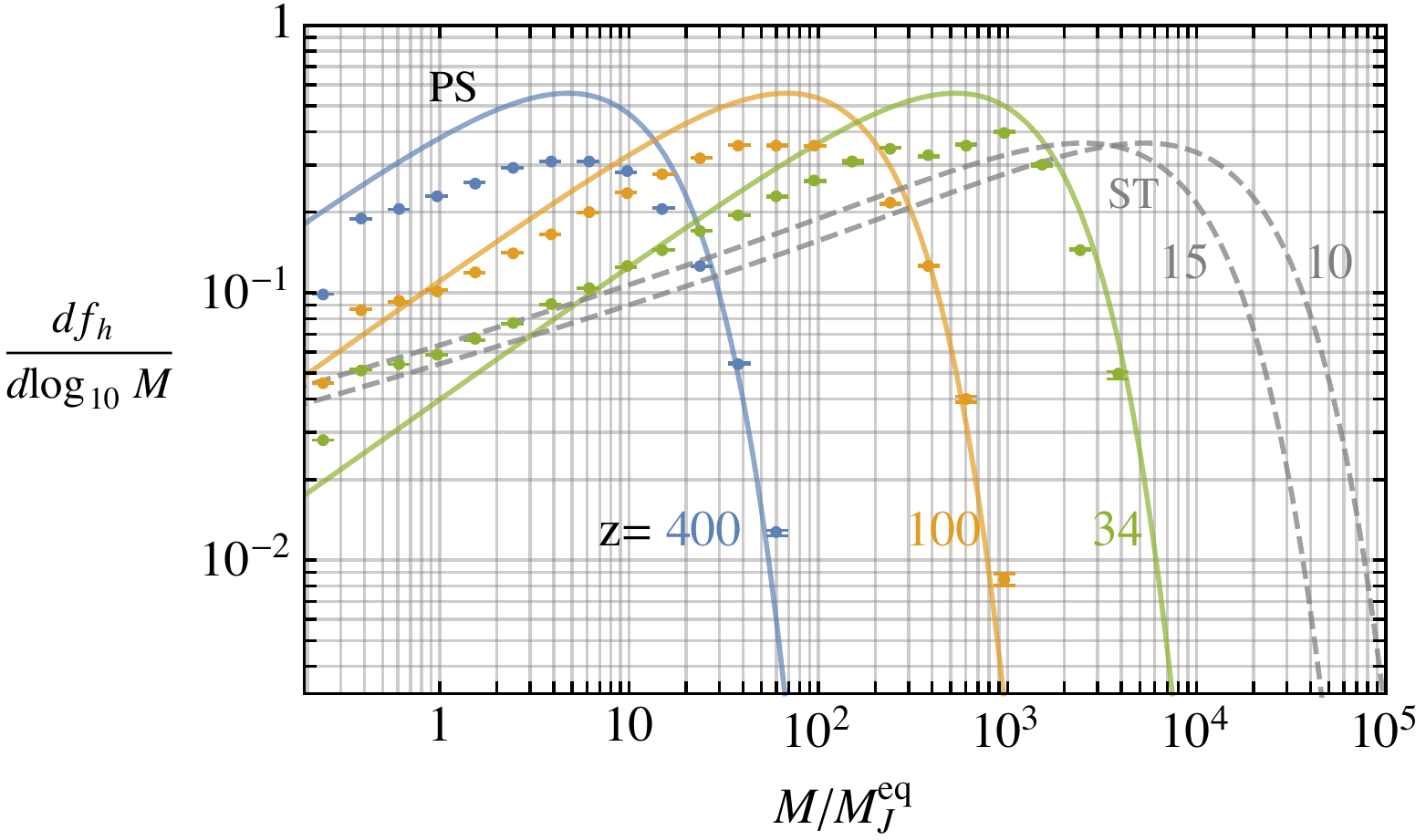}
	\end{center}
	\caption{
The time evolution of the mass distribution of compact halos, which arise from the collapse of modes in the $k^3$ part of the power spectrum $\mathcal{P}_\delta$ in Figure~\ref{fig:Pspectra}. Data points are the results of N-body simulations,  starting from a realisation of the full dark photon energy density field at $a\ll a_{\rm eq}$. Given that simulations are carried out in the absence of quantum pressure, only the dynamics of the most massive objects with $M\gtrsim 50 M_J^{\rm eq}$ are expected to be correctly captured. The Press--Schechter analytic prediction in eq.~\eqref{eq:dfhdlogM} is also plotted (PS), and reproduces the dynamics remarkably well. We also show the halo mass distribution at $z=10,15$ (around when compact halos stop evolving and get incorporated to adiabatic halos) reconstructed with a Sheth--Tormen inspired model (ST), which reproduces the data accurately at earlier redshifts (not plotted). The masses are given in units of $M_J^{\rm eq} =5.2\cdot10^{-23} M_\odot \left(\eV/m\right)^{3/2}$.
		\label{fig:HMF_white_noise_halo}} 
\end{figure} 

To check the validity of this analysis, in Figure~\ref{fig:HMF_white_noise_halo} we compare the halo mass distribution in eq.~\eqref{eq:dfhdlogM} using $\delta_c=1.7$ (as predicted by the so-called spherical collapse model~\cite{1972ApJ...176....1G}) with results of N-body simulations, at different times, i.e. redshifts. The simulation starts at $a/a_{\rm eq}=0.01 \ll 1$ with initial conditions given by the full (non-Gaussian) density field in eq.~\eqref{eq:deltax} -- converted to a particle configuration -- and vanishing initial velocity $\vec{v}\simeq 0$. These can be run until the most IR modes in the finite box start to become non-linear, around $z=30$, when finite volume systematic errors become significant.\footnote{Our N-body simulations have comoving box size of $80 \lambda_\star$.} From the plot, it is evident that the PS method reproduces the peak and the high mass cutoff of the halo mass distribution at all times. Although the height of the peak is slightly different between the two, the overall agreement is impressive.

We can also use N-body simulations to estimate how large a mass a compact halo must have in order for it to be unaffected by the dynamics of the most UV modes, near the $k_\star$ peak, which are influenced by quantum pressure. To do so, in Figure~\ref{fig:HMF_cut_vs_real} (left) of Appendix~\ref{app:analytic_prediction} we compare the halo mass distribution from N-body simulation with initial conditions given by a Gaussian field with the power spectrum of Figure~\ref{eq:Pdelta}, but with a UV cutoff at $k>0.5k_\star$ (for this momentum range, $\mathcal{P}_\delta \lesssim 0.05$, so the initial fluctuations on such scales are indeed very close to Gaussian, and for them the quantum pressure is irrelevant). The two distributions coincide for $M\gtrsim 50 M_J^{\rm eq}$. This is only a rough test because, even with initial conditions with a UV cut, once they become non-perturbative the IR modes will source higher $k$ modes, which will still, incorrectly, evolve as in the absence of quantum pressure. Nevertheless, the difference between the two data points in Figure~\ref{fig:HMF_cut_vs_real} (left) gives a feeling of the impact of high $k$ modes and suggests that eq.~\eqref{eq:dfhdlogM} can indeed  be trusted for large enough $M$.

The evolution of compact halos according to eq.~\eqref{eq:dfhdlogM} continues up until $z\simeq 10\div15$, when the adiabatic modes also become nonperturbative.\footnote{After they reenter the horizon, the adiabatic perturbations grow logarithmically in radiation domination because of the gravitational potential generated by the photon perturbation, see e.g. \cite{mo2010galaxy}  (this does not happen for the isocurvature ones). The logarithmic increase is the largest for the highest adiabatic modes (because they have been subhorizon the longest), and this changes the slope of the adiabatic spectrum in Figure~\ref{fig:Pspectra}, making it larger than 0. Standard structure formation therefore proceeds similarly to the `primordial' one induced when the $k^3$ modes become non-perturbative, but with a power spectrum that is much flatter and is initially much smaller.} 
Unfortunately our computational resources do not allow us to simulate the evolution of the adiabatic part and the $k^3$ modes at the same time, therefore we cannot directly investigate how compact halos are incorporated into the much larger adiabatic halos. We estimate the distribution of compact halos today (if not disrupted) by evaluating $d f_{\rm h}/d\log M$ at $z\simeq 10\div15$, when most of the DM is bound in adiabatic halos.
To improve accuracy, instead of using the PS analysis we fit the N-body simulation data with a Sheth--Tormen (ST) inspired form~\cite{Sheth:1999mn}, which provides a  better fit than eq.~\eqref{eq:dfhdlogM} (see Appendix~\ref{app:compact_halos} for details) to the data points in Figure~\ref{fig:HMF_white_noise_halo}. We indicate the mass distribution reconstructed at $z=10,15$ from the ST fit with gray lines in Figure~\ref{fig:HMF_white_noise_halo}. The peak occurs at around 
\begin{equation}\label{eq:halomass}
M_{\rm h}=(3\div 5)\cdot 10^3 M_J^{\rm eq}\simeq (1.5\div3) \cdot 10^{-19} M_\odot (\eV/m)^{3/2} ~.
\end{equation}

\begin{figure}%[h!]
	\begin{center}
		\includegraphics[width=0.54\textwidth]{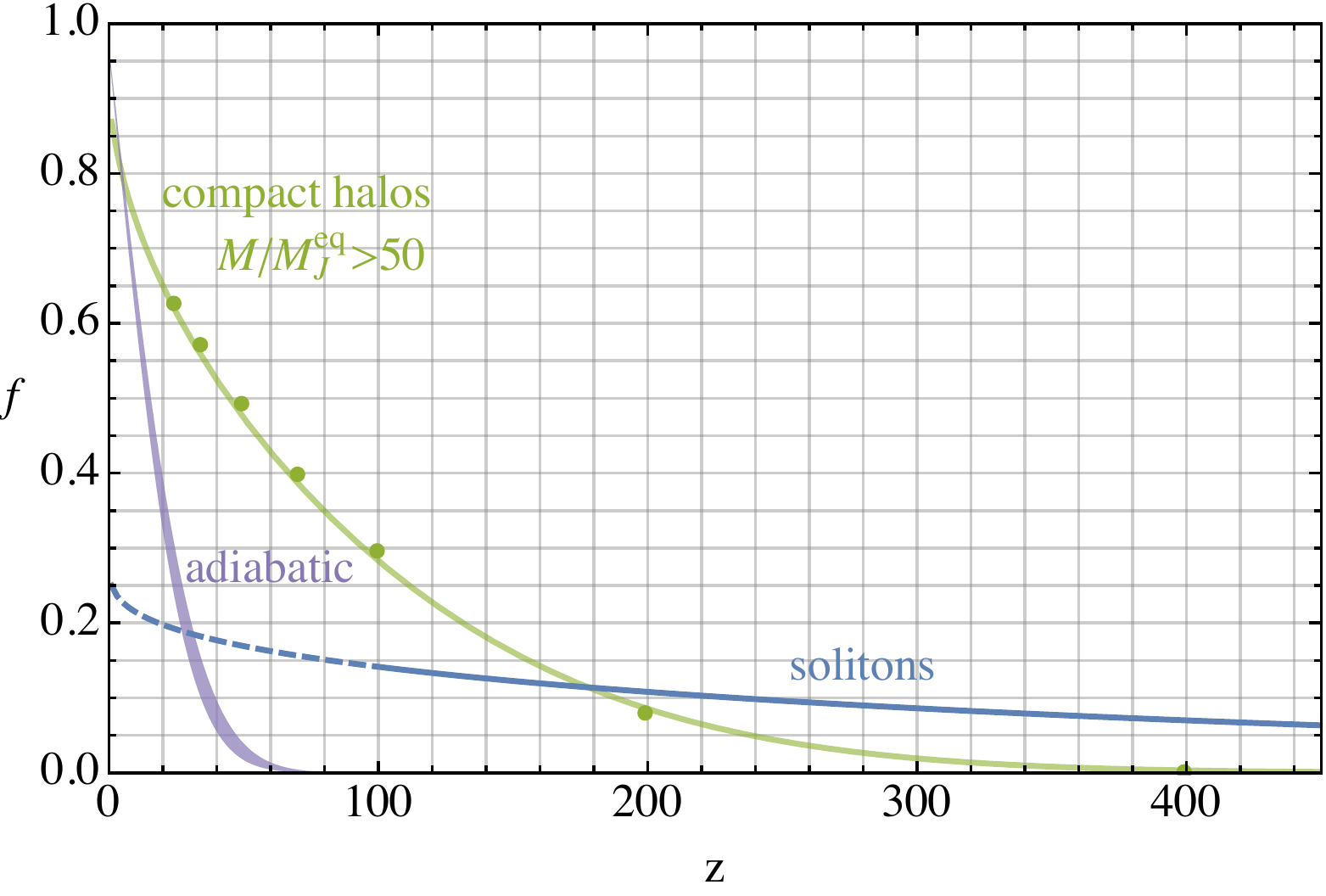}
	\end{center}
	\caption{The fraction of dark matter collapsed in solitons (blue), compact halos (green) and adiabatic halos (purple) as a function of redshift. The fraction of mass in solitons is calculated from the analytic analysis of Section~\ref{ss:solitonmassf} %, fit to the simulation data,
	 as in Figure~\ref{fig:HMF_soliton}. The fraction of DM bound in compact halos is evaluated from N-body simulations (data points) and from the analytic Sheth--Tormen analysis (line), with parameters fit to the data. The fraction collapsed in adiabatic perturbations is calculated using the Press--Schechter method from an extrapolation of observed power spectrum. 
	  This plot ignores the fact that the dynamics of objects on larger distance scales might halt the formation of structure on smaller scales: Soliton formation is prevented after $z\simeq 100$ when most of the DM is bound into compact halos (indicated by the predicted soliton fraction being plotted dashed for $z<100$); similarly, compact halos are expected to stop forming at $z\simeq 15$.	\label{fig:collapsef}	} 
\end{figure} 

Finally, in Figure~\ref{fig:collapsef} we plot the fraction $f_i(a)\equiv \int d M\frac{df_i(a,M)}{d M}$ of dark matter collapsed in soliton $i={\rm s}$ (from the analytic estimate in eq.~\eqref{eq:dfFMdMgaussian}), in compact halos $i={\rm h}$ and in adiabatic halos $i={\rm a}$, as a function of time. The fraction of DM bound in compact halos is obtained from  the ST analysis in Appendix~\ref{app:Nbody} (with data points from simulations also plotted) excluding objects of mass $M<50M_J^{\rm eq}$ which will be strongly affected by quantum pressure and not correctly described by our analysis. The fraction in adiabatic halos is estimated from a PS analysis, extrapolating the observed power spectrum to smaller scales.\footnote{In particular, we put a UV cut the power spectrum at the point where the isocurvature spectrum dominates, which depends on $m$. In Figure~\ref{fig:collapsef} the width of the adiabatic collapse fraction line corresponds to the, barely noticeable, effect of varying $m$ from $10^{-5} \eV$ to $1\eV$.}

%%%%%%%%%%%%%%%%%%%%%%%%%%%%%
\subsection{Compact halo profiles} \label{ss:profile_compact}

We can also study the density profiles of the compact halos. To do so we use an argument normally employed to reconstruct the density of adiabatic halos produced during canonical structure formation (e.g. \cite{Ludlow:2013vxa}), which has been used in the context of axion miniclusters by \cite{Dai:2019lud,Kavanagh:2020gcy,Ellis:2020gtq}. This is based on three simple assumptions.

(1) We assume that all the compact halos that form at a given time have a particular mass, determined parametrically by peak value PS distribution in eq.~\eqref{eq:dfhdlogM}, i.e. halos with mass $M$ are created at $a_c(M)=(8\pi^{3/2}/3^{7/4})(\delta_c/\nu_c)\sqrt{M_\star/M}a_{\rm eq}$ with $\nu_c$ an order one parameter. Halos are assumed to remain subsequently undisturbed. Clearly this is a rough approximation, since at a given time halos with some range of masses will form, and also because a halo of a certain mass could form from mergers of lighter halos.

(2) Since halos grow slowly from the low-density fluctuations (as opposed to from the collapse of already large scale fluctuations), it is reasonable to expect that they all have  a fixed overdensity with respect to the background density at their moment of formation. We will therefore assume that the mean density of the halos, which we call $\rho_M$, is $\Delta$ times larger than the average DM density at the time of formation, i.e. $\rho_M=\Delta\, \bar{\rho}(a_c(M))$, with $\Delta>1$ a universal time-independent number for all the halos.

As we will see from N-body simulations, the halo profiles are well described by the NFW form \cite{Navarro:1995iw,Navarro:1996gj}
\begin{equation}\label{eq:NFWhalo}
\rho(r)=  \frac{\rho_0}{r/r_0 \left(1 + r/r_0 \right)^2  }~,
\end{equation}
with scale radius $r_0$ and density $\rho_0$. As is well known, the total mass in such a halo $4\pi \int_0^{\infty} dk  k^2\rho(r)$ is logarithmically divergent. The halo extends up to an edge defined by the virial radius $R_\Delta$, related to the scale radius $r_0$ via the so-called `concentration' parameter $c_\Delta>1$ as $R_\Delta=c_\Delta r_0$ ($c_\Delta$ parameterises how large the halo is with respect to $r_0$). The average density is therefore $\rho_M=4\pi \int_0^{R_\Delta} dk  k^2\rho(r)/(4\pi R_\Delta^3/3)$. Note that the halo is completely specified by the three parameters $\{\rho_0,r_0,c_\Delta\}$.

(3) Given that the formation of halos is self-similar during the evolution, we assume that all the halos are formed with a universal concentration parameter, irrespectively of the time when they are created. 

The assumptions above allow the parameters $\rho_0$ and $r_0$ of the NFW profiles to be calculated in terms of $\nu_c$, $\Delta$ and $c_\Delta$. 
We report the full analytic formulas in Appendix~\ref{app:compact_halos}. The results for $\nu_c\simeq 0.67$, $c_\Delta\simeq 30\div 50$ and $\Delta\simeq 300\div 500$, which we will see fits the numerical results well, are\footnote{For canonical structure formation, the parameters are $\nu_c\simeq O(1)$,  $c_\Delta\simeq 4$, $\Delta\simeq 200$ \cite{Zhao:2003jf}.}
\begin{align}\label{eq:rho0compact}
\rho_0\simeq &\,\,  0.7\left[\frac{10^3M_J^{\rm eq}}{M}\right]^{3/2}\bar{\rho}^{\rm eq} \ 
 \simeq 0.3\, {\rm eV}^4\left[\frac{10^3 M_J^{\rm eq}}{M}\right]^{3/2}  ,\\
r_0\simeq &\,\, 5.4 \left[\frac{M}{10^3M_J^{\rm eq}}\right]^{5/6}\lambda_J^{\rm eq}\ \simeq 7\cdot 10^5 \, {\rm km}\left[\frac{M}{10^3 M_J^{\rm eq}}\right]^{5/6} \left[\frac{1\, {\rm eV}}{m}\right]^{1/2} \ ,\label{eq:r0compact}
\end{align}
The lighter halos are denser, since they are created at earlier times when the background density is larger. Note that $\rho_0$ represents the density of the halo at $r\simeq r_0$. Since the halo extends a factor $c_\Delta>1$ beyond $r_0$ the average density is smaller than suggested by eq.~\eqref{eq:rho0compact} and, for $c_\Delta \gg 1$, is approximately\footnote{Note that the average density of compact halos with mass $M>5000M_J$ is smaller than the local dark matter density.}
\beq
\left< \rho \right>\simeq\frac{3\rho_0}{c_\Delta^3} \left(\log c_\Delta-1\right) ~.
\eeq

\begin{figure}
	\begin{center}
	\includegraphics[width=0.475\textwidth]{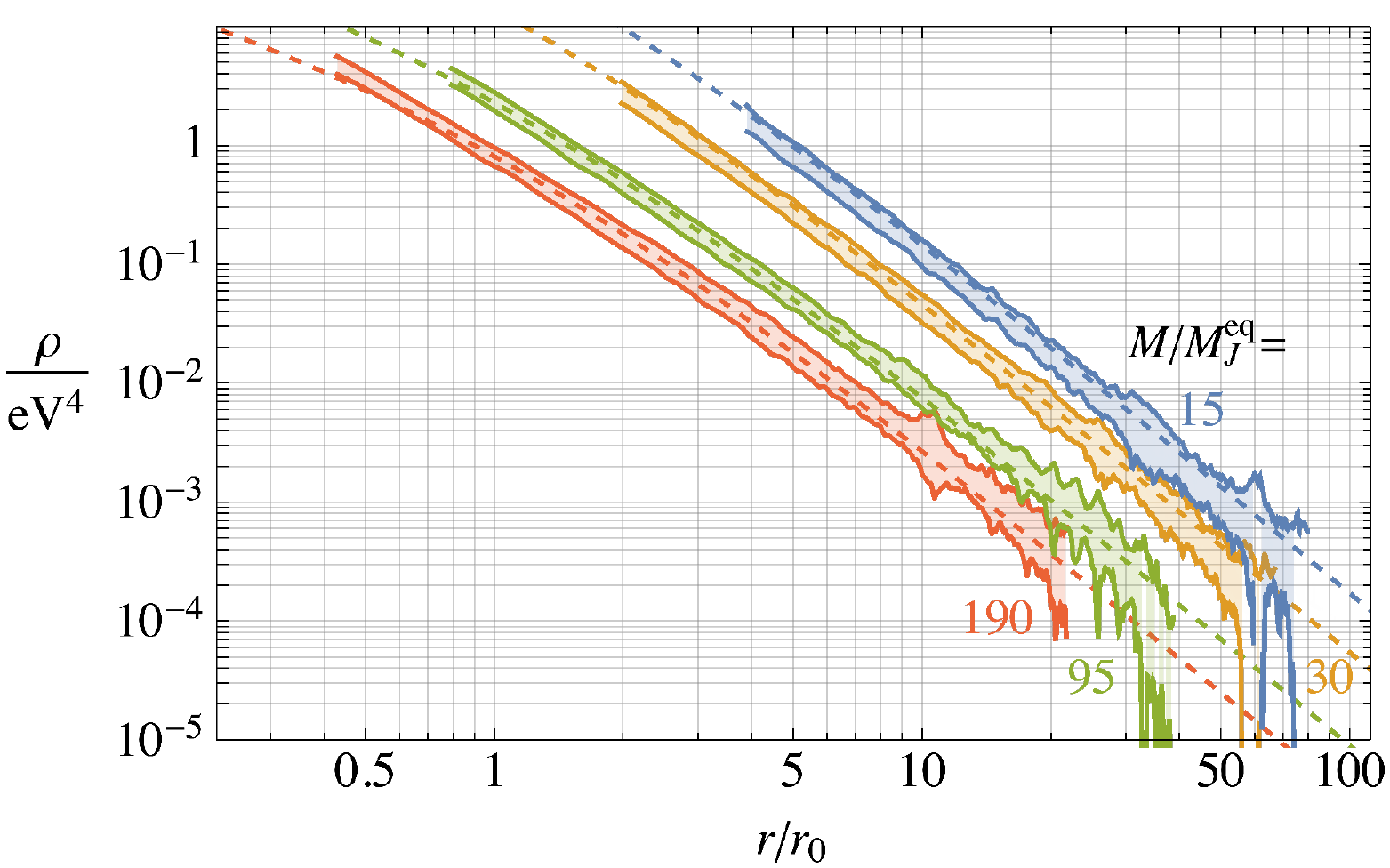}
		~~~				\includegraphics[width=0.475\textwidth]{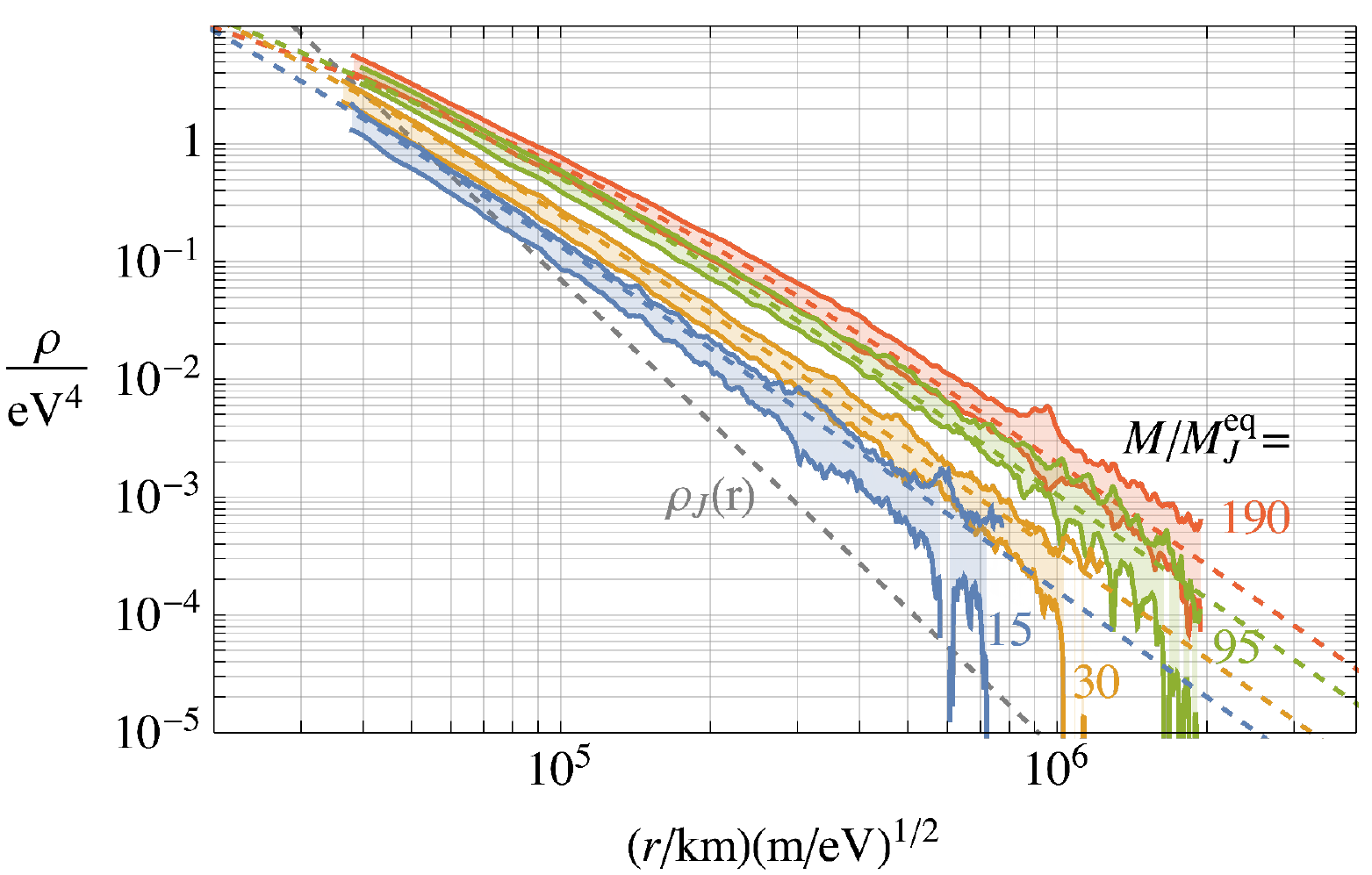}
	\end{center}
	\caption{The (spherically averaged) density profile of compact halos of different masses at $z= 70$, where all halos within $1\%$ of the labeled mass are averaged. Also plotted is the analytic expectation of the halo profiles given in eqs.~\eqref{eq:rho0compact} and~\eqref{eq:r0compact}. The profiles are plotted in terms of the scale parameter $r_0$ of the NFW profile (which depends on the halo mass), left, and the physical distance from the center, right. Note that lighter halos, which are produced earlier, have larger density parameter $\rho_0$ as can be seen in the left panel, despite having smaller density at a fixed physical distance from their centres. In the right panel, we also plot the density $\rho_J(r)$, defined by $\lambda_J(\rho_J)= r$. Quantum effects are relevant for densities $\rho \lesssim \rho_J$, since $\lambda_J \propto \rho^{-1/4}$.
	\label{fig:profile_white_noise}} 
\end{figure}

The energy density profile of the compact halos can be calculated in the same N-body simulations discussed in Section~\ref{ss:form_compact}. In Figure~\ref{fig:profile_white_noise} we show the (spherically averaged) profile of halos of different masses at $z=70$, together with the predicted analytic form in eqs.~\eqref{eq:NFWhalo},~\eqref{eq:rho0compact} and~\eqref{eq:r0compact}. Albeit not from first principles, we see that the analytic analysis is in remarkable agreement with the data, which indeed follow an NFW profile. Although in Figure~\ref{fig:profile_white_noise} we only show the profiles of the halos present at a fixed redshift, we checked that the analytic expectation in eqs.~\eqref{eq:rho0compact} and~\eqref{eq:r0compact} matches simulation results at other accessible values of the redshift, for approximately the same values of $\nu_c$,  $c_\Delta$, $\Delta$. Given the rough nature of our analytic analysis (e.g neglecting that some compact halos will merge, be bound inside bigger halos, accrete mass) we do not attempt a global fit, and the values given above are sufficiently accurate for our purposes.

We also note that, similarly to the halo mass distribution, the halo profile should be trusted only for compact halos with mass $M\gtrsim 50M_J^{\rm eq}$, which come from modes $k \ll k_\star$ so are unaffected by quantum pressure (at least during their linear growth). 
Additionally, the density profiles of compact halos in N-body simulations are only reliable in regions where the quantum pressure is negligible.\footnote{In Figure~\ref{fig:profile_white_noise} we plot the data only down to radius $r_{\rm soft}$ below which the N-body results do not reproduce the equations of motion of even cold dark matter, for numerical implementation reasons. Further details are given in Appendix~\ref{app:Nbody}.} In Figure~\ref{fig:profile_white_noise}  (right) we plot the curve $\rho_J(r)\propto r^{-4}$ defined by $\lambda_J(\rho_J) = r$. Since $\lambda_J \propto \rho^{-1/4}$, $\rho_J(r)$ is the minimum  average density a region of size $r$ can have before quantum pressure becomes relevant.\footnote{If in a region of size $r$ the density is $\rho<\rho_J(r)$, due to quantum pressure the configuration will relax (into a soliton) with radius larger than $r$ (and density even smaller than $\rho$).}  Therefore, regions of the halos for which $\rho(r) > \rho_J(r)$ are expected to be mostly unaffected by quantum pressure.   
We see that the majority of the compact halo mass is self-consistently in a region where quantum pressure is negligible, especially for relatively heavy halos, although this fails at the centre of the halos.  For instance, for $M/M_J^{\rm eq} \simeq 200$ less than $5\%$ of the mass of the NFW profile is inside $r_c$.

\section{Late Time Evolution} \label{s:late_time}

Possible observational and experimental signals of the solitons, their fuzzy halos, and the compact halos (which we collectively refer to as dark matter clumps) rely on them surviving undisrupted to the present day. The resulting discovery potential will depend on how often collisions between a clump and e.g. an observer on Earth occurs.

\subsection{Survival of the clumps} \label{ss:destroy}

There are various processes that could destroy a clump, the most important of which occur when it is bound inside a larger structure, e.g. a larger clump or the Milky Way itself. As previously discussed, the cosmological history of the clumps is extremely complicated. The majority of solitons (and also the relatively low mass compact halos) will be bound in a series of compact halos of increasing mass and decreasing average density. Subsequently the dark matter clumps will fall into adiabatic halos. For direct detection we are interested in clumps that are eventually bound in the Milky Way, with orbits that cross the neighbourhood of the Earth.  Such orbits could have a range of forms (mostly in the disk, on a rosette orbit, etc. \cite{Gnedin:1998bp}), which will affect the survival probability. 

Given this complexity we do not attempt to directly track the survival probability of a clump from when it formed to the present day in full detail.  Instead we estimate whether typical clumps survive by calculating the rate of the processes that are most likely to lead to their destruction. Most of our analysis will be done treating the  clumps as being made up of classical particles, although this will not be accurate for the solitons. Despite these approximations, our analysis will be sufficient to indicate whether or not solitons and compact halos are likely to survive in various environments. On the other hand, further dedicated study would certainly be worthwhile, especially of the possible destruction during hierarchical structure formation, which (although challenging to study) might be important.

Among  the mechanisms that can lead to destruction are tidal forces from a central potential, dynamical friction and tidal shocking by a central potential. For an object in the Milky Way there is also destruction by encounters with stars and by tidal forces from crossing the galactic disk. Destruction by tidal shocks is also possible when a soliton or compact halos falls into a larger compact halo (or a relatively small halo from a small scale adiabatic fluctuation) during hierarchical structure formation. We analyse each of the destruction processes in detail Appendix~\ref{app:survival}. Here we summarise the main conclusions: First, the only property of a clump that the rate of destruction depends on is its mean density $\bar{\rho}(r) = \int_0^r \rho(r) 4\pi r^2dr/\left(4/3\pi r^3 \right)$ (for all the important processes). Consequently the disruption is independent of the dark photon mass.  Regions of clumps within which the mean dark matter density $\bar{\rho}(r)$ is larger than about $0.05 \eV^4$ are likely to survive to the present day if they end up on a typical orbit that passes through the Earth. 

The solitons have central densities $\rho_s \simeq \left( 0.1 \div 100 \right) \eV^4$ (see Figure~\ref{fig:HMF_soliton}). From Figure~\ref{fig:profile_log}, the density profile switches from soliton-like to the fuzzy halo NFW form at radius of $r_{\rm edge} \simeq 2 \lambda_J(\rho_s)/3$. At this radius $\bar{\rho}(r_{\rm edge}) \simeq 0.2 \rho_s$ (with local density $\rho(r_{\rm edge})\simeq 0.05 \rho_s$). Consequently, the majority of the solitons, and especially the relatively heavy ones, which form at the times accessible in the simulations in Section~\ref{section:SPsection}, are likely to survive.

Given the profiles of the fuzzy halos around solitons, $\bar{\rho} \simeq  0.05\eV^4$ corresponds to regions of these with local densities $\rho(r) \simeq  10^{-3} \eV^4$.  Therefore, the outer part of the fuzzy halos around solitons is likely to be destroyed. This is to be expected given that the halos extend out to densities barely greater than the dark matter density in the neighbourhood of the Earth. However, the central part of the fuzzy halos are likely to survive. The density in the part of the fuzzy halos that survives corresponds to an enhancement over the local dark matter density (which for definiteness we fix to $0.5 \GeV/\cm^3$) of $\sim 10^3$.

The compact halos are far less dense than the solitons and much more likely to be destroyed. Compact halos with masses in the range $\left( 10^2 \div 10^4 \right) M_J^{\rm eq}$ (see Figure~\ref{fig:HMF_halo}) have typical mean densities $\bar{\rho} \simeq (10^{-6} \div 10^{-3}) \eV^4$ and the local densities at their edges are $\left(10^{-7} \div 10^{-4}\right) \eV^4$ (see Figure~\ref{fig:destruction_profile_2} right), which at the lower end is even smaller than the local dark matter density. Consequently, the majority of such halos, which contain $\simeq 70\%$ of the dark matter at $z\simeq 10$, are likely to be destroyed. The relatively dense core of the compact halo might survive, but these will contain a smaller fraction of the dark matter. Additionally, less massive (more dense) compact halos that are subsequently bound in the most massive compact halos might survive.

\subsection{Collision rate} \label{ss:collision_rate}

We now analyse the rate at which a point-like observer collides with a compact object. Given the uncertainty about whether the compact halos survive to the present day, we focus on collisions with the solitons and the parts of their surrounding fuzzy halos with local density $\rho \gtrsim 0.01\eV^4$ (corresponding to a mean density $\bar{\rho} \gtrsim 0.05\eV^4$), which are likely to persist to today. With an eye to future analysis of direct detection signals, we present our results assuming a local dark matter density $\rho_{\rm local} = 0.5\GeV/\cm^3$ (this is subject to significant uncertainties, see e.g. \cite{Buch:2018qdr}), corresponding to the Earth's local environment, but they can easily be rescaled to other densities.

For convenience, we note that a soliton of mass $M$ has central density, given by eq.~\eqref{eq:rhos}, that corresponds to
\begin{equation}
	\rho_{s} \simeq 1.51\cdot 10^4 \eV^4 \left( \frac{M}{M_J^{\rm eq}} \right)^{4} ~,
\end{equation}
and a size set by $\lambda_J(\rho_{s})$, which, from eq.~\eqref{eq:quantumJeans}, is
\beq \label{eq:lambdaJkm}
\begin{aligned} 
	\lambda_J(\rho_{s}) 
	&= 4.6 \cdot 10^3~ {\rm km} \left(\frac{\eV}{m}\right)^{1/2} \left(\frac{M_J^{\rm eq}}{M}\right) ~,
\end{aligned}
\eeq
where $M_J^{\rm eq}$ is given in terms of solar masses, as a function of $m$, in eq.~\eqref{eq:MJeq}.

We can easily obtain an analytic estimate of the collision rate. Approximating that the solitons all have a single mass $M$, their local number density is
\beq
\begin{aligned}
	n & = f_s \bar{\rho}(t_0)/M 
	& \simeq  10^{20} {\rm pc}^{-3} \left( \frac{f_s}{0.05} \right) \left( \frac{\rho_{\rm local}}{0.5 \GeV/{\rm cm}^3 } \right) \left( \frac{0.1 M_J^{\rm eq}}{M} \right) \left( \frac{m}{ \eV} \right)^{3/2} ~,
\end{aligned}
\eeq
where, as in Section~\ref{section:SPsection}, $f_s$ is the fraction of DM in solitons.\footnote{See~\cite{Banerjee:2019epw} for a similar calculation.} The number of collisions per unit time between a point in space (e.g. a dark matter detector) and solitons is therefore 
\beq \label{eq:rateapprox}
\begin{aligned}
	\Gamma & \simeq n \pi R^2 v_{\rm rel}  \\
	& \simeq \frac{0.1}{{\rm yr}} \left(\frac{m}{ \eV}\right)^{1/2}  \left( \frac{0.1 M_J^{\rm eq}}{ M} \right)^3 \left( \frac{v_{\rm rel}}{10^{-3}} \right)  \left( \frac{f_s}{0.05} \right) \left( \frac{\rho_{\rm local}}{0.5 \GeV/{\rm cm}^3 } \right) ~,
\end{aligned}
\eeq
where $R$ is the maximum radius that the soliton profile extends to, which in the second line we have set to $\lambda_J(\rho_s)$ given the results of Section~\ref{ss:fuzzy_halo}.

Apart from the explicit dependence, all the factors in the second line of eq.~\eqref{eq:rateapprox} are independent of the dark photon mass $m$. Consequently the interaction rate is larger when the dark photon mass is larger. This has a straightforward interpretation:  the fraction of DM in solitons, $f_s$, and the densities of the solitons are independent of $m$; therefore the fraction of time that a point spends inside a soliton is independent of $m$. However, the physical size of the solitons (with fixed $M/ M_J^{\rm eq}$) is inversely proportional to $m^{1/2}$ (see eq.~\eqref{eq:lambdaJkm}), so each encounter with a soliton lasts less time when $m$ is larger. In particular,  a collision with impact parameter $b$ with a soliton lasts for roughly
\beq
t_{\rm collision} \simeq 10^2 ~{\rm s} \left( \frac{0.1 M_J^{\rm eq}}{ M} \right) \left(\frac{ \eV}{m}  \right)^{1/2} ~.
\eeq
Consequently the collision lasts roughly a  minute for $m\simeq \eV$.  From Figure~\ref{fig:HMF_soliton} we see that during the collision the dark matter density is typically enhanced by a factor $\sim 10^4 \div 10^7$ compared to the local density. Given that the halos around solitons might survive out to approximately $\simeq 10 \lambda_J(\rho_{s})$, the interaction rate with these is expected to be substantially larger and the interaction time significantly longer (although the typical enhancement over the local density will be smaller).

We can make more accurate predictions using the simulation and analytical results for the soliton mass function from Section~\ref{ss:solitonmassf}.  We continue to consider the rate at which a single point collides with a clump. Given the size of the solitons eq.~\eqref{eq:lambdaJkm} this will be appropriate for direct detection signals, and also e.g. neutron stars provided $m\lesssim 10^5 \eV$. It is straightforward to repeat our calculations to determine the rate of collision between an object with size comparable to $\lambda_J(\rho_s)$ and clumps.

\begin{figure}[t!]
	\begin{center}
		\includegraphics[width=0.75\textwidth]{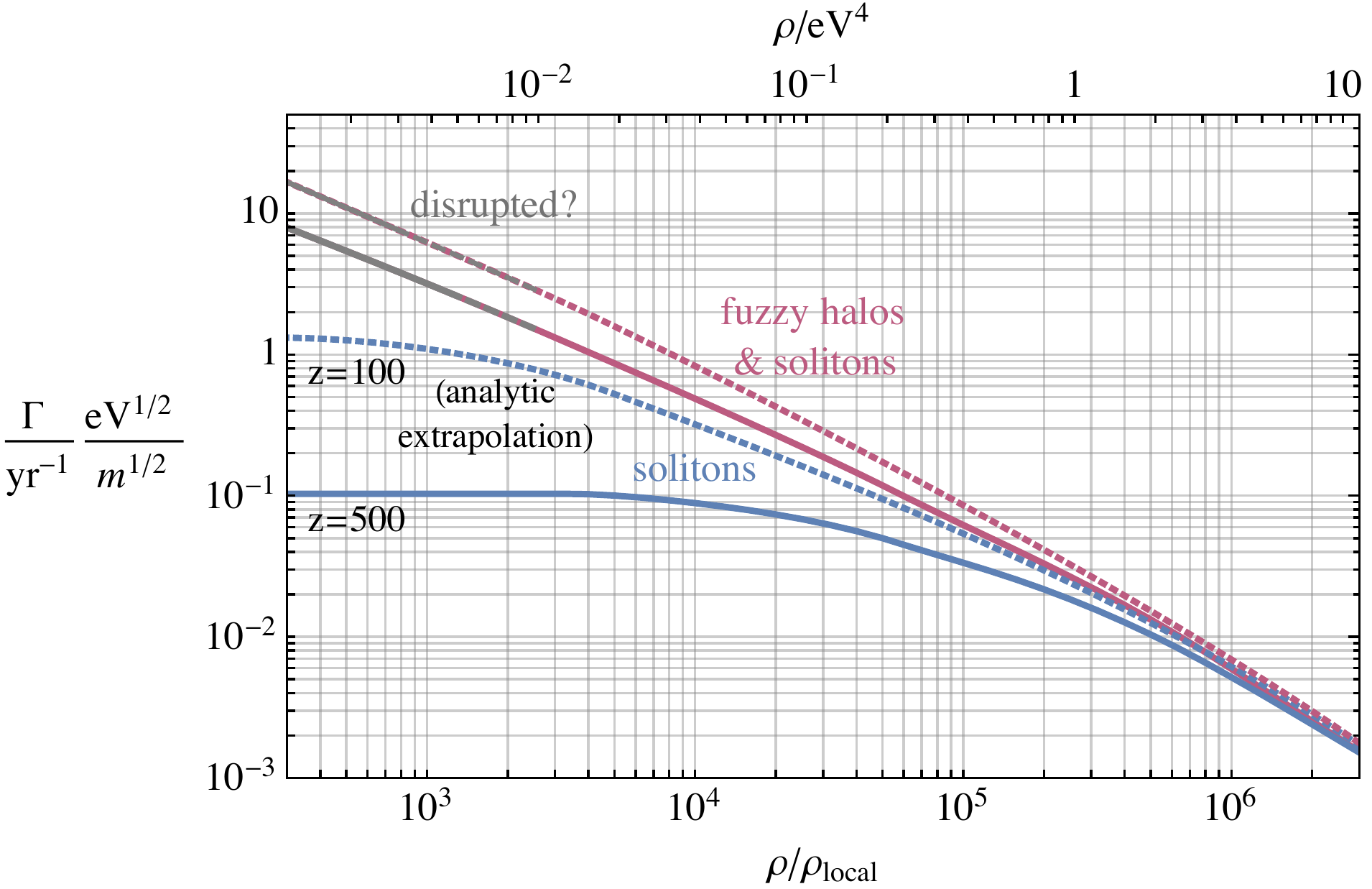}
	\end{center}
	\caption{The expected rate $\Gamma$ at which a point-like observer collides with a dark photon clump in such a way that the (maximum) resulting dark matter density enhancement seen is  $\geq \rho/\bar{\rho}$. Results are shown considering collisions with the central regions of solitons, $r/\lambda_J(\rho_s)<0.6$, (blue) and due to colliding with either a soliton or its surrounding halo (purple). As discussed in the main text, $\Gamma$ is proportional to $m^{1/2}$. Results are plotted using the soliton mass function at $z=500$ obtained directly from simulation data  (solid) and from the analytic extrapolation of the soliton mass function to $z=100$ when the collapse of the $k^3$ fluctuations is expected to prevent further formation of solitons (dashed). The fuzzy halos around the solitons are assumed to have the NFW form seen in simulations. The local dark matter density is fixed to $0.5 \GeV/{\rm cm}^3$ and the corresponding physical densities that the observer passes through are shown on the upper axis. Regions of the halos surrounding solitons with densities $\lesssim 10^{-2} \eV$ are likely to be disrupted in the local environment. The values of $\rho/\rho_{\rm local}$ where this will affect $\Gamma$ are plotted in grey. The majority of the solitons contributing to the rate are expected to survive to the present day. \label{fig:direct_rate}} 
\end{figure}

In Figure~\ref{fig:direct_rate} we plot the rate  $\Gamma$ at which collisions that result in a dark matter density enhancement of at least $\rho/\rho_{\rm local}$ occur. In other words, a point is expected to experience a collision that  results in a density enhancement (at its peak) of at least $\rho/\rho_{\rm local}$ roughly once per $\Gamma^{-1}$ time. In this plot we assume $\rho_{\rm local} = 0.5 \GeV/\cm^3$, and a mean relative velocity of $10^{-3}$ between the solitons/fuzzy halos and the clump. The scaling with $\Gamma \propto m^{1/2}$ in the estimate of eq.~\eqref{eq:rateapprox} is exact also for the full analysis, so we factor this out on the vertical axis.

Figure~\ref{fig:direct_rate} shows $\Gamma$ considering only collisions with the solitons (given Figure~\ref{fig:profile_log}, we define this as the region $r< \lambda_J(\rho_s)$) and allowing collisions with the halos surrounding the solitons. Additionally we plot results obtained from the soliton mass function at the final simulation time and from the analytic extrapolation, see Figure~\ref{fig:HMF_soliton}. The rate of collision with the halos surrounding the solitons are obtained by assuming these take the form of the NFW halo plotted in Figure~\ref{fig:profile_log}.\footnote{As discussed in Section~\ref{ss:fuzzy_halo}, for relatively heavy solitons at the end of the SP simulations the NFW  halos already extends to roughly the $\rho$ that are likely to survive tidal disruption. Meanwhile for lighter solitons the halo has not yet extended this far, but it is likely that it will continue growing.}  We also indicate the values of  $\rho/\rho_{\rm local}$ such that some of the halos that would contribute to the rate are likely to have been destroyed. For a dark photon mass of $m\simeq \eV$ collisions with solitons leading to enhancements in the dark matter density of a factor of $10^{3} \div 10^4$ occur on reasonable experimental timescales, and collisions with the surrounding halos are even more frequent.

\begin{figure}[t]
	\begin{center}
				\includegraphics[width=0.69\textwidth]{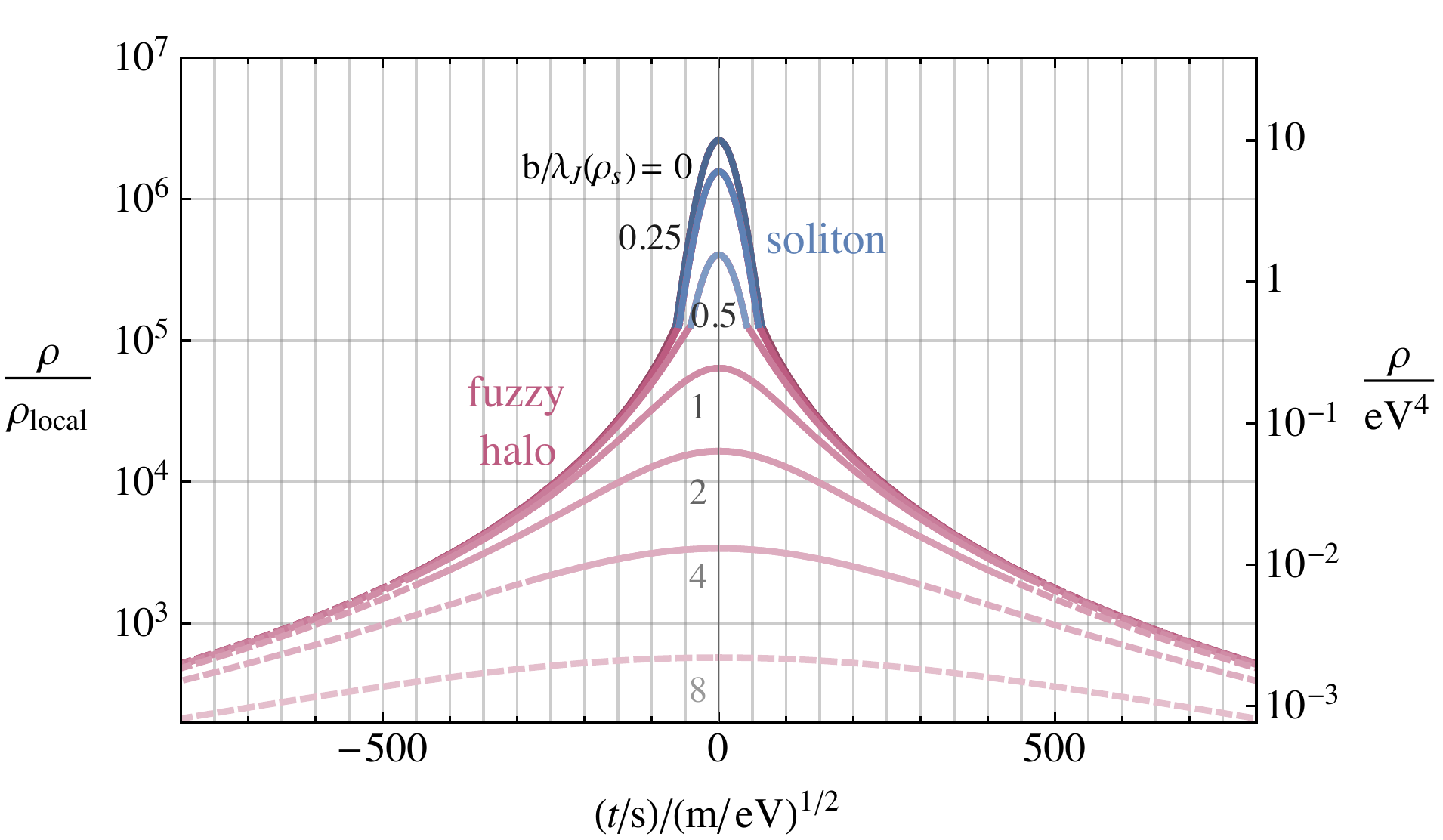}
	\end{center}
	\caption{
		The dark matter density as a function of time encountered by a point-like observer colliding with a soliton, and its fuzzy halo. The soliton's central density is fixed to $\rho_s=10\eV^4$. We show results for impact parameter $b$ varying from $0$ (passing through the centre of the soliton) to those that only pass through the fuzzy halo, measured in units of $\lambda_J(\rho_s)$, the quantum Jeans distance at the soliton core (the soliton-like part of the clump extends to roughly $\lambda_J(\rho_s)$). We also indicate the parts of the fuzzy halo that are likely to have been disrupted prior to the collision, in the Earth's local environment. 
		 \label{fig:direct_time}} 
\end{figure}

As well as the maximum dark photon density experienced, the time that an enhancement lasts for and the density profile experienced is also potentially important for detection and observation. This can easily be extracted from the density profile of the solitons/fuzzy halos (and likewise the compact halos if they are assumed to survive). In Figure~\ref{fig:direct_time} we plot an example of the density seen as a function of time during a collision with a soliton/fuzzy halo of central density $10\eV^4$ (corresponding to soliton mass of  $0.16 M_J^{\rm eq}$, in the main part of the soliton mass function of Figure~\ref{fig:HMF_soliton}), for different impact parameters that both lead to collisions with the soliton-like part of the object and that are just with the fuzzy halo. We also indicate the part of the signal corresponding to regions of the fuzzy halo that are likely to have been disrupted prior to the collision. Such disruption could lead to more diffuse streams of dark matter around the object \cite{Tinyakov:2015cgg}, resulting in interesting features in the density as a function of time when an observer passes through the edge of a dark photon clump, although we do not investigate this further in our current work.

An additional potentially important feature is that the dark photon field is coherent over the entire soliton core (since the field's de Broglie wavelength inside the core coincides with the size of the soliton). This is far larger than for unbound dark matter, for which it is $2\cdot 10^{-4}\,{\rm m} \left(\eV/m\right) \left(10^{-3}/v \right)$. Even in the fuzzy halo the typical velocity (from the NFW profile) is tiny, and the de Broglie wavelength is typically less than a factor of $10$ smaller than in the soliton core, typically in the range $10^2 \,{\rm km} (\eV/m)^{1/2}$ (see eq.~\eqref{eq:lambdaJkm}), still many orders of magnitude larger than it would otherwise be. 
Finally, we note that we have assumed that the soliton retains its form throughout the collision. 
In reality, the tidal forces experienced by a clump during a collision with the Earth are 
likely to lead to its eventual destruction. However, the clump's profile remains mostly unchanged until long after passing the Earth and this effect will not alter the density profile seen by a detector \cite{Arvanitaki:2019rax}.\footnote{Moreover, note that for $m$ sufficiently large, the encounter of solitons could modify the expectation of gravitational focusing of ultralight DM, studied in~\cite{Kim:2021yyo} if the field is made of waves.}

\section{Discussion and Future Directions} \label{sec:Conclusion}

There are a plethora of potential observational and experimental implications of the solitons and their fuzzy halos, which we will study in detail in a companion paper. For $m \gtrsim \eV$, a typical observer in the local environment collides with solitons regularly, and during each collision the dark matter density is boosted by a factor of up to $10^6$ and is far more coherent than it is outside solitons. For smaller $m$, collisions are rarer, but the clumps contain more mass and are larger in size. Although less relevant for direct detection, there is a range of possible striking astrophysical and indirect signals in this case. If the vector has self-interactions, the solitons might be unstable and could explode, leading to further potential signals, similar to the analysis for axion stars in \cite{Eby:2021ece}. Moreover, compact halos have a macroscopic mass (of the order of the mass of a planet for $m \lesssim 10^{-5}$ eV), and -- being abundant in our Galaxy -- can in principle be observed and constrained via gravitational measurements. 

Notice that the classical description of the field is valid when the number of particles $\mathcal{N}$ within a de Broglie wavelength $\lambda_{\rm dB}\equiv 2\pi/(mv)$ (of the order of the inverse typical momentum) is much greater than 1: i.e. $\mathcal{N} =  \bar{\rho} \lambda_{\rm dB}^3 / m \gg 1$. As is well known, for this to be valid today requires $m\lesssim 10$~eV if the field is made out of waves. However, inside the solitons the occupation number is much larger, and therefore -- as we now show -- the validity of our results extends to much larger $m$. 

The crucial condition for the solitons to form is that the vector bosons are classical at around MRE when the overdensities start collapsing, as described in Section~\ref{section:SPsection}. Since the soliton energy density and occupation number remain constant, solitons never exit the classical regime afterwards. Classicality at MRE requires
%If the vector bosons are classical at around MRE when the overdensities collapse, solitons will form at that time (as described in Section~\ref{section:SPsection}) and since their energy density and occupation number remain constant they never exit the classical regime afterwards. The classicality at MRE requires %$\left( 2\pi / k_\star \right)^3  \rho_{\rm eq}/m \simeq 90 \rho_{\rm eq}^{1/4}/ (G^{3/4} m^{5/2}) \gg 1 $
\begin{equation} \label{eq:NoccMRE}
\mathcal{N} \simeq \left( \frac{2\pi}{k_\star/a_{\rm eq}} \right)^3  \frac{\rho_{\rm eq}}{m} \simeq 90 \frac{\rho_{\rm eq}^{1/4}}{ G^{3/4} m^{5/2}} \simeq 30 \left( \frac{10^{17} \eV}{m} \right)^{5/2} \gg 1 ~.
\end{equation}

%Since the de Broglie wavelength is of order the size of the soliton, the occupation number in the core of the soliton is large if
%\begin{equation} \label{eq:NoccS}
%	N_{\rm occ} \simeq \frac{ \rho_s R^3}{m}  \simeq \frac{M}{m} \simeq 20 \left( \frac{10^{17} \eV}{m} \right)^{5/2} \gg 1 ~,
%\end{equation}
%using eqs.~\eqref{eq:rhos} and \eqref{eq:MJeq}.  The constraint on $m$ in eq.~\eqref{eq:NoccS} coincides with the condition for the classical description to be valid around MRE, which is required for our analysis of Section~\ref{section:SPsection} to hold, i.e. 
% $\left( 2\pi / k_\star \right)^3  \rho_{\rm eq}/m \simeq 90 \rho_{\rm eq}^{1/4}/ (G^{3/4} m^{5/2}) \gg 1 $.
This condition is parametrically the same as requiring that the occupation number in the core of the soliton is large ($\rho_s R^3/m  \simeq M/m\gg1$), since the density and size of the solitons match the typical density of the field and the size of the fluctuations at MRE. 
Eq.~\eqref{eq:NoccMRE} implies that our results will apply for all the masses in the range $ m\lesssim 10^{8}$ GeV% (note that this also includes heavy gauge bosons)
. The upper limit happens to coincide with the largest $m$ that does not exceed the Hubble parameter during inflation, requiring that the dark photon makes up the full relic abundance,  see eq.~\eqref{eq:OmegaDM}.

For  $m \gtrsim 10 \eV$ in most of the Universe one needs to consider the heavy gauge bosons as quantum objects (similarly to WIMPS), however the solitons are perfectly captured by the classical description.\footnote{We expect the solitons to be stable, even if surrounded by a non-classical background (indeed they might accrete via gravitational relaxation), although a detailed analysis would be useful.}  
In this regime, the soliton radius is tiny, but the number density of solitons and their encounter rate with the Earth (and other astrophysical objects) is huge. There are potentially dramatic signatures, e.g. at direct detection experiments that aim to detect the collision of single particles.

\subsection{Possible improvements and extensions}

There are a number of ways that our analysis could be refined or extended. One direction is to better test, and if necessary improve, our analytic prediction of the soliton mass function from the density power spectrum.\footnote{Given the local (quadratic) non-Gaussianities in the initial energy density field and the similarity with primordial black holes discussed in Section~\ref{ss:SP_simulations}, it could be useful to follow approaches already developed in this case~\cite{Riccardi:2021rlf}.} This would be particularly valuable when studying similar dynamics in a theory where the initial power spectrum changes as the theory's parameters vary, or in the context of a theory where the initial power spectrum is uncertain. It would also be interesting to study the fuzzy halos that surround the solitons in detail.  For example, simulations of ultra-light axions seem to show a relationship between the masses of solitons that form at the centre of galaxies in such models and their halos  \cite{Bar:2018acw}, and it would be nice to understand if our fuzzy halos follow a similar relationship. 
Additionally, it would be useful to study  the transition between collapse to objects supported by quantum pressure and collapse to objects well approximated as being composed of cold DM more systematically. Although we have analysed the compact halos assuming that quantum pressure is negligible, which is self-consistent over most of their profile especially for the most massive objects, solitons are likely to still form at their cores, and these could be relevant for observation or detection signals. It would also be interesting to study the effect of the vector having self-interactions, or interactions with the SM or new fields. For example, this could lead to the solitons decaying on cosmological timescales, potentially leading to observational signals~\cite{Hertzberg:2010yz,Eby:2017azn,Croon:2018ybs,Eby:2020ply}, and it could also affect the rate at which solitons form or gain mass by dynamical relaxation~\cite{Chen:2021oot}. 

%Additionally, the solitons already present might increase their mass by accreting the background DM via gravitational relaxation, or solitons might merge together. This lead to an uncertainty on their mass. We do not try to study these potentially important issues in our present paper.

A key direction for future work is to determine the probability that the solitons, fuzzy halos and compact halos survive to the present-day more accurately. In our analysis we had to make multiple approximations and focus only on particular processes. It would certainly be valuable to better understand the probability of destruction or possible accretion during hierarchical structure formation and also to determine whether wave-like effects alter the probability of the solitons or fuzzy halos surviving. One could also carry out a more detailed study of the probability of destruction by collisions with stars, e.g. along the lines of \cite{Delos:2019tsl,Kavanagh:2020gcy,Facchinetti:2022sai}. The solitons, as mentioned in Section~\ref{ss:SP_simulations}, could also merge in the late Universe or when they are bounded into halos, which could change their mass distribution \cite{Diamond:2021dth}. %Additionally, as mentioned in Section~\ref{ss:SP_simulations}, it would be interesting to understand whether solitons might merge when they are bounded into halos.

On the numerical side, there are several directions in which our approach could be developed, which could help address some of the issues above. Our simulations of the SP system could only reach $a/a_{\rm eq}=7$ and it was impossible for us to use them when analysing the compact halos in Section~\ref{s:compact_halos}. An adaptive mesh approach, e.g. as employed in \cite{Schive:2014dra,Schive:2014hza} could enable a much greater range in $a/a_{\rm eq}$. One could also employ a hybrid SP - N-body approach, evolving the SP equations only in regions where quantum pressure is relevant, or modify the equations of motion in N-body simulations to attempt to reproduce the effects of quantum pressure. With such work, one might be able to see whether the solitons survive if they are bound in a compact halo.

It would be valuable to understand whether vector solitons could form with other production mechanisms, e.g. by gravitational condensation analogous to the way that QCD axion stars are thought to form in the post-inflationary scenario \cite{Levkov:2018kau,Eggemeier:2019jsu}. We have focused on a very minimal theory, but it would also be interesting to study the impact of additional interactions. As mentioned at the end of Section~\ref{s:initial}, these could lead to changes at all stages, from the production of inflationary fluctuations through to the dynamics around $H_\star$, MRE and the present day (see also \cite{Arvanitaki:2021qlj}). Additionally, there could be new types of compact object that form when the vector has interactions \cite{Zhang:2021xxa}.  Last, it would certainly be worthwhile to analyse what changes if the dynamics responsible for the vector boson's mass lie below the scale of inflation.

\section*{Acknowledgements}

We thank Giovanni Villadoro for valuable discussions, and Josh Eby, David J.E. Marsh and Chen Sun for comments on a draft. EH thanks David J.E. Marsh and the Liverpool John Moores N-body simulation group for useful discussions. 
MG is grateful to Kfir Blum, Jovan Markov, Gilad Perez and Marko Simonovic for stimulating discussions and insights. 
EH, JMR, NS and SW would like to thank the UK Science and Technology Facilities Council (STFC) for funding this work through support for the Quantum Sensors for the Hidden Sector (QSHS) collaboration under grants ST/T006102/1, ST/T006242/1, ST/T006145/1, ST/T006277/1, ST/T006625/1, ST/T006811/1, ST/T006102/1 and ST/T006099/1. EH is also supported by UK Research and Innovation Future Leader Fellowship MR/V024566/1 and STFC grant ST/T000988/1. JMR also thanks the STFC Quantum Sensors for Fundamental Physics and Society initiative ST/S002227/1, the MAGIS-100 (Gordon and Betty Moore Foundation and DOE QuantISED program) and AION (ST/T006633/1) collaborations, and support from STFC grant ST/T000864/1.  We acknowledge use of the University of Liverpool Barkla HPC cluster.

\appendix

\section{Details of the Initial Conditions from Inflation} \label{app:initial_conditions}

In this Appendix we discuss some more details on the production of the vector field during inflation, which was first analysed in \cite{Graham_2016} (see also \cite{kolb2021completely}). In particular, we give the derivation of eq.~\eqref{eq:PAt} and report the useful analytic approximation eq.~\eqref{eq:Pdelta}.

To derive eq.~\eqref{eq:PAt}, we estimate the solution $A_L(t,k)$ of eq.~\eqref{eq:eomAL} in the limits $k\ll k_\star$ and $k\gg k_\star$. While relativistic and superhorizon, for all modes eq.~\eqref{eq:eomAL} is approximated by $(\partial_t^2+H\partial_t)A_L=0$, with obvious solution $A_L\simeq A_{L,0}$ (i.e. the modes are frozen) and $\rho\simeq m^2A_L^2/a^2\propto a^{-2}$. However:
\vspace{-1.5mm}{}
\begin{enumerate}[label={(\arabic*)},itemsep=-0.4ex,leftmargin=0.25in]
	\item Even after they become nonrelativistic, the modes with $k<k_\star$ while superhorizon still follow $(\partial_t^2+H\partial_t)A_L=0$. Therefore $A_L\simeq A_{L,0}$, and their energy density decreases as $\rho\simeq m^2A_L^2/a^2\propto a^{-2}$. This behaviour of $\rho$ is crucially different than for a scalar field, for which $\rho$ is frozen for nonrelativistic superhorizon modes.\footnote{As mentioned in the main text, the difference is due to the form of the mass terms of such particles, which controls the energy density of nonrelativistic superhorizon modes: for a scalar, $\frac12m^2\varphi^2$ does not change during the universe expansion, while $\frac12m^2g^{ij}A_iA_j\propto a^{-2}$ for a vector.} The solution $A_L\simeq A_{L,0}$ is valid until $H=m$ (i.e. $a=a_\star$), after which eq.~\eqref{eq:eomAL} is approximated by $(\partial_t^2+H\partial_t+m^2)A_L=0$, with solution $A_{L}\propto a^{-1/2}$ and $\rho \propto a^{-3}$, which is the usual matter behaviour.
	
	\item On the other hand, the modes with $k>k_\star$ follow $(\partial_t^2+3H\partial_t+k^2/a^2)A_L=0$ when they enter the horizon at $k/a=H$ (while still relativistic), and have solution $A_L\propto a^{-1}$ so $\rho\propto a^{-4}$. When they become nonrelativistic at $k/a=m$, they follow $(\partial_t^2+H\partial_t+m^2)A_L=0$, as before with solution $A_{L}\propto a^{-1/2}$ and $\rho \propto a^{-3}$.%, which is the usual matter behaviour. 
\end{enumerate}
Summarising, after they become nonrelativistic and subhorizon, all modes behave like matter. Approximating the transition between the different regimes as immediate, we have for low frequency modes $A_L/A_{L,0}\simeq (a_\star/a)^{1/2}%=(H/m)^{1/4}
$, while for high frequency modes $A_L/A_{L,0}\simeq (a_{\rm e}/a_{\rm nr})(a_{\rm nr}/a)^{1/2}=(k_\star/k)^{3/2}(a_\star/a)^{1/2}$, where $k/a_{\rm e}\equiv H(a_{\rm e})$ and $k/a_{\rm nr}\equiv m$ and we assumed radiation domination, for which $H\propto a^{-2}$. This allows us to conclude that in radiation domination $A_L$  has a Gaussian power spectrum given by (using eq.~\eqref{eq:PAt})
\begin{equation}\label{eq:PAt1_b}
	\mathcal{P}_{A_L}(t,k)=\left(\frac{ k_\star H_I}{2\pi m}\right)^2\left(\frac{a_\star}{a}\right) F^2_{A_L}\left[k/k_\star\right]\simeq \left(\frac{ k_\star H_I}{2\pi m}\right)^2\left(\frac{a_\star}{a}\right) \frac{(k/k_\star)^2}{1+(k/k_\star)^3}\, ,
\end{equation}
where the exact form of $F_{A_L}[x]$ (see Figure~\ref{fig:Pspectra} left) can be by extracted by solving numerically eq.~\eqref{eq:eomAL}, but from the discussion above $F_{A_L}[x]\to \{x,x^{-{1/2}}\}$ in the asymptotic limits $x\to \{0,\infty\}$. A very good analytic approximation is $F_{A_L}[x]\simeq x/\sqrt{1+x^3}$, reported in the right hand side of eq.~\eqref{eq:PAt1}.

Finally, using the fact that $\partial_t A_L$ and $A_L$ are independent Gaussian fields (and that $\mathcal{P}_{\partial_t A_L}=m^2\mathcal{P}_{A_L}$), it is straightforward to get an expression for the density power spectrum \cite{Graham_2016},
\begin{equation}\label{eq:Pdelta_2}
	\mathcal{P}_{\delta}(t,k)=\frac{k^2}{8\langle A^2_L\rangle^2}\int_0^{\infty} dq\int^{q+k}_{|q-k|} dp \frac{(k^2-q^2-p^2)^2}{q^4p^4}\mathcal{P}_{A_L}(t,p)\mathcal{P}_{A_L}(t,q)\simeq \frac{\sqrt{3} (k/k_\star)^3}{\pi  \left((k/k_\star)^{3/2}+1\right)^{8/3}} \ ,
\end{equation}
where the second equality is the approximate expression reported in eq.~\eqref{eq:Pdelta}. Note that the power $\mathcal{P}_{\delta}$ in~\cite{Graham_2016} is a factor of $2$ larger, because that reference neglected the first term in eq.~\eqref{eq:rho}.

\section{Evolution of Overdensities} \label{app:over}

The evolution of small density fluctuations during radiation and matter domination can be analysed perturbatively, with and without quantum pressure. Although this is not applicable to the large density perturbations at the scales  $k\simeq k_\star$ (see Figure~\ref{fig:1dslice}), the perturbative analysis still gives a useful hint towards the type of effects quantum pressure might have.

In the absence of quantum pressure, the evolution of a density perturbation $\delta(k)$ (in momentum space) as a function of time $t$, with scale factor $a(t)$, is well known \cite{coles2003cosmology} 
\beq \label{eq:pertEv}
\ddot{\delta}+\frac{2}{a}\dot{a}\dot{\delta} - 4\pi Gf\rho_{\rm nr} \delta=0 ~,
\eeq
where here $f= \Omega_{A}/\Omega_{\rm M}$ with $\Omega_{\rm M}$ the total matter content (this differs from e.g. $f_s$ used in the main text, which was normalised to $\Omega_{\rm DM}$). Eq.~\eqref{eq:pertEv} is obtained by combining eqs.~\eqref{eq:cont},~\eqref{eq:Euler} and~\eqref{eq:Poisson} (with $\Phi_Q=0$) in the limit of small overdensity and velocity. Defining $y= a/a_{\rm eq}$, eq.~\eqref{eq:pertEv} simplifies to
\beq
\frac{\partial^2\delta}{\partial y^2}+\frac{2+3y}{2y(1+y)}\frac{\partial \delta}{\partial y} - \frac{3 f}{2y(1+y)} \delta=0 ~.
\eeq
If $f=1$ this leads to the usual growing solution $\delta = \delta_0 (1+\frac{3}{2} \frac{a}{a_{\rm eq}})$, where $\delta_0$ is the overdensity at an early time, deep inside radiation domination: overdensity are frozen during radiation domination, and grow linearly in matter domination. For $f=0.84$ the (growing) solution is slightly modified, but the perturbation is still almost completely frozen before MRE and then grows. At large $a/a_{\rm eq}$ we have $\delta \propto (a/a_{\rm eq})^{\sqrt{1+24 f}/4}$.

Quantum pressure leads to an additional term, and the perturbation evolves according to
\beq
\frac{\partial^2\delta}{\partial y^2}+\frac{2+3y}{2y(1+y)}\frac{\partial \delta}{\partial y} - \frac{3 f}{2y(1+y)} \left( -1 +  \left( \frac{k}{k_{J}} \right)^4 \right) \delta =0 ~,
\eeq
where $k_J$ is the quantum Jeans momentum, and $k/k_J \propto y^{-1/4}$. For $k \gtrsim k_J$ the solutions oscillate, both in radiation domination ($y\ll1$, when they would otherwise be frozen) and in matter domination ($y\gg 1$, when they would otherwise grow). For $k \ll k_J$ the evolution is as in the absence of quantum pressure, as expected. A detailed analysis, including a study of the applicability of the perturbative expansion and the effects of the next terms in the expansion can be found in \cite{Li:2018kyk}.

\section{Solving the Schr\"odinger--Poisson equations} \label{app:SP}

\subsection{At matter radiation equality} \label{aa:SPMRE}

Here we summarise how the Schr\"odinger--Poisson (SP) system of equations eq.~\eqref{eq:sp1a} can be written in a form suitable for the numerical evolution and how we implement a realisation of the initial conditions from inflation.  

Around MRE the scale factor $a(t)$ satisfies $\dot{a}=a H(a)$ with
\begin{align}\label{hubble}
	H^2=\frac{8\pi G}{3}\rho_{\rm tot}=\frac{8\pi G}{3}\frac{\bar{\rho}_{\rm eq}}{2}\left[\left(\frac{a_{\rm eq}}{a}\right)^{3}+\left(\frac{a_{\rm eq}}{a}\right)^{4}\right]\equiv \frac{H_{\rm eq}^2}{2}\left[\left(\frac{a_{\rm eq}}{a}\right)^{3}+\left(\frac{a_{\rm eq}}{a}\right)^{4}\right] ~,
\end{align}
where $H_{\rm eq}^2\equiv 8\pi G\bar{\rho}_{\rm eq}/3$, so that (neglecting changes in the number of degrees of freedom, which is appropriate for the Standard Model) 
\begin{align}\label{eq:scalea}
	\frac{da}{dt}=a\frac{H_{\rm eq}}{\sqrt2}\sqrt{\left(\frac{a_{\rm eq}}{a}\right)^{3}+\left(\frac{a_{\rm eq}}{a}\right)^{4}}.
\end{align}
Note that the previous equation can be integrated exactly in conformal time. Similarly to \cite{Schive:2014hza,Paredes:2015wga,Edwards:2018ccc}, we rewrite the Schr\"odinger--Poisson system as
\begin{align}\label{sp2a}
	\left(i\partial_{\tilde{t}}+\frac{\nabla'^2}{2}-\Phi'\right)\psi'_i=0\\
	\nabla'^2\Phi'= a\sum_i\left(   |\psi'_i|^2 -\langle |\psi'_i|^2   \rangle\right)~,\label{sp2b}
\end{align}
where $t'=t/T$, $x'=x/\sqrt{T/m}$ (we call $L=\sqrt{T/m}$), $\Phi'=a^2\Phi/(mT)^{-1}$ and $\psi'_i=\psi_i/(\sqrt{4\pi G}T)^{-1}$, where $T$ is any inverse mass scale. We also defined $\tilde{t}=\int dt'/a^2$ (so that $d\tilde{t}/dt'=1/a^2)$.

Given the form of the initial conditions, it is natural to choose the typical length to be $L=2\pi(m a_\star/a_{\rm eq})^{-1}$%, where $t_\star$ is defined to be the time when $H=m$
. As in Section~\ref{ss:kstarkJ}, in the limit $a_\star/a_{\rm eq}\ll1$, $a_\star/a_{\rm eq}=(H_{\rm eq}/(\sqrt{2}m))^{1/2}g_R^{-1}$. Therefore we have 
$L=2\pi(\sqrt{2} H^{-1}_{\rm eq}g_R^2/m)^{1/2}$. 

Our choice of $L$ fixes the typical time to be $T=\sqrt{2}   \beta H_{\rm eq}^{-1}$ where $\beta\equiv (2\pi)^2g_R^{2}$. 
With this choice, eq.~\eqref{eq:scalea} simply becomes
\begin{align}\label{eq:scalea1}
	\frac{da}{d\tilde{t}}=\beta a^3\sqrt{\left(\frac{a_{\rm eq}}{a}\right)^{3}+\left(\frac{a_{\rm eq}}{a}\right)^{4}},
\end{align}
and the dependence on $H_{\rm eq}$ (as well as any numerical input except for $\beta$) have dropped out from eqs.~\eqref{sp2a}, ~\eqref{sp2b} and~\eqref{eq:scalea1}. From now on we will set $a_{\rm eq}=1$.

Next we discuss how we generate the initial conditions (at $a\ll a_{\rm eq}$). The field configuration $\psi_i$ is obtained from $A_i$ using the definition $A_i=\frac{1}{\sqrt{2 m^2 a^3}}(\psi_i e^{-imt}+c.c.)$, which implies $\text{Re}[\psi_i]=(a^3/2)^{1/2}(m  \cos(mt)A_i-\sin(mt)\dot{A}_i)$ and $\text{Im}[\psi_i]=-(a^3/2)^{1/2}(m \sin(mt) A_i+\cos(mt)\dot{A}_i)$. In the initial conditions the field $A_i$ and $\dot{A}_i$ have only longitudinal component ($A_L$ and $\dot{A}_L$). This is generated, using standard algorithms for a random field, according to the power spectra, $\mathcal{P}_{A_L}$ and $\mathcal{P}_{\partial_tA_L}$, defined by eq.~\eqref{eq:PX}. 
We use the expression of $\mathcal{P}_{A_L}$ in eq.~\eqref{eq:PAt1}. The same expression holds also for $\mathcal{P}_{\partial_t A_L}$, modulo a factor of $m^2$. 

The above procedure fixes the initial conditions except for the overall amplitude of $\psi_i$, that depends on the amount of vector dark matter present (ultimately linked to $H_I$). We normalise it such that $\sum_i\langle|\psi_i|^2\rangle=f\bar{\rho}_{\rm eq}/2$, where as before $f=\Omega_{\rm A}/\Omega_{\rm M}$,\footnote{This is because $\frac12\dot{A}_i^2+\frac12m^2A_i^2=|\psi_i|^2$ at the time when $a=q_{\rm eq}=1$.} which we set to $f= 0.84$, so that the vector is all of the observed dark matter.\footnote{Note that given the power $a^{-3/2}$ in the definition of $\psi$ from $A$, $\sum_i\langle|\psi_i|^2\rangle$ is constant in time; moreover, since $a=1$ at MRE, $\sum_i\langle|\psi_i|^2\rangle$ is the dark matter energy density at MRE.} Consequently the amplitude of $\psi'$ is set such that
\begin{align}\label{psiampl}
	\langle|\psi'|^2\rangle=\frac{\langle|\psi|^2\rangle}{(\sqrt{4\pi G}T)^{-2}} =\frac{f\bar{\rho}_{\rm eq}/2}{(\sqrt{4\pi G}\sqrt{2}\beta H_{\rm eq}^{-1})^{-2}}=%\frac{2 \beta^2f}{2}\frac{4\pi G\rho_{\rm eq} }{H_{\rm eq}^2}=
	\frac{3\Omega_{\rm A}/\Omega_{\rm M}}{2}\beta^2.
\end{align}
Integrating eq.~\eqref{sp2b} with the initial configuration of $\psi'$ just described gives the corresponding initial configuration of $\Phi'$.

Consequently, the SP system at around MRE boils down to eqs.~\eqref{sp2a} and~\eqref{sp2b} with $a$ given by integrating eq.~\eqref{eq:scalea1} with initial amplitude in eq.~\eqref{psiampl}.\footnote{Relativistic corrections to the SP equations have been studied \cite{Salehian:2021khb}, but these are negligible for our purposes.} Note that the only free parameters on which these equations depend are $g_R$ and $\Omega_{\rm A}/\Omega_{\rm M}$. Additionally, from eqs.~\eqref{sp2a} and~\eqref{sp2b} and~\eqref{eq:scalea1} with $a_{\rm eq}=1$ it follows that $\Phi'$ is of order one (which is the size of $\nabla'$ and $\partial_{\tilde t}$ when applied to $\Phi$) when $a\simeq 1/\delta$ from eq.~\eqref{sp2b}, and at this time the gravitational potential becomes relevant in the dynamics and $\psi'$ starts evolving nonlinearly (though, as we saw, perturbations do not collapse due to quantum pressure).

\subsection{Numerical solution} 

We solve eqs.~\eqref{sp2a} and~\eqref{sp2b} (together with eq.~\eqref{eq:scalea1}) on a finite lattice of constant comoving lattice spacing and in a periodic box of constant comoving length.  We use a 6th order pseudo-spectral algorithm, developed in \cite{Yoshida:1990zz} and applied to study the cosmology of axion stars in \cite{Levkov:2018kau} and fuzzy dark matter in \cite{Schwabe:2020eac} (see also e.g. \cite{Du:2018qor} for other similar implementations). 
Given the fields $\psi_i$ at time $t$, the fields at time $t+\Delta t$ are obtained by evaluating
\beq \label{eq:SP_ev}
\psi_i (t+\Delta t)= \left(\prod_{\alpha =1}^8 e^{-i d_\alpha \Delta \tilde{t} \Phi({\bf x})} e^{-i c_\alpha \Delta \tilde{t} k^2/2} \right) \psi_i (t) ~,
\eeq
where the product is ordered from right to left so the $\alpha=1$ part is applied to $\psi_i(t)$ first, and the evolution with the $k^2$ operator is understood to happen in momentum space. Here $k$ denotes the momentum in units of $2\pi/L$. The potential $\Phi$ is re-calculated after every step with the momentum operator, by solving the Poisson equation, eq.~\eqref{sp2b}, numerically (by transforming to momentum space and back).

As discussed in  \cite{Levkov:2018kau},  this  pseudo-spectral algorithm has several beneficial features and advantages over other approaches. It automatically conserves particle number and it has a high degree of stability (i.e. there are no spurious growing modes). Compared to lower order pseudo-spectral algorithms, such as used in \cite{Edwards:2018ccc}, much larger $\Delta t$ are allowed without introducing significant systematic uncertainties, so simulations are faster (detailed analysis comparing algorithms of different orders can be found in \cite{Schwabe:2020eac}).

\subsection{Systematic uncertainties} \label{aa:systematics}

\begin{figure}[t]
	\begin{center}
		\includegraphics[width=0.45\textwidth]{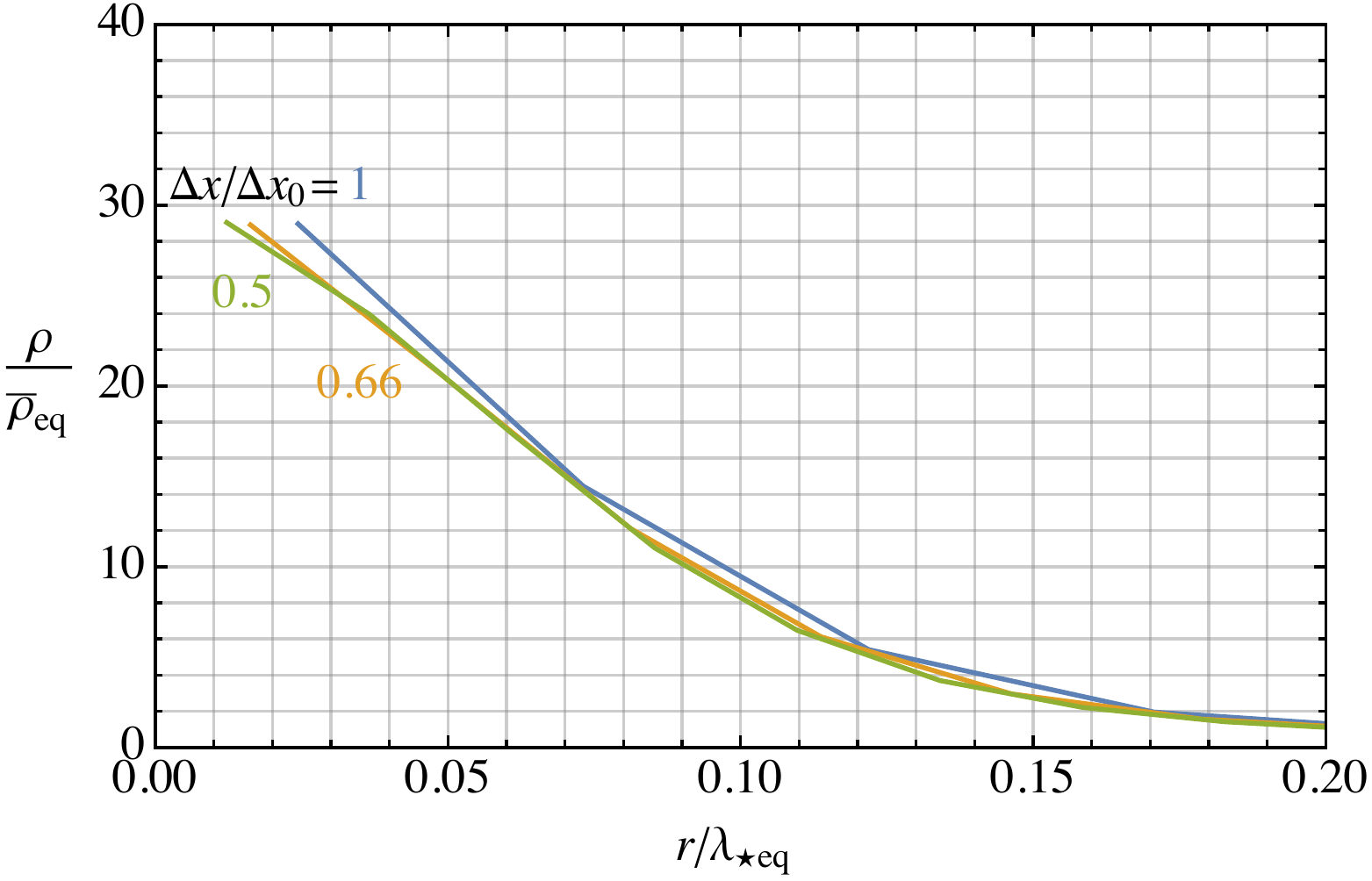}
		\includegraphics[width=0.54\textwidth]{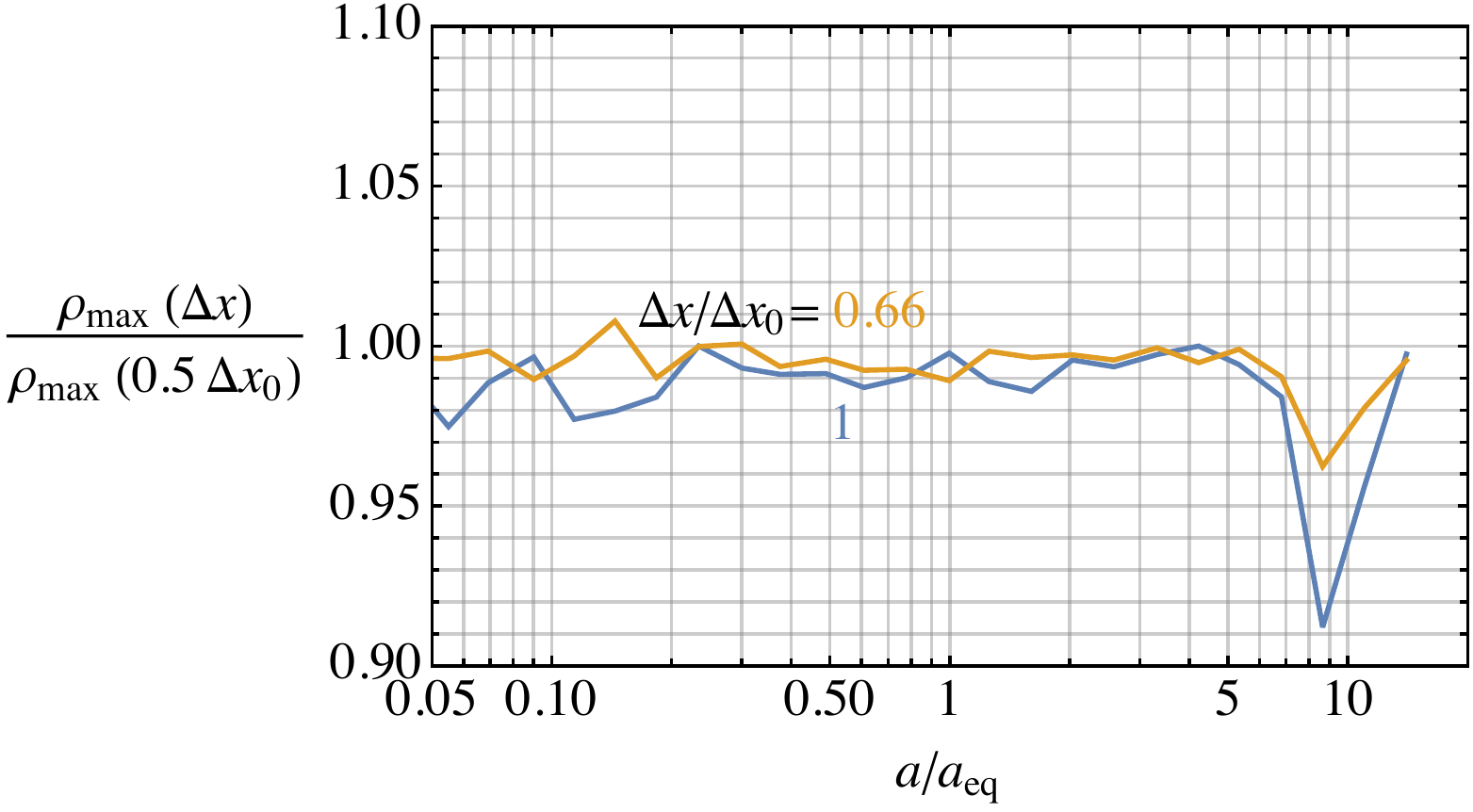}
	\end{center}
	\caption{ {\bf \emph{Left:}} Comparison between the spherically averaged density profile around the densest object in a simulation at $a/a_{\rm eq}=14$ with the space-step we use for our main runs $\Delta x_0$ and with finer resolutions $0.66 \Delta x_0$ and $0.5 \Delta x_0$ (starting from identical initial conditions). Although resolution of the soliton core is starting to be lost in the $\Delta x_0$ run, the central density and soliton mass are still accurate to few percent level. {\bf \emph{Right:}} The maximum density in a simulation obtained evolved with the space-step used for our main runs $\Delta x_0$, and with finer resolution $0.66 \Delta x_0$, compared to that obtained  with a $0.5 \Delta x_0$ (starting from identical initial conditions). At $a/a_{\rm eq}$ our choice of space-step introduces less than a $\sim 10\%$ systematic uncertainty. \label{fig:finitegrid}} 
\end{figure}

\begin{figure}[t]
	\begin{center}
		\includegraphics[width=0.43\textwidth]{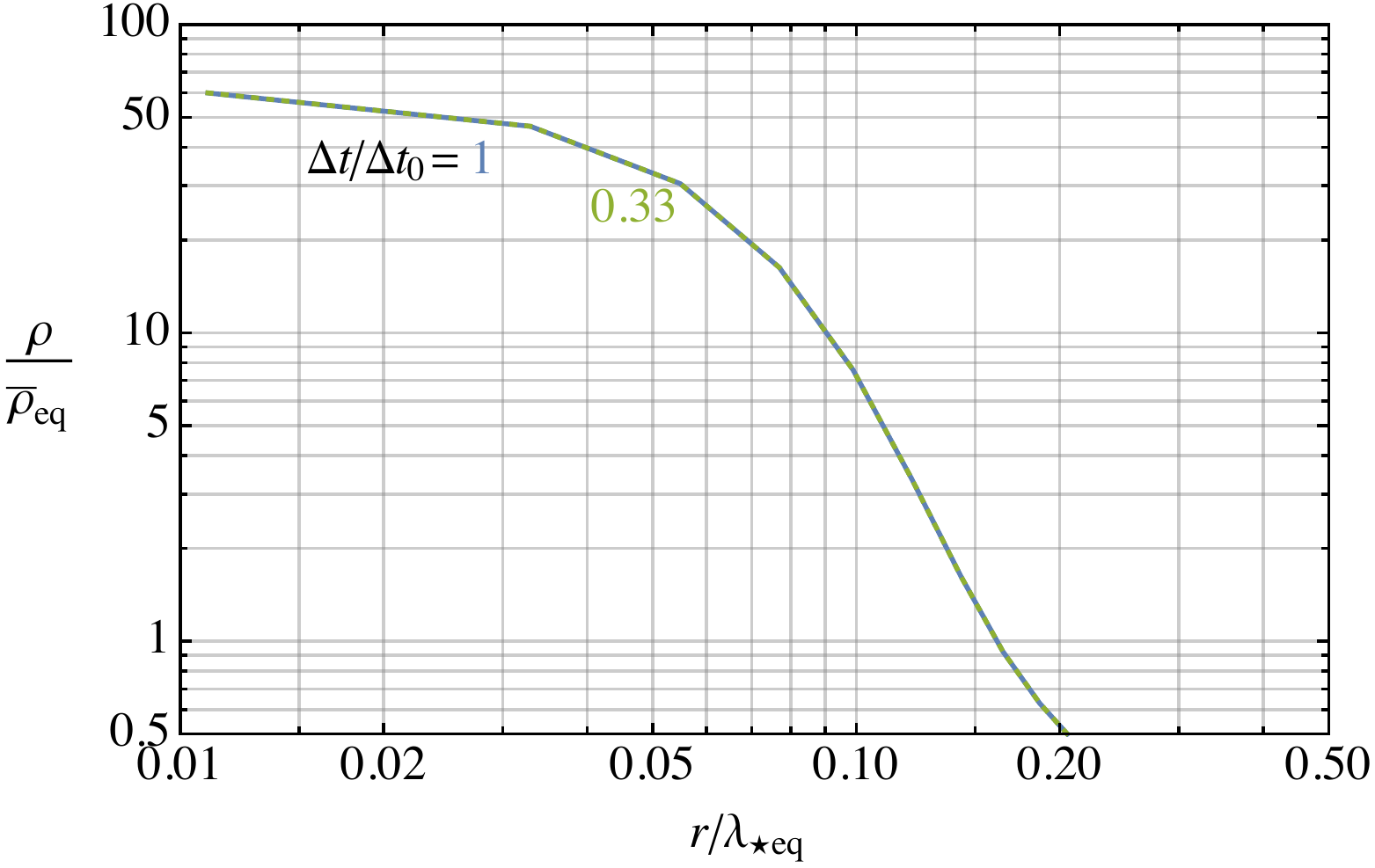}
		\includegraphics[width=0.54\textwidth]{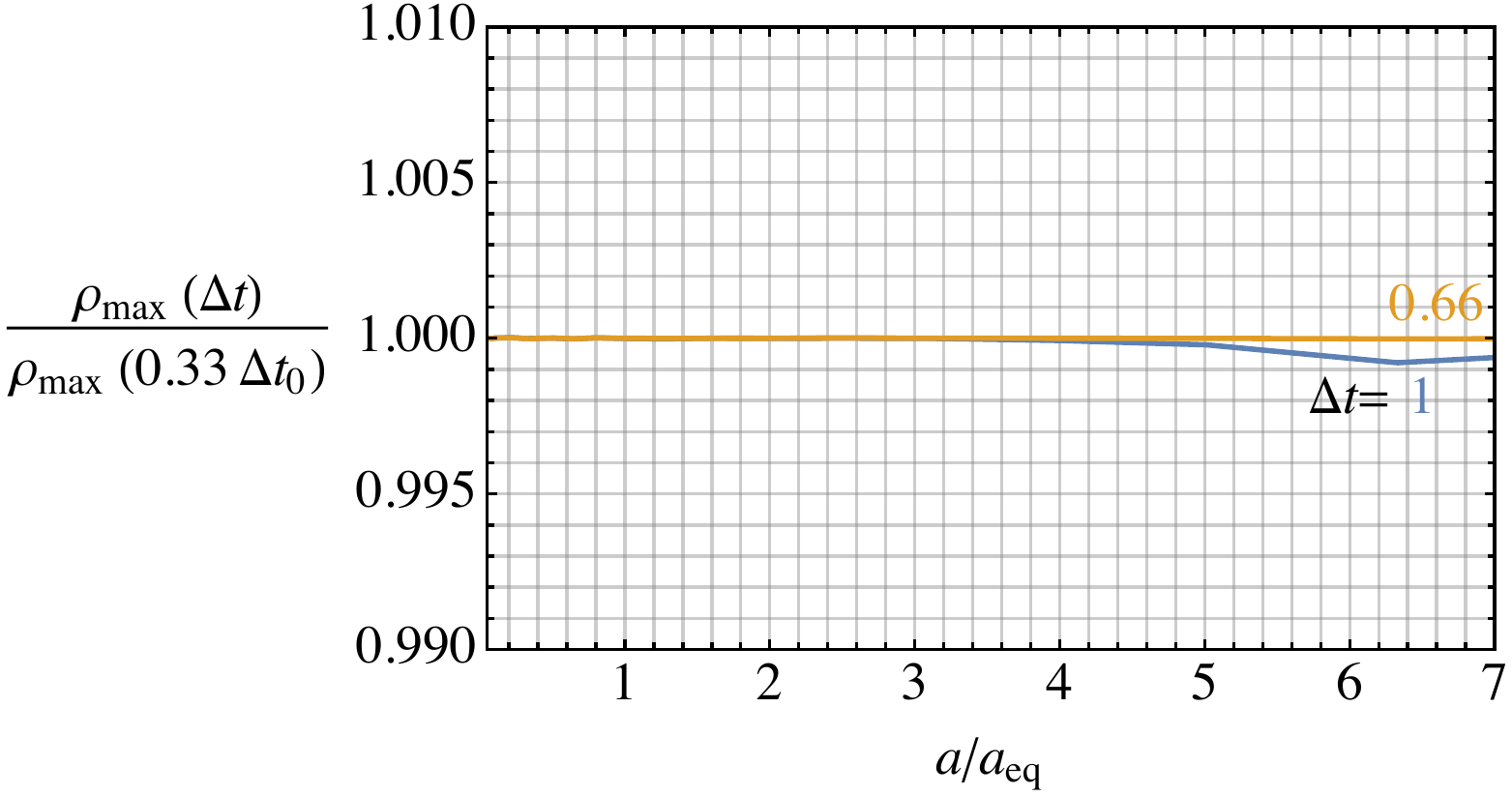}
	\end{center}
	\caption{ {\bf \emph{Left:}} Comparison between the spherically averaged density profile around the densest object in a simulation at $a/a_{\rm eq}=10$ with the time-step we use for our main runs $\Delta \tilde{t}_0$ and with a time-step of $0.33 \Delta \tilde{t}_0$ (starting from identical initial conditions). The almost perfect agreement indicates that the systematic uncertainties from this source are negligible. {\bf \emph{Right:}} The maximum density in a simulation obtained evolved with the time-step used for our main runs $\Delta \tilde{t}_0$, and with a smaller time-step $0.66 \Delta \tilde{t}_0$, compared to that obtained evolving with a time-step of $0.33 \Delta \tilde{t}_0$ (starting from identical initial conditions). At $a/a_{\rm eq}=7$ our choice of time-step introduces only much less than $\sim 1\%$ systematic uncertainty compared to the smaller time-step. \label{fig:finiteT}} 
\end{figure}

\begin{figure}[t]
	\begin{center}
		\includegraphics[width=0.54\textwidth]{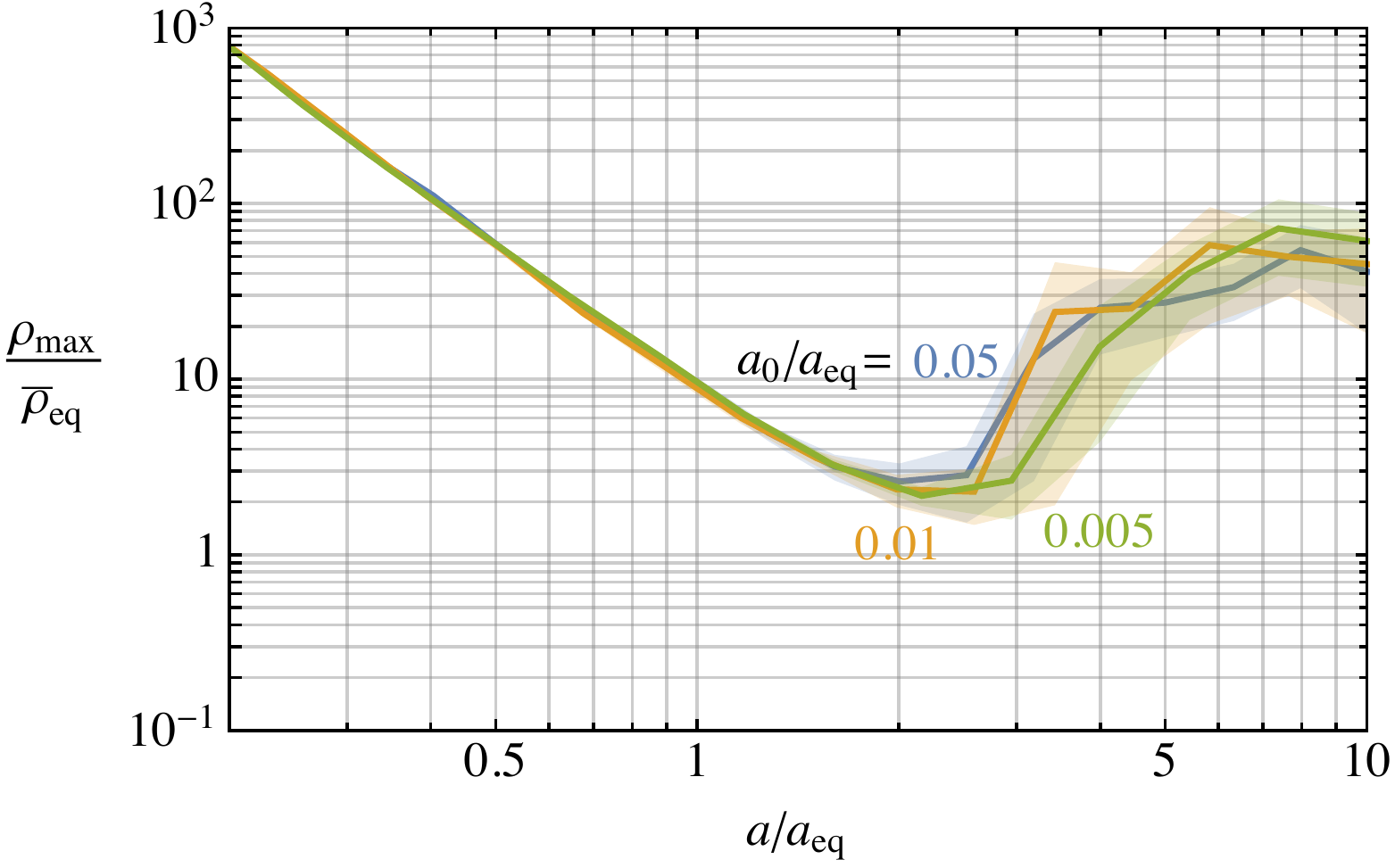}
	\end{center}
	\caption{  The mean maximum density in simulations starting from different initial times. Due to the oscillation of modes $k>k_J$ even during radiation domination and the non-linear dynamics during gravitational collapse, simulations with initial conditions starting at different $a_i$ can differ significantly. However, averaged over multiple runs, there is no systematic effect from starting at $a_i/a_{\rm eq} =0.01$ (as we do in our main runs) rather than earlier.
	\label{fig:finiteV}} 
\end{figure}

For our main data collection runs we use a lattice of  $N^3=1080^3$ points, with a  comoving box length of $ \ell = 3.75 \lambda_\star$. Evolution is started at $a/a_{\rm eq}=0.01$, and stopped at $a_f/a_{\rm eq}=7$.  We vary the timestep throughout a simulation as $\Delta \tilde{t} = 0.00025 a_{\rm eq}/a$.\footnote{This is because the influence of the gravitational potential is the only thing that prevents the evolution in eq.~\eqref{eq:SP_ev} being exact for finite $\Delta \tilde{t}$ and which therefore leads to a condition on $\Delta \tilde{t}$ to not be too large. For a static field configuration, the influence of the potential increases proportionally to $a$, see eq.~\eqref{sp2a}. Therefore our rescaling of $\Delta \tilde{t}$ is a reasonable choice to avoid a large amount of CPU time being spent evolving with small $\Delta t$ at values of $a$ such that this is not necessary.}  We now show that these choices lead to negligible systematic uncertainties, which arise from various sources: 
\begin{enumerate}
	\item \emph{Finite lattice spacing.} Once formed the solitons have constant physical size, whereas the lattice spacing increases $\propto a$. The size of the resulting systematic uncertainties is controlled by the hierarchy of the physical distance between lattice points at $a_f$ and the size of the soliton cores. For negligible systematic uncertainties we expect that  $a_f \ell/ N  \ll \lambda_J(\rho_s)$ is required, where $\rho_s$ is the central density of a soliton. Since $\lambda_J$ is a decreasing function of $\rho_s$, we require that this inequality is satisfied for the  densest, i.e. largest mass, solitons that can form. $\lambda_J(\rho_s)$ is parametrically set by $\lambda_\star$.
	
	To confirm that our choice of parameters leads to negligible systematics from this source, we compare results from a single simulation with our main value of $\Delta x=\ell/N = \Delta x_0$ with results starting from identical initial conditions, but with $\Delta x$ a factor of $2/3$ and $1/2$ smaller. To have sufficient computing power to do the finer resolution tests we carry out this test in boxes with smaller $\ell$. The initial conditions that we choose happen to have a largest soliton mass  $\sim 0.17 M_J(a_{\rm eq})$. Since $k_J(\rho_s)\propto 1/M$ we run our lattice spacing tests to $a_f=14$, which gives the same value of $a_f \ell/ \left( N   \lambda_J(\rho_s)\right)$ as in a large simulation with a soliton of mass $0.34 M_J(a_{\rm eq})$ at $a_f=7$, which covers the vast majority of the solitons that form in our main simulations. Results are shown in Figure~\ref{fig:finitegrid}. One can see that the soliton core is just about resolved with our main value of $\Delta x_0$. The central density of the soliton, and therefore the inferred soliton mass, is still accurately reproduced to about $5\%$ level, which is sufficient for our purposes.
	
	\item \emph{Finite time step.} As mentioned, owing to the gravitational potential, the evolution in eq.~\eqref{eq:SP_ev} is exact only in the limit $\Delta \tilde{t} \to 0$. To test the importance of systematic uncertainties with our choice of $\Delta \tilde{t}$ we compare results when identical initial conditions are evolved with our main choice $\Delta \tilde{t}_0$ and with $\Delta \tilde{t}$ reduced by a factors of $2$ and $3$ smaller. Results are shown in  Figure~\ref{fig:finiteT}, where it can be seen that our choice of $\Delta \tilde{t}_0$ is sufficient for percent level accuracy.
	
	\item \emph{The initial time of simulations.} Even though density perturbations are not expected to evolve much during radiation domination, there is still a numerical question of small $a_i/a_{\rm eq}$ must be so that significant systematic uncertainties are not introduced. An estimate of a suitable $a_i$ can be obtained from the $f=1$, $k_J\to \infty$ perturbative prediction $\delta(a)= \delta(0)(1+\frac{3}{2} a/a_{\rm eq})$, which suggests that starting that $a_i/a_{\rm eq}=0.01$ is enough for percent level accuracy. To test this, we plot the maximum density in simulations as a function of time for different $a_i$ in Figure~\ref{fig:finiteV} (left). Because of the 
	non-linear dynamics of the system after MRE small changes in the initial conditions can lead to large changes at late times, so for each $a_i$ plot results averaged over a set of $5$ different initial conditions (identical between different $a_i$).  We see that the results with $a_i/a_{\rm eq}=0.01$ agree with those with $a_i/a_{\rm eq}= 0.005$ (and even $a_i/a_{\rm eq}=0.05$ would be sufficient).
	
	\item Finally, the \emph{finite box size} can lead to systematic uncertainties. These are expected to be negligible as long as the most IR modes in a simulation have power spectrum $\mathcal{P}_\delta(k) \ll 1$, indicating that they are still perturbative, and dynamics on the scale of the box size are not affecting the evolution on smaller scales. With our choices of box size and $a_f$ in our main runs, the most IR modes have $\mathcal{P}_\delta(k) \simeq 0.5$	 (see Figure~\ref{fig:mode_growth}, right) at the final time, suggesting that this is not a source of major systematics. 

\end{enumerate}

\section{Further Results from Schr\"odinger--Poisson Simulations} \label{app:moreSP}

In this Appendix we collect some additional discussion and results from our numerical simulations of the SP equations, supporting the analysis of Section~\ref{section:SPsection}.

\begin{figure}[t]
	\begin{center}
		\includegraphics[width=0.48\textwidth]{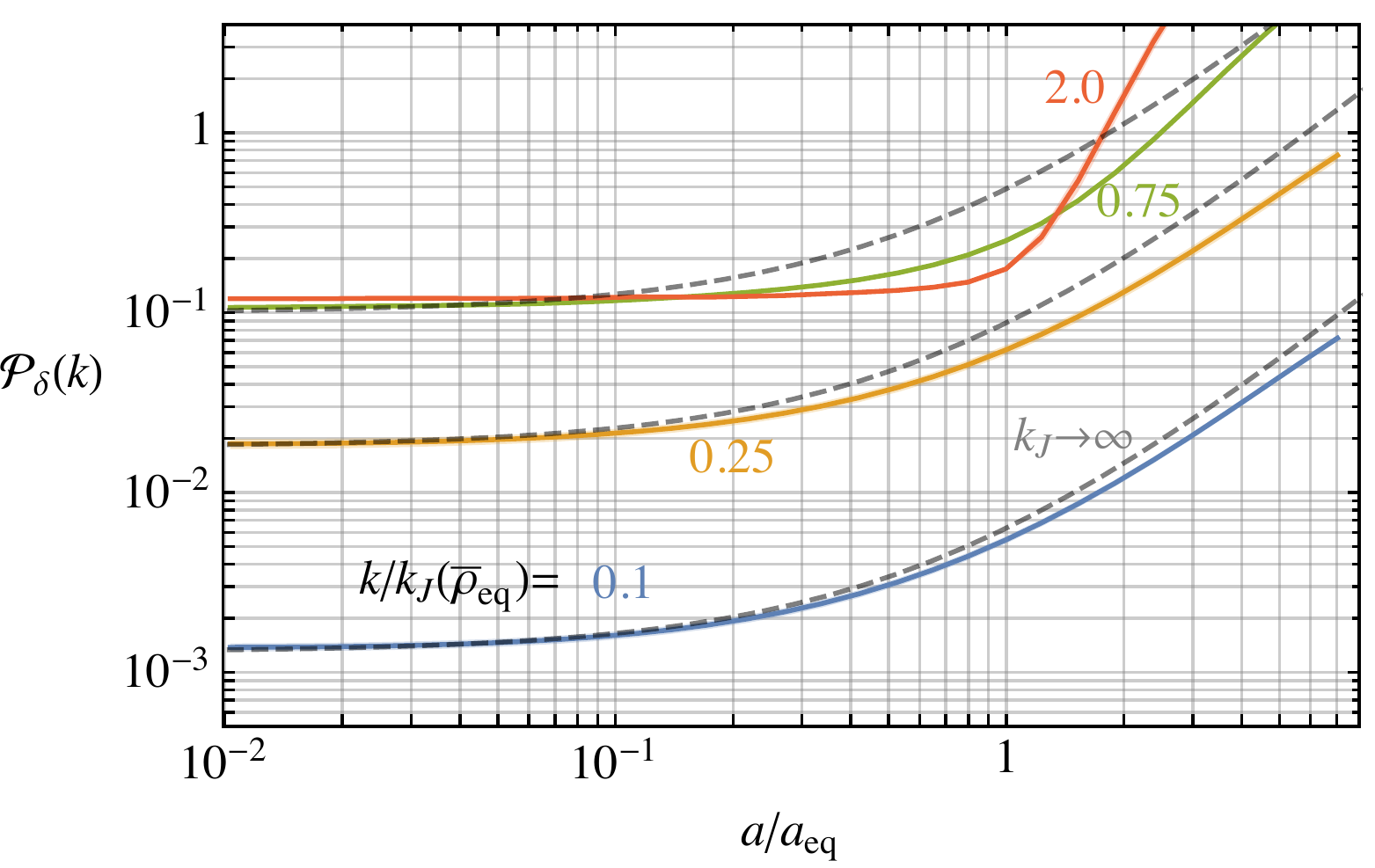}
		~~		\includegraphics[width=0.48\textwidth]{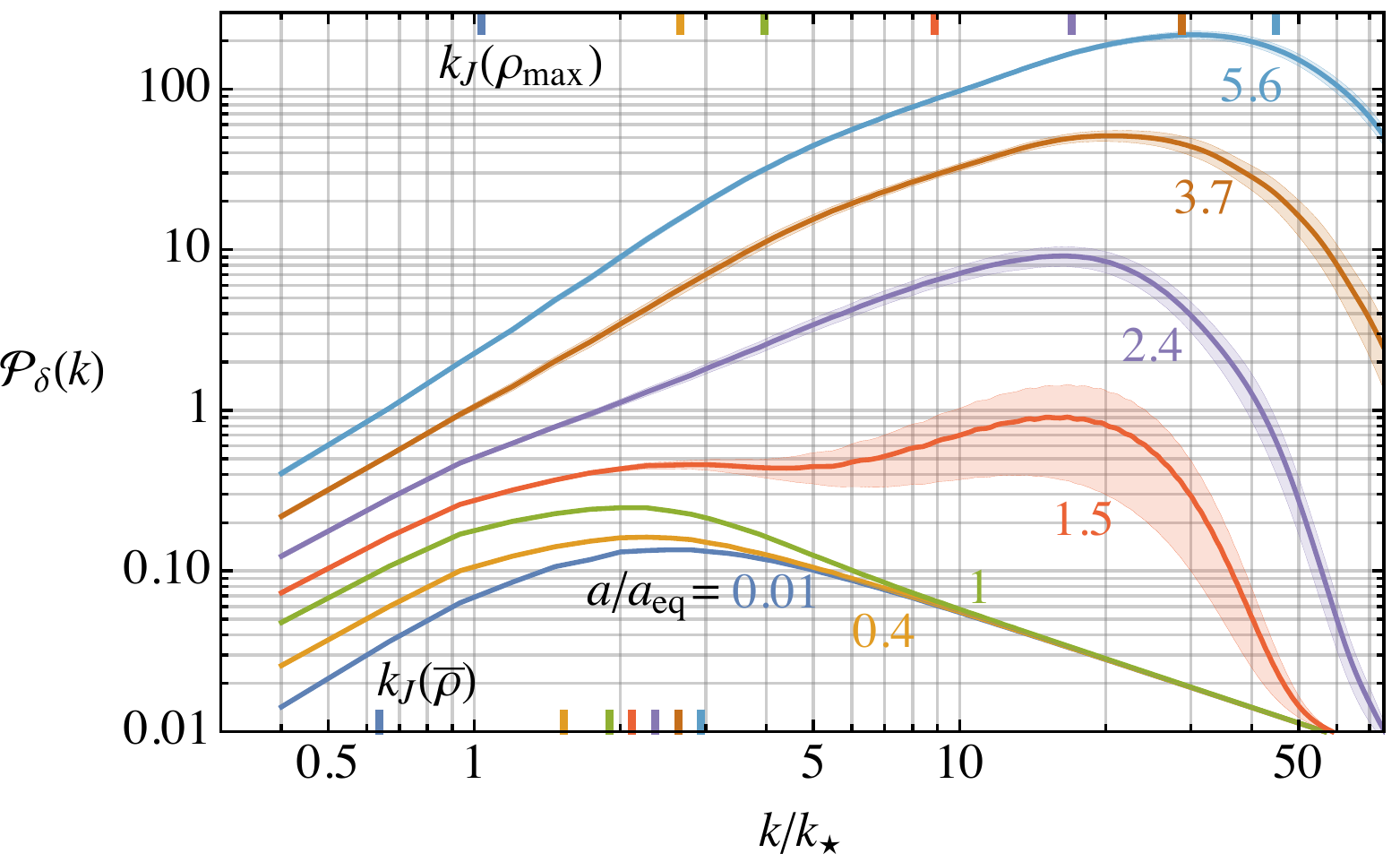}			
	\end{center}
	\caption{{\bf \emph{Left:}} The average growth of different momentum modes as a function of time. The predicted growth in the absence of quantum pressure, $k_J \rightarrow \infty$, is also plotted. The growth of modes $k/ k_J(\bar{\rho}_{\rm eq})=0.1$ is only slightly affected, while the higher momentum modes have more suppressed growth. Modes $k/ k_J(\bar{\rho}_{\rm eq})\gtrsim 1$ are prevented from growing almost entirely until the quantum Jeans momentum in some region $k_J(\rho)$ is comparable to $k$. {\bf \emph{Right:}} The density power spectrum in simulations at different times averaged over multiple realisations of the initial conditions, with statistical uncertainties. The quantum Jeans momenta corresponding to the mean dark matter density $k_J(\bar{\rho})$ and to the (averaged) maximum density in the simulations $k_J(\rho_{\rm max})$ are indicated. The IR modes grow unaffected by quantum pressure, following the standard expectation. Quantum pressure delays the growth of modes on smaller length scales. Once solitons form the power spectrum is peaked at scales $k_J(\rho_{\max})$, set by the density at their cores. The peak tracks the typical size of the solitons. \label{fig:mode_growth}} 
\end{figure}

First we consider the growth of density perturbations in more detail. In Figure~\ref{fig:mode_growth} (left) we plot the evolution of the power spectrum $\mathcal{P}_{\delta}$ at different momenta (averaged over multiple simulations) and compare this to the prediction from Appendix~\ref{app:over} in the limit that quantum pressure is unimportant, $k_J\to \infty$. It can be seen that modes with $k=0.1 k_J^{\rm eq}$ evolve basically as they would in the absence of quantum pressure, and the growth of modes with $k=0.25 k_J^{\rm eq}$ at MRE is slightly hindered by quantum pressure. In contrast, as expected, quantum pressure prevents modes close to the peak of the initial power spectrum (see Figure~\ref{fig:Pspectra} right) from growing until around MRE. The same dynamics can be seen in Figure~\ref{fig:mode_growth} (right), where we plot the density power spectrum at different times (note the solitons are not screened when calculating this). At a given time, the growth of modes with $k$ greater than the quantum Jeans momentum are suppressed. Since the quantum Jeans momentum is larger at larger density, an overdensity on a scale $k^{-1}$ can grow provided $k^{-1} < k_J^{-1}(\rho)$ where $\rho$ is the local dark matter density, which might be greater than the mean density. Indeed, at late times there is a clear UV cutoff at $\mathcal{P}_\delta \simeq k_J(\rho_{\rm max})$, corresponding to the inverse size of the most massive solitons produced.

\begin{figure}[t]
	\begin{center}
				\includegraphics[width=0.5\textwidth]{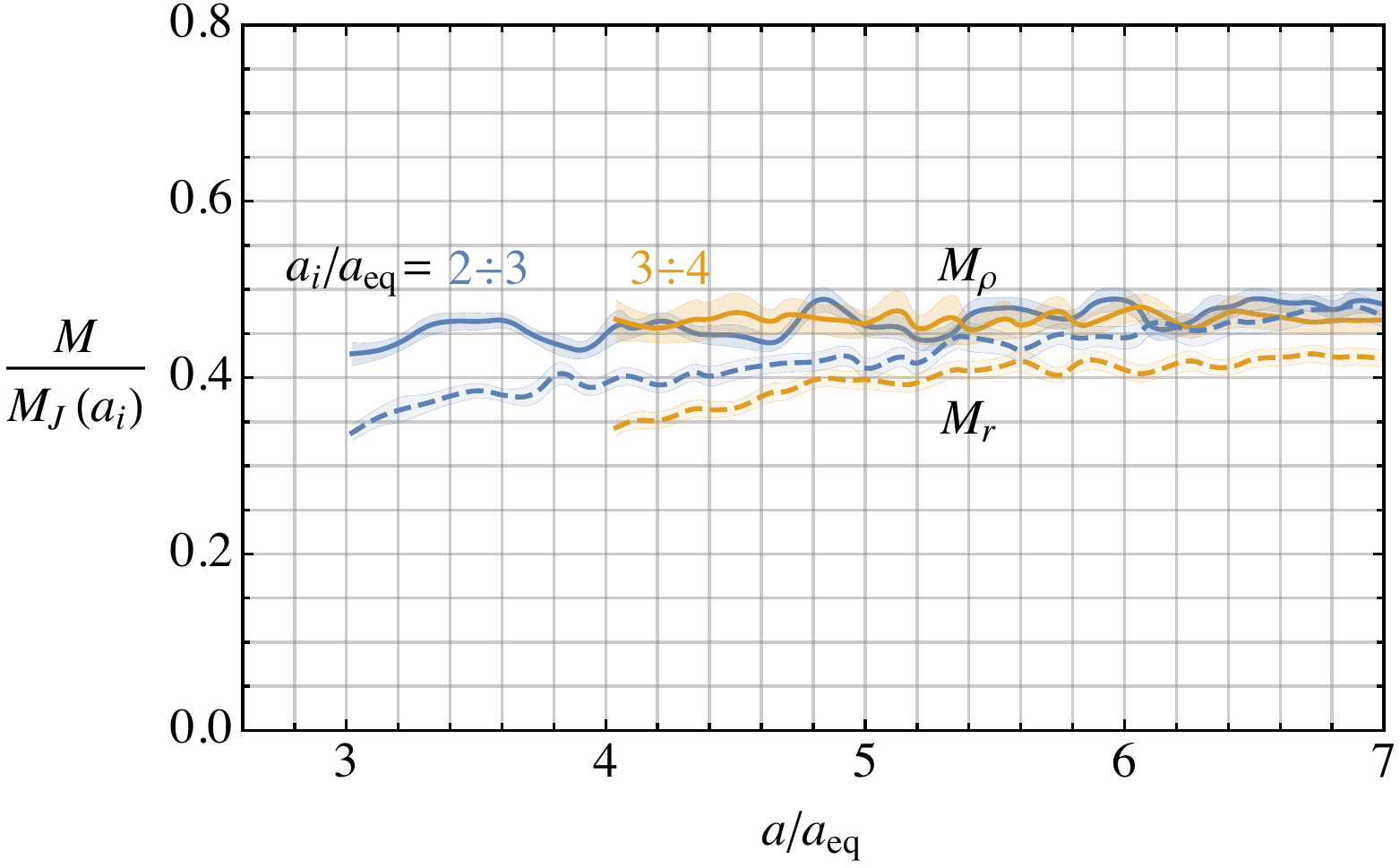}
		\includegraphics[width=0.48\textwidth]{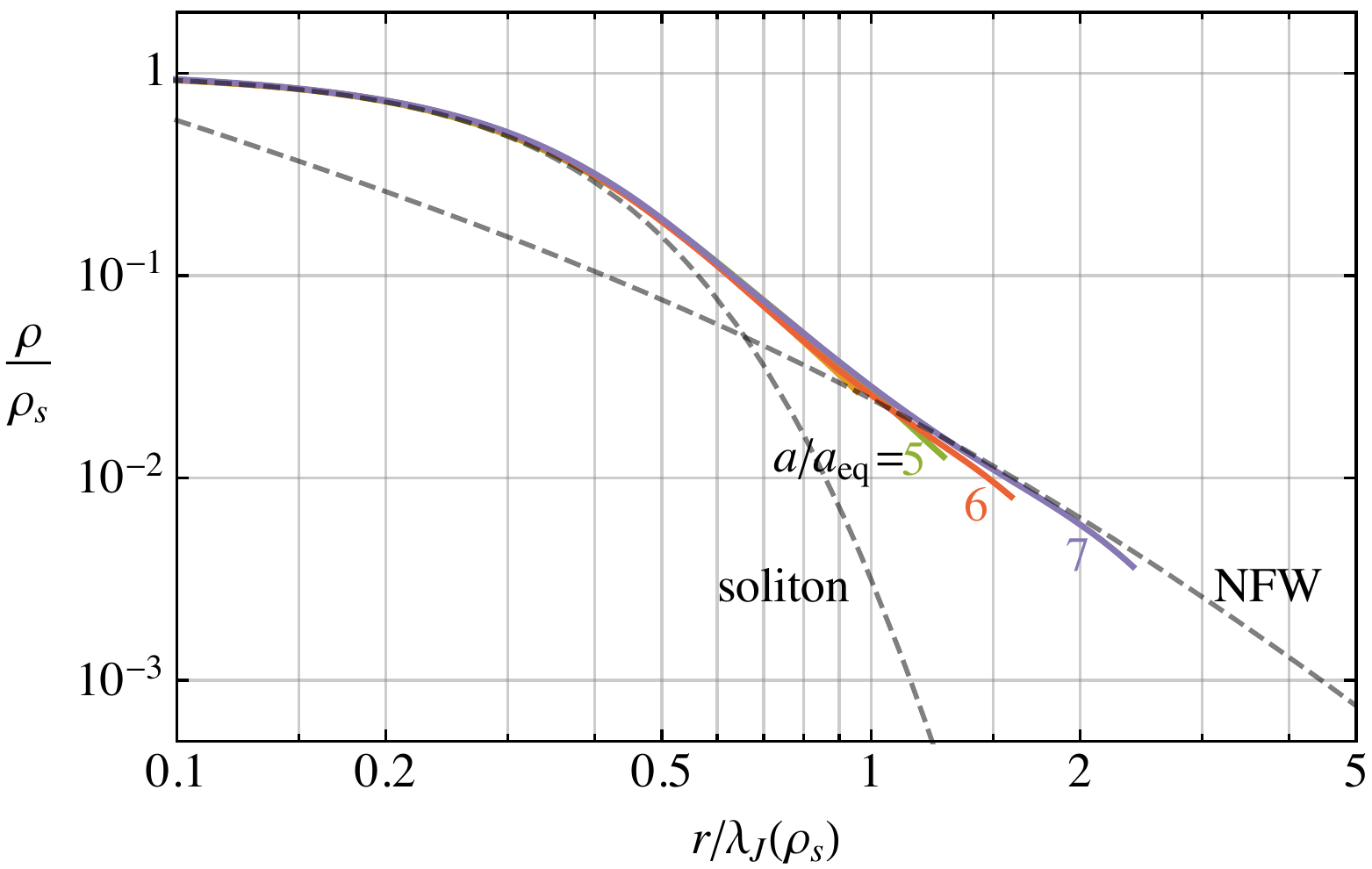}
	\end{center}
	\caption{
        \textbf{\emph{Left:}} The masses of solitons as a function of time, with the solitons grouped based on when they are produced, with statistical error bars. The masses are normalised as in Figure~\ref{fig:massa}. The soliton masses are measured in two ways: first from the central densities ($M_{\rho}$), and second by integrating over their density profiles ($M_r$). The mass measured from the central densities are approximately constant, while though measured from the density profile increase slightly and approach those inferred from the central density.
		\textbf{\emph{Right:}} The spherically averaged density profile around solitons, as in Figure~\ref{fig:profile_log}, but selecting only solitons with medium masses  $(0.15 \div 0.18) M_J(a_{\rm eq})$.	The NFW profile plotted is that obtained by fitting the profile around heavy solitons, as described in Section~\ref{ss:fuzzy_halo}. The fuzzy halos around the relatively light solitons are approaching the same profile.
		\label{fig:HMF_halo}} 
\end{figure}

Next we clarify the choice made in Section~\ref{ss:SP_simulations} to determine the soliton masses in simulations from their central densities. In Figure~\ref{fig:HMF_halo} (left) we plot the average soliton mass as a function of time, grouping the solitons by the time when they formed $a_i$, as in Figure~\ref{fig:massa} (right), but showing results for the masses calculated from the central densities $M(\rho_s)$ and from integrating the density profile. In particular, we evaluate the mass inside the radius $r_{s}$ where the spherically averaged density $\rho(r)$ drops a factor of $10$ from the central value, i.e. $M_r=  \int_0^{r_{s}} \rho(r) 4\pi r^2 dr$, since over this region that profile is (on average) soliton-like, see Figure~\ref{fig:profile_log}. The mass calculated from integrating the density profile is close to that inferred from the central density, as expected since we know that on average the profile is close to the soliton one. However, at least at early times there are a small differences and the mass measured by integrating the density profile initially increases slightly approaching that predicted from the central density. We interpret this as the soliton density profile taking some time to settle down to its eventual form. Meanwhile, apart from oscillations due to quasinormal modes, the solitons' central densities reach their late time values fast (see also Figure~\ref{fig:denslice} left), so are a good way to infer the solitons' eventual masses.

We also note that the time of soliton formation is ambiguous. For the purposes of Figure~\ref{fig:massa} (right) and Figure~\ref{fig:HMF_halo} (left), we fix $a_i$ by the first time when the object has a central density $\geq 200 \bar{\rho}$ and the averaged density profile does not drop by more than $25\%$ within one lattice spacing from the centre. Given that the gravitational collapse happens fairly fast (see Figure~\ref{fig:denslice} left), the value of $c_M$ we extracted is not particularly sensitive to this choice.

In Section~\ref{ss:fuzzy_halo} we studied the fuzzy halo around solitons by considering the halos around relatively heavy solitons in Figure~\ref{fig:profile_log}.\footnote{We note that there are several choices required to make such a plot. We define a collapsed object as having maximum density $\rho \geq 200 \bar{\rho}$. Having identified such an object, we label the entire region out to the radius where the spherically averaged density profile drops to $\rho = 20 \bar{\rho}$ as being part of the same object (and so not considered further in finding other compact objects). We also discard objects for which the spherically averaged density profile drops by more than $25\%$ one lattice space away from the core. This removes a few isolated points, which would otherwise be identified as collapsed objects and are generally on the outskirts of a genuine collapsed object, just outside their cutoff. These choices do not make a substantial difference to our results.} 
In Figure~\ref{fig:HMF_halo} (right) we show the analogous plot for medium mass solitons, which form approximately between $5\lesssim a/a_{\rm eq}\lesssim 7$ (see Figure~\ref{fig:HMF_soliton}). We further restrict to objects that have a halo extending to at least $r/ \lambda_J(\rho_s)= (0.95, 1.25, 1.6, 2.4)$ respectively at the four times. We also plot the NFW profile obtained fitting to the fuzzy halo around the relatively heavy solitons. Not surprisingly, at a fixed $a/a_{\rm eq}$ the fuzzy halos extend less far from the medium mass solitons than from the heavy solitons. However,  the halos are growing outwards from the medium mass solitons fast, and are approaching the NFW form of the fuzzy halos around the heavy solitons, with the same halo parameters. This is a nice feature, which makes it reasonable for us to assume that the halos around all solitons will eventually take this form, at least out to some cut-off. We do not know what this cut-off is, however fortunately this is not a major source of uncertainty. The typical central densities of the medium mass solitons is about $10\eV^4$, so from Figure~\ref{fig:HMF_halo} (right) the fuzzy halo is likely to be destroyed outside the radius $(10\div 15)\lambda_J(\rho_s)$. This is approximately the distance that the fuzzy halos around the massive solitons already reach in simulations, so it is reasonable to estimate that fuzzy halos around all solitons will reach this size (which is what we assume when studying the rate of collision with an observer in Section~\ref{ss:collision_rate}). 

\begin{figure}[t]
	\begin{center}
		\includegraphics[width=0.54\textwidth]{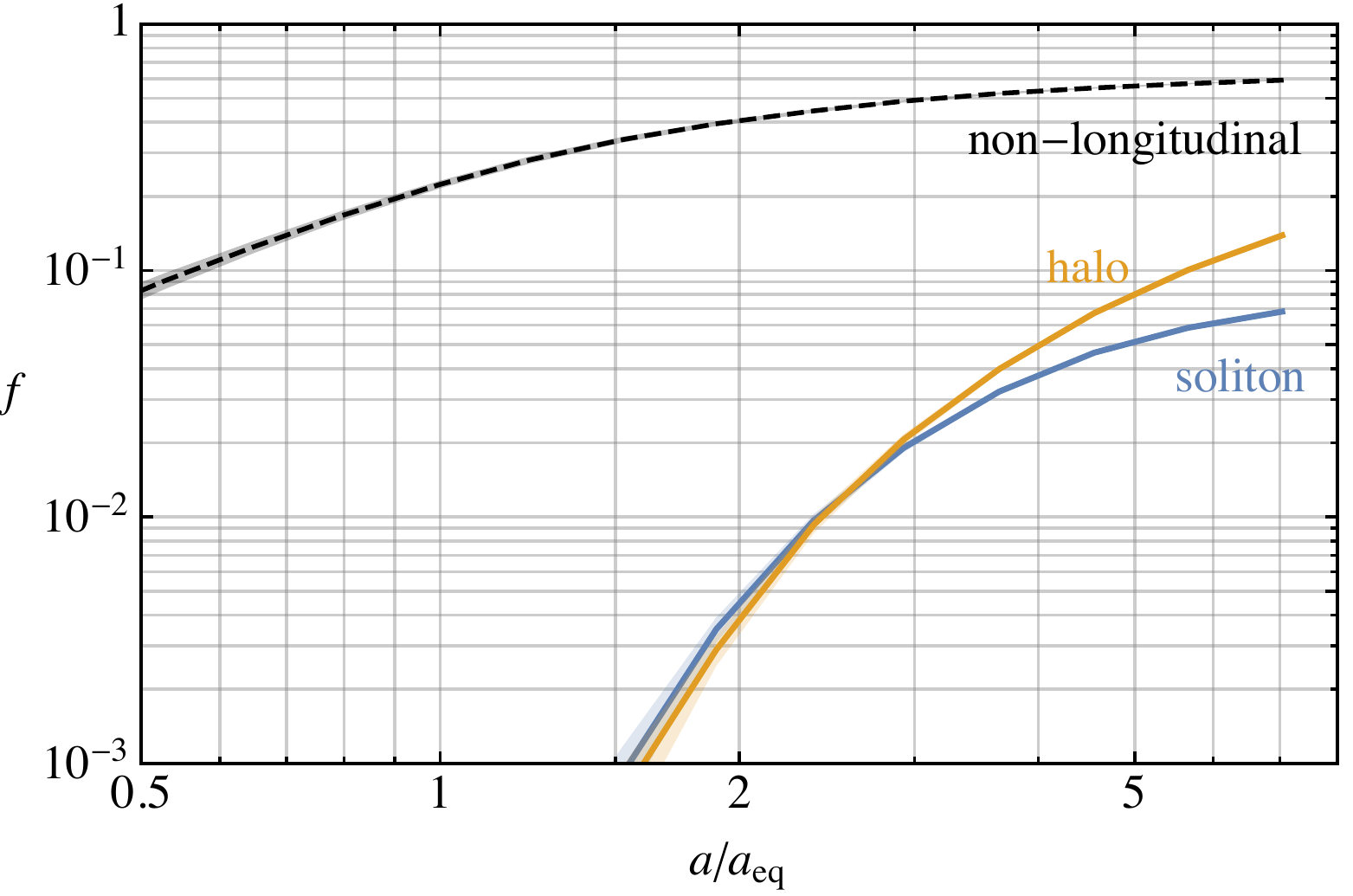}
	\end{center}
	\caption{
		The fraction of dark matter in the solitons and their surrounding halos as function of time. We also plot the fraction of the dark photon energy that is in non-longitudinal modes, defined by $1-f= \left<(\hat{k}.A(k))^2\right>/\left<A(k)^2\right>$ where $\left<\ldots \right>$ denotes the spatial average. The field's energy is initially in purely longitudinal modes, but gravitational interactions that become relevant around matter radiation equality transfer energy into transverse modes (e.g. the soliton solution is not pure longitudinal).
		\label{fig:bound_frac}} 
\end{figure}

In Figure~\ref{fig:bound_frac} we plot the fraction of dark matter in solitons and the halos around them (with the edges of the halos defined as the point where the spherically averaged density drops to $20 \bar{\rho}$, as before). We also plot the fraction of the dark photon energy that is in non-longitudinal modes, which, due to gravitational interactions, increases from $0$ at the start of the simulation and becomes a substantial fraction around MRE.

\section{Analytic Prediction of the Soliton Mass Distribution} \label{app:analytic_prediction}

As discussed in Section~\ref{ss:solitonmassf}, our analytic estimate of the soliton mass distribution is based on two simple assumptions:\vspace{-2mm}% (more precisely at $a=a_{\rm eq}$)
\begin{enumerate}[label={\arabic*)},leftmargin=0.2in] \setlength\itemsep{0.15em}
	\item The solitons produced at any time have a mass $M=M(a)$ given by eq.~\eqref{eq:MJ}, with $c_M\simeq 0.45$.
	
	\item  As soon as the comoving Jeans scale drops below the size of a fluctuation, this -- if already of order one -- should collapse into a soliton. Therefore, the number of solitons produced   is expected to be proportional to the `frequency' that the amplitude of such fluctuations in the initial conditions is larger than a critical value of order one, that we call $\delta_c$.

	 More precisely phrased: Between $a_1\simeq a_{\rm eq}$ and $a$, the fraction of dark matter collapsed into solitons  
	 is equal to the probability $\Pi _{\delta >\delta_c}(a_1,a)$ that the field $\delta_s$ smoothed between $k_J(a_1)$ and $k_J(a)$ is larger than $\delta_c$. The smoothed field is $\delta_s(x)\equiv(2\pi)^{-3}\int d^3k \exp(i\vec{k}\cdot\vec{x}){\delta}(k) W(k;k_J(a_1),k_J(a))$, where $W(k;k_1,k_2)$ is a window function that vanishes outside $k_1<k<k_2$, which therefore selects the component of the field within these momenta.  Using 1), the number density of solitons produced per unit time is therefore simply $\frac{d\Pi_{\delta >\delta _c}(a_1,a)}{da}\bar{\rho}(a)/M(a)$. Note that, contrary to the usual PS argument, the field does not increase with $D[a]\propto1+(3/2)a/a_{\rm eq}$, since -- as mentioned -- at those scales the overdensities oscillate without growing.

\end{enumerate}
\vspace{-1mm}

\noindent By integrating over $a_1<a<a_2$, a crude estimate for the soliton mass distribution at $a=a_2$ in eq.~\eqref{eq:dfDMdlog} is
\begin{equation}\label{eq:dfDMdManalytic}
	\frac{d f_{s}(a_2,M)}{d\log M}=\frac{M^2}{\bar{\rho}(a_2)}\int_{a_1}^{a_2}da\frac{d\Pi_{\delta >\delta _c}(a_1,a)}{da}\frac{\bar{\rho}(a)}{M(a)}\left(\frac{a}{a_2}\right)^3\delta[M-M(a)] = \frac{d\Pi_{\delta >\delta _c}(a_1,a_M)}{d\log M}	\, ,
\end{equation}
where the last equality is valid for $M<M(a_2)$ (for $M>M(a_2)$ the mass distribution vanishes).
Here $a_M$ is the inverse of the function $M(a)$ in eq.~\eqref{eq:MJ}, and corresponds to the scale factor when solitons with mass $M$ are produced. Note that, aside from the freedom in the choice of the window function $W$, the only free parameter in eq.~\eqref{eq:dfDMdManalytic} is $\delta_c$.

Unfortunately it is not possible to compute the probability $\Pi_{\delta>
	\delta_c}(a_1,a)$ rigorously starting from the power spectrum $\mathcal{P}_\delta$ alone, since, as mentioned, $\delta(x)$ it is not Gaussian at $k\gtrsim k_\star$. We leave this for a future work. Here we limit ourselves to the Gaussian approximation. For a Gaussian distributed field, this probability is fully determined by the field's variance $\sigma_{\delta}^2(a_1,a)\equiv\langle \delta_s^2(x)\rangle=\int dk/k \mathcal{P}_\delta(k) W^2(k;k_J(a_1),k_J(a))$ (see eq.~\eqref{eq:PX}) and is simply $\Pi _{\delta >\delta _c}(a_1,a)=\frac{1}{2} \text{erfc}[\delta _c/\sqrt{2} \sigma _{\delta }(a_1,a)]$. Therefore, fixing $a_1=a_{\rm eq}$ (since the first overdensities collapse at MRE given the remarkable coincindence in eq.~\eqref{eq:kJoks})
\begin{equation}\label{eq:dfFMdMgaussian}
	\frac{d f_{s}}{d\log M}
	\simeq
	\frac{\delta_c}{\sqrt{2\pi}\sigma(M)}e^{-\frac{\delta_c^2}{2\sigma^2(M)}}\left|\frac{d\log \sigma(M)}{d\log M}\right| ~,
\end{equation}
where $\sigma(M)\equiv \sigma_\delta(a_{\rm eq},a_M)$ is the variance of the field smoothed between $k_J^{\rm eq}$ and $k_J(a_M)=k_J^{\rm eq}(c_M M_J^{\rm eq}/M)^{1/3}$. One can think of this variance evaluated at smaller and smaller masses as corresponding to smoothing the field at larger and larger comoving momenta, starting from $k_J^{\rm eq}$.

\begin{figure}[t]
	\begin{center}
		\includegraphics[width=0.54\textwidth]{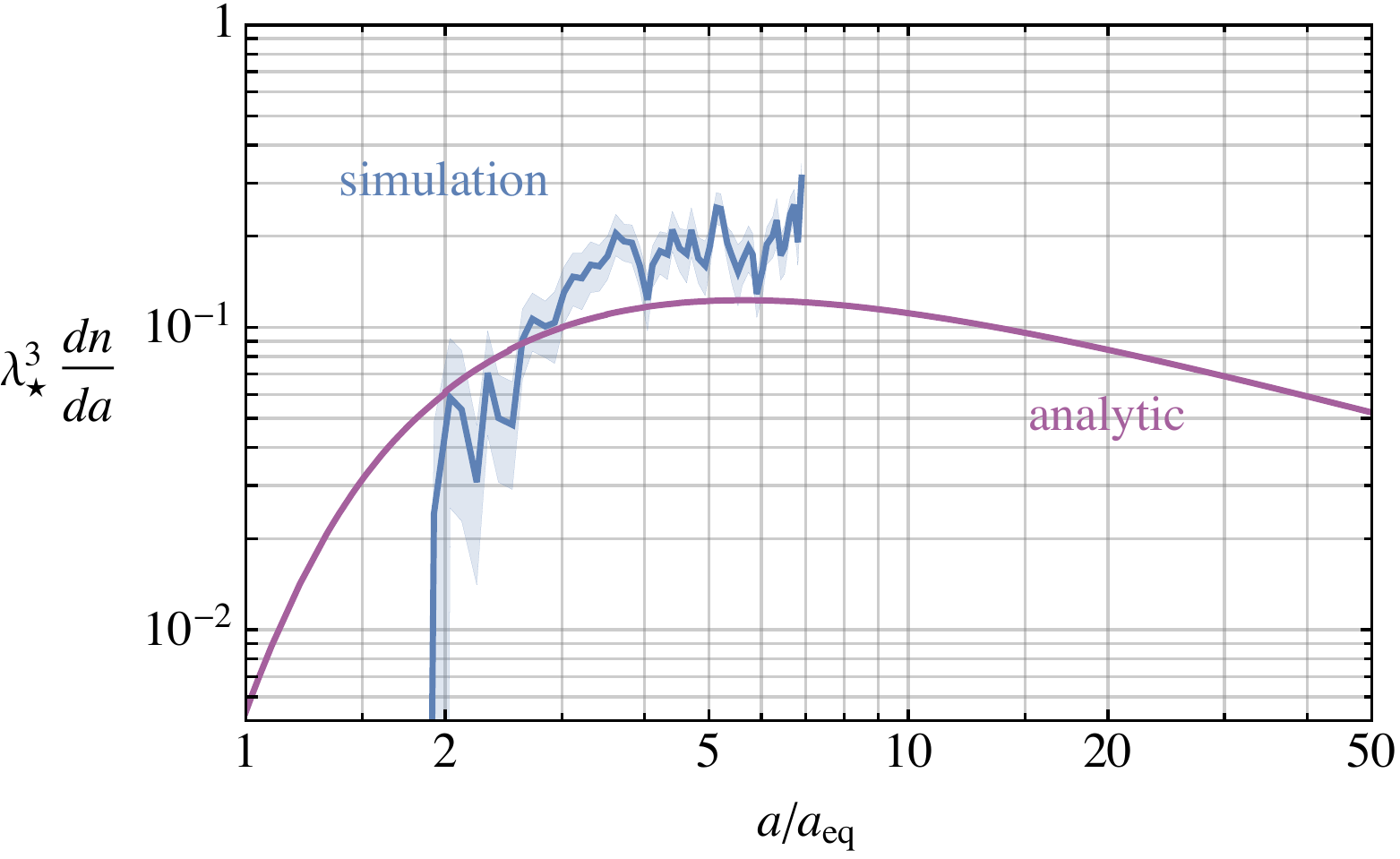}
	\end{center}
	\caption{
		Soliton production rate, defined as the rate of change of the total number of solitons $n$ per comoving $\lambda_\star^3$ volume with respect to the scale factor $a$. Results are shown from simulations and from the analytic prediction. \label{fig:production}} 
\end{figure}

Clearly $\sigma(M)=0$ for $M=c_M M_J^{\rm eq}$ and, from the approximate form of $\mathcal{P}_\delta$ in eq.~\eqref{eq:PAt1}, it is easy to see that $\sigma(M)$ increases at $M<c_M M_J^{\rm eq}$ until reaching the asymptotic value $\sigma(M)\simeq 0.35$ at $M/(c_M M_J^{\rm eq})= 0$.
Therefore from eq.~\eqref{eq:dfFMdMgaussian} the soliton mass distribution is exponentially suppressed at $M\to c_M M_J$, it increases at smaller $M$ and for $M/(c_M M_J)\ll 1$ it tends to zero as $df_{s}/d\log M\propto d\log\sigma/d \log M\propto (M/(c_MM_J))^{2/3}$ (i.e. at small $M/M_J$ lighter and lighter solitons are less and less produced -- this is related to the suppression of $\mathcal{P}_\delta$ at large momenta as $1/k$).\footnote{These behaviours can be seen analytically by calculating $\sigma(M)$ with a crude step window function $W(k;k_1,k_2)=1$ if $k_1<k<k_2$, and $W(k;k_1,k_2)=0$ otherwise.} Interpolating between these two behaviours, there is a peak whose position and amplitude depend on the value of $\delta_c$. These features can be seen in Figure~\ref{fig:HMF_soliton}, where we show the estimate in eq.~\eqref{eq:dfFMdMgaussian} for $\delta_c=0.22$ and $c_M=0.45$ with $\sigma(M)$ calculated using the smooth window function $W(k,k_1,k_2)=W_s(k_1/k)W_s(k/k_2)$, with $W_s(x)\equiv \frac12 (1-\tanh(2(x-1)))$. In Figure~\ref{fig:production} we show the prediction for the comoving number density of soltions produced per unit time, $%dn(a)/da=
	\frac{d\Pi_{\delta >\delta _c}(a_1,a)}{da}\bar{\rho}(a)/M(a)$, for the same $W$ and $\delta_c$ and we compare it with the simulation data.

As we observed, our analytic argument reproduces the data in Figure~\ref{fig:HMF_soliton} remarkably well. Moreover, as mentioned, it predicts that at late times the soliton production is suppressed, with a peak in $df_{s}/\log M$ at $M/M_J^{\rm eq}\simeq 0.1$ (not captured by the numerical simulation, because of the limited time range). In any case, given that the field is not Gaussian and the exponential sensitivity to $\delta_c$ in eq.~\eqref{eq:dfFMdMgaussian} (present in the Gaussian approximation), we refrain from a precise fit of the parameter $\delta_c$, or of the window function, and regard the agreement as mostly qualitative.

\begin{figure}[t]
	\begin{center}
		\includegraphics[width=0.65\textwidth]{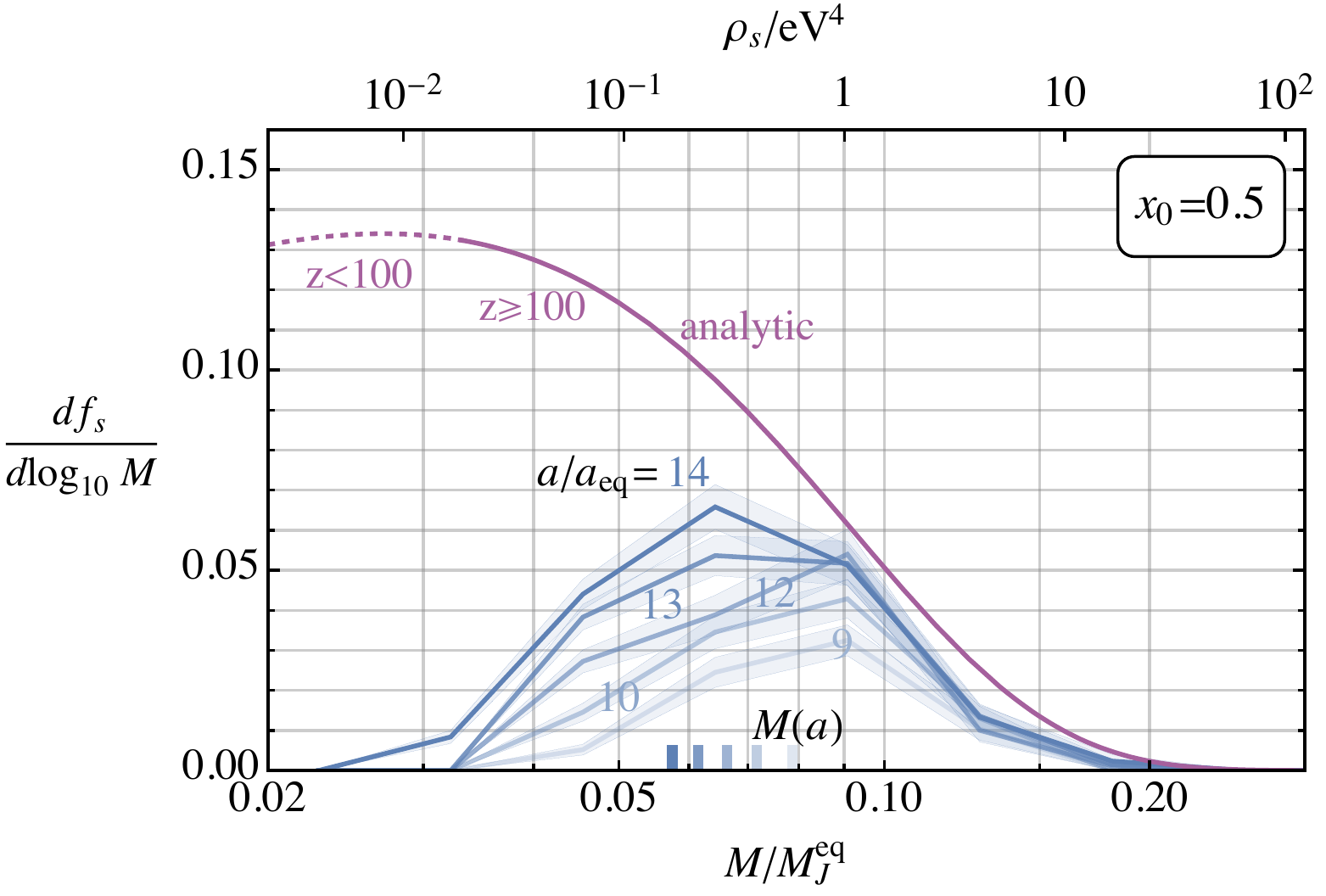}
	\end{center}
	\caption{ The soliton mass function when starting from artificially rescaled initial conditions $k_\star \to k_\star /2$, to test our analytic analysis. We find that our analytic approach works well also with these initial conditions (with a slightly different value of $\delta_c$).
		\label{fig:x02}} 
\end{figure}

\begin{figure}[t]
	\begin{center}
		\includegraphics[width=0.48\textwidth]{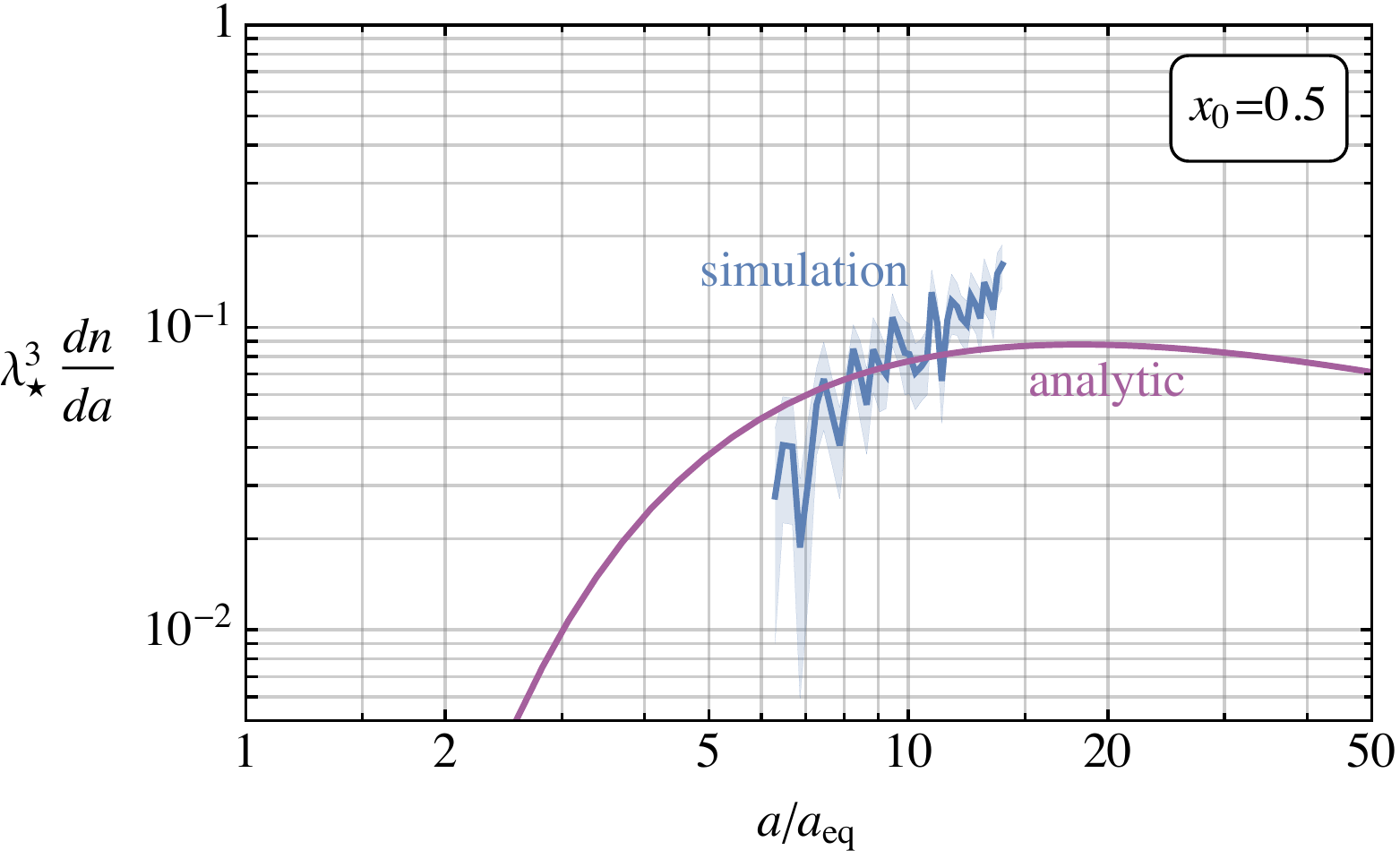}
		\includegraphics[width=0.48\textwidth]{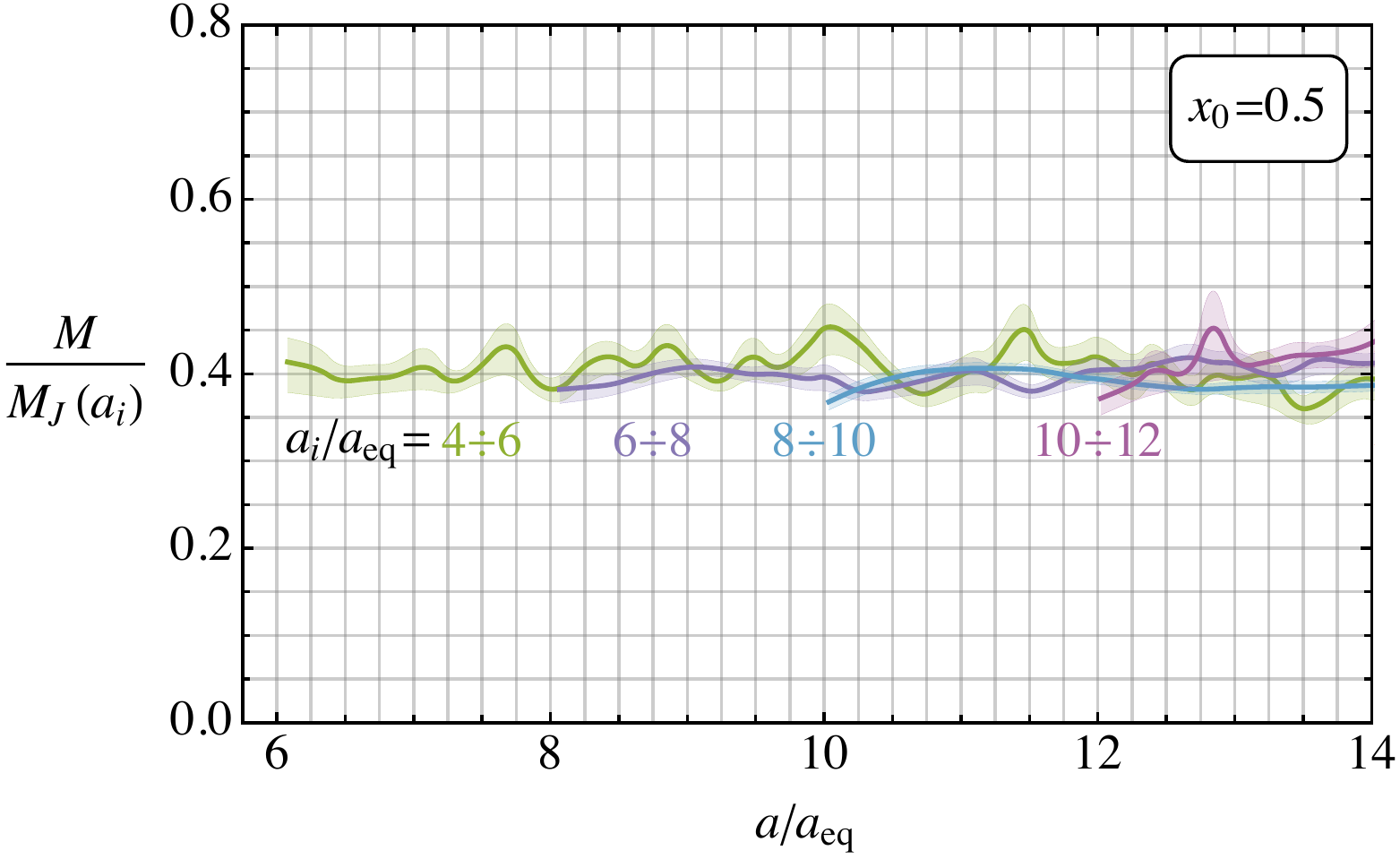}
	\end{center}
	\caption{ {\bf \emph{Left:}} The soliton production rate with artificially rescaled initial conditions compared to the analytic prediction, analogous to Figure~\ref{fig:production} for the physical initial conditions. {\bf \emph{Right:}} The average mass of solitons produced in different time intervals compared to the predicted parametric dependence $M_J(a_i)$ with the rescaled initial conditions. The results are similar to those obtained starting from the physical initial conditions, as plotted in Figure~\ref{fig:massa}.
		\label{fig:x02B}} 
\end{figure}

To check the reliability of our analytic method, we have carried out simulations with the same initial conditions as the power spectrum in eq.~\eqref{eq:PAt1} but with $k_\star\to x_0 k_\star$ with $x_0 = 1/2$. As expected, this change to the initial power spectrum leads to a different (smaller) soliton mass function, plotted in Figure~\ref{fig:x02}, and production rate, plotted in Figure~\ref{fig:x02B} (left). It also leads to a slightly smaller value of $c_M \simeq 0.4$, which determines the masses of solitons produced at a given time. We find that our analytic approach, with the same window function, also matches the data well in this case, albeit with a slightly different value of $\delta_c \simeq 0.3$. This gives us confidence that our analysis is capturing the main aspects of the underlying physics, although a further refinement would be useful in the future.

\section{Details of N-body Simulations}  \label{app:Nbody}

We carry out our N-body simulations using the code Gadget 4 \cite{Springel:2020plp}, modified to include the effect of radiation on the Universe's expansion. For simplicity, we follow the standard convention of including only radiation, dark matter and vacuum energy in simulations (i.e. neglecting baryons). We fix the cosmological parameters $\Omega_0=0.311$, $\Omega_\Lambda = 0.689$ and $\Omega_{\rm rad}= 0.0000924$ (corresponding to $z_{\rm eq} = 3370$) and the present day Hubble parameter $H=67.7\km {\rm s}^{-1} {\rm Mpc}^{-1}$. Note however that most of our results can be rephrased in terms of $a/a_{\rm eq}$ in which case these numerical choices do not matter. The absence of baryons (which, as discussed, do not collapse into clumps around MRE but might be relevant later) will affect the growth of density perturbations slightly, but to the level of accuracy that we are aiming for this is not a major uncertainty. As discussed in Section~\ref{s:compact_halos}, we start our simulations prior to MRE, at $a/a_{\rm eq} = 0.01$ (as with our Schr\"odinger-Poisson simulations the particlar numerical value has no effect provided $a/a_{\rm eq} \ll 1$). The simulations are pure N-body with just gravitational interactions, so cannot capture the effects of quantum pressure. We set Gadget's time-stepping parameter to $0.02$, which has been found to be sufficiently accurate in similar simulations \cite{Xiao:2021nkb}. Halos are identified using Gadget's friend-of-friend algorithm, with dimensionless link length $0.2$, so particles are linked if their spacing is less than $0.2$ of the mean particle spacing, and fix the minimum group length for a set of particles to be classed as a halo to $32$.

We use a box size of $80 \lambda_\star$, and obtain our results averaging over three simulation runs. The comoving gravitational softening length is chosen to be $0.003 \lambda_{\star}$. This is smaller than $\lambda_J$ in the cores of the densest objects that form, at which scales N-body simulation results will not reproduce the true physics anyway, see Section~\ref{ss:profile_compact}. E.g. at $z=70$, corresponding to the density profiles plotted in Figure~\ref{fig:profile_white_noise}, the gravitational softening length is about $2\cdot 10^4 {\rm km} (m/\eV)^{1/2}$ and $\lambda_J(5\eV^4) \simeq 3\cdot 10^4 \km$.  Simulations are run from $a/a_{\rm eq}=0.01$ until $z=24$ (i.e. $a/a_{\rm eq}\simeq 140$). At this time the most IR modes have $\mathcal{P}_\delta(k) \simeq 0.05$, and still closely match the values predicted by a linear analysis, indicating that finite volume systematics are not yet becoming important (we could run to slightly later times, but the statistical uncertainty is already starting to limit our results anyway).\footnote{We are able to reach relatively small $z$ compared to other similar analysis in the literature, e.g. \cite{Eggemeier:2019khm}, because we do not attempt to resolve the structures that collapse around MRE in our simulations and can therefore use a comparatively large box size.}

\begin{figure}[t]
	\begin{center}
		\includegraphics[width=0.48\textwidth]{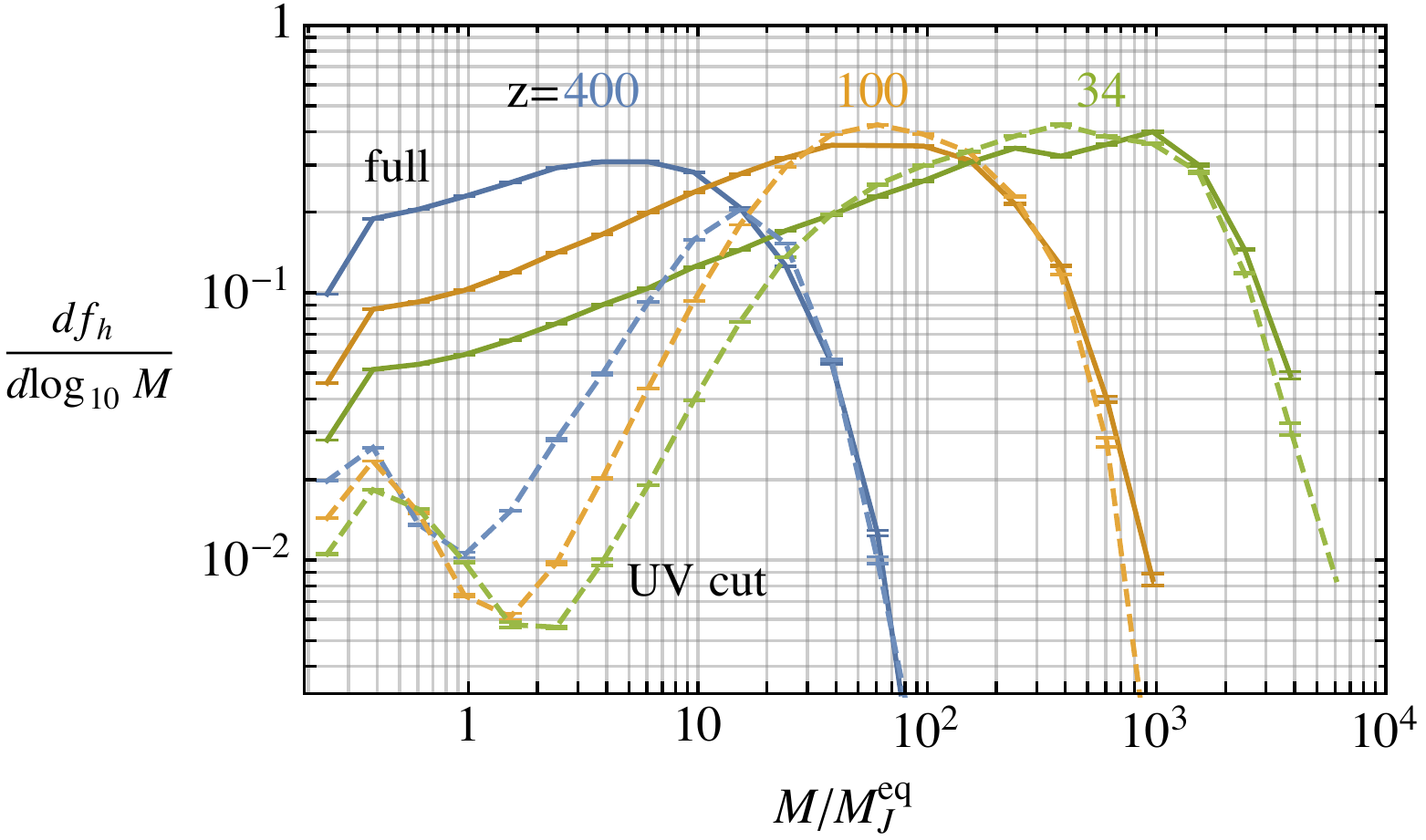}
				\includegraphics[width=0.48\textwidth]{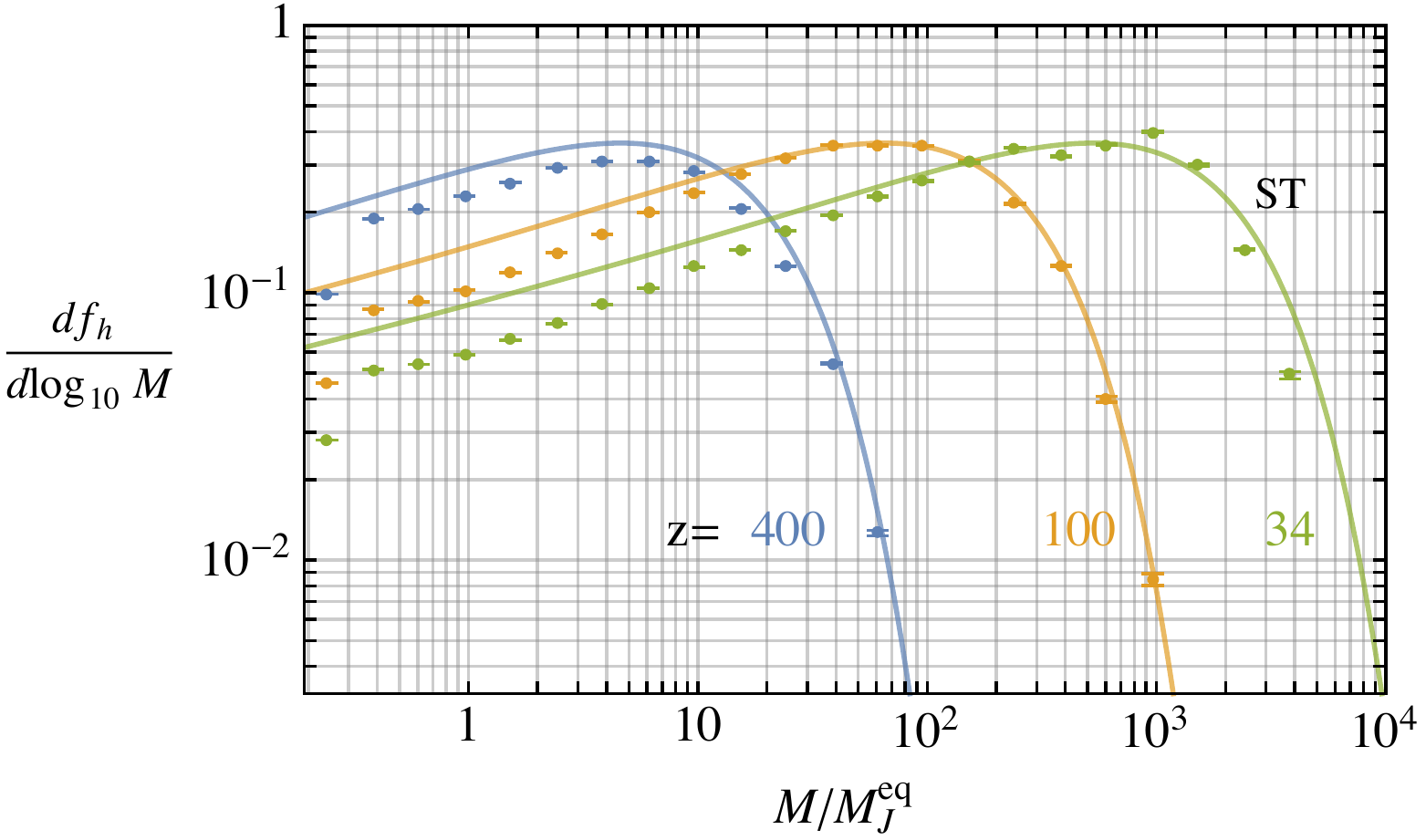}
	\end{center}
	\caption{ {\bf \emph{Left:}} Comparison between the halo mass function obtained in N-body simulations starting from a realisation of the full dark photon initial conditions (solid) and from initial conditions including only perturbations on scales $k/k_\star < 0.5$, which are unaffected by quantum pressure (dashed). The difference between the results from these two sets of initial conditions gives an estimate of the importance of dynamics at small scales for the system's evolution, and so an estimate of the scales on which the results of N-body simulations without quantum pressure can be trusted. The number of objects with small mass $M/M_J^{\rm eq} \lesssim 50 $ differ widely, but properties of heavier objects broadly agree.
		{\bf \emph{Right:}} The mass function of compact halos at different redshifts from simulations and from the Sheth-Tormen  inspired fit to the data (ST), which reproduces the results well. 
		 \label{fig:HMF_cut_vs_real}} 
\end{figure}

In Figure~\ref{fig:HMF_cut_vs_real} we plot the mass function of compact halos from N-body simulations comparing results from initial conditions with a UV cut and from realistic initial conditions. The initial conditions with a UV cut are a realisation of Gaussian density field with  the power spectrum of Figure~\ref{fig:Pspectra} (right) for $k< 0.5 k_\star$ and $0$ otherwise, converted to particles (for the $m$ of interest only the isocurvature part of the initial spectrum is relevant given our box sizes). This choice is made to exclude the density fluctuations on scales that will collapse to solitons around MRE (and which will be affected by quantum pressure), see Figure~\ref{fig:mode_growth} (right). The full initial conditions are generated from a realisation of the density field, generated from the power spectrum of $A$ in Figure~\ref{fig:Pspectra} (left). Of course in this case the dynamics of the small scale density perturbations are not correctly reproduced. As discussed in the main text, the agreement of the mass functions for $M\gtrsim 50 M_J^{\rm eq}$ is an indication that the mass function in this interval is unaffected by the dynamics of modes at small scales.

\section{Analytic Analysis of the Compact Halos} \label{app:compact_halos}

In the Press-Schechter analysis of Section~\ref{ss:form_compact} we smooth the field over a distance $R$ using a window function $W(k, R)$ that vanishes at $k\gtrsim 2\pi/R$, i.e. $\delta_s(t,\vec{x})=(2\pi)^{-3}\int d^3k \delta(t,\vec{k})\exp(i\vec{k}\cdot\vec{x})W(k,R)$. We choose the top-hat window function $W(k,R)=W_{\rm t}(kR)$ with $W_{\rm t}(x)\equiv(3/x^3)(\sin x - x \cos x)$. As mentioned in the main text, since $\delta$ is Gaussian at scales larger than $\lambda_\star$, the probability that the field smoothed over $R=(3M/4\pi\bar{\rho})^{1/3}$ is larger than $\delta_c$ is determined only by the variance $\Pi_{\delta>\delta_c}=\frac12 \text{erfc}[\delta _c/\sqrt{2} \sigma _s(M)]$ where
$\sigma _s^2\equiv\langle \delta_s^2(t,\vec{x})\rangle=D^2[a]\int dk/k \, \mathcal{P}_\delta(t\ll t_{\rm eq},k)W^2(k,R)$, with $\mathcal{P}_\delta$ at $t\ll t_{\rm eq}$ as in eq.~\eqref{eq:Pdelta}.

To obtain eq.~\eqref{eq:dfhdlogM} we estimate $\sigma_s^2(M)\simeq 3^{3/2} D^2[a]M_\star/(16\pi^3 M)$ using a crude approximation $\mathcal{P}_\delta(k)\simeq (\sqrt{3}/\pi)(k/k_\star)^3 $ at $k<k_\star$ and otherwise zero (since modes with $k\gtrsim k_\star$ are affected by quantum pressure, and form solitons).\footnote{This is effectively equivalent to considering a field smoothed between $k_\star$ to $2\pi/R$, motivated because the collapse of the other momenta forms solitons.} 

As mentioned in the main text, although the PS result captures the most important features of the dynamics, it is not precisely accurate. 
Instead, to reconstruct the halo mass function at $z\simeq 15$, we fit the data with an empirical Sheth--Tormen (ST) inspired form, which reproduces the data at early times more accurately than the PS prediction in eq.~\eqref{eq:dfhdlogM} (a similar fit was carried out in the context of axion miniclusters in \cite{Xiao:2021nkb}). The ST form for the halo mass function is
\begin{equation}
	\frac{df_{\rm h}}{d\log M}=A(p) (1+(q\nu^2)^{-p}) \left( q \nu^2 \right)^{1/2} e^{-q\nu^2/2}\left|\frac{d\log \nu}{d\log M}\right| ~,
\end{equation}
where $\nu(M)= \delta_c/\sigma_s(M)$ as before, $q$ and $p$ are parameters to be fit, and $A^{-1}(p)=  1 +  \frac{1}{\sqrt{\pi}} 2^{-p} \Gamma\left(\frac{1}{2} - p \right)$ is required by the normalisation condition $\int dMdf_{\rm h}/dM=1$. We obtain $q=0.696$ and $p=0.305$ (with $\delta_c=1.7$ fixed), which fit the N-body results at early times well (see Figure~\ref{fig:HMF_cut_vs_real} right). These parameter values are close to those that come out of the original analytic analysis \cite{Sheth:1999mn}.

Finally, we give complete formula for the NFW parameters of the compact halo profiles, described in Section~\ref{ss:profile_compact}:
\begin{align}\label{eq:rho0an}
	\rho_0=&\,  \frac{ 4.4 \cdot 10^{-3}\, c_\Delta^3  \nu_c^3 \Delta }{(c_\Delta+1)^{-1}+\log (c_\Delta+1)-1}\left[\frac{M_J^{\rm eq}}{M}\right]^{3/2}\bar{\rho}^{\rm eq}  %\rightarrow \
	%\simeq 0.2\left[\frac{10^3 M_J^{\rm eq}}{M}\right]^{\frac32}\bar{\rho}^{\rm eq}
	 \, ,\\
	r_0=&\, \frac{4.2}{c_\Delta \nu_c\Delta^{1/3}} \left[\frac{M}{M_J^{\rm eq}}\right]^{5/6}\lambda_J^{\rm eq} \ ,\label{eq:r0compactaB}
\end{align}
which, once the values of $c_\Delta$, $\Delta$ and $\nu_c$ mentioned in Section~\ref{ss:profile_compact} are put in, lead to eqs.~\eqref{eq:rho0compact} and  \eqref{eq:r0compact}.

\section{Survival of the Dark Matter Substructure} \label{app:survival}

\subsection{Disruption mechanisms} \label{s:destroy}

In this Appendix we study the survival of the solitons, the fuzzy halos that surround them, and the compact halos from the $k^3$ part of the initial power spectrum.   Given the possible importance for direct detection, we concentrate on the survival of objects that today are on orbits that pass through the neighbourhood of Earth. 

There has been extensive analysis of destruction of dark matter substructure in the literature, and we can adapt known results (the derivations of which can be found in the references). As mentioned in Section~\ref{ss:destroy}, we mostly make the approximation that dark matter clumps can be approximated as being composed of classical particles. This is reasonable for the fuzzy halos around the solitons and especially for the compact halos. There could be significant changes to our results for solitons, since the de Broglie wavelength is comparable to the size of these (we will briefly comment on the possible corrections). Additionally, as discussed in the main text, we focus on particular processes that could destroy the clumps rather than attempting to follow a clump's full history. Nevertheless, our analysis will be sufficient to show that it is plausible that the majority of solitons, and a substantial fraction of the fuzzy halos around them, survive to the present day. 

We first describe the various mechanisms that could destroy an object, and then apply these to the different types of objects.

\subsubsection*{Tidal forces from a central potential}

If a dark matter clump is bound inside a larger gravitationally bound object, which we call the host, the clump experiences tidal forces that could destroy it. We define the tidal radius as the distance from the centre of the clump beyond which the tidal force is greater than the gravitational attraction to the centre of the clump. The part of the clump outside the tidal radius is expected to be destroyed. For simplicity we assume the clump has an approximately circular orbit inside the host, in which case the tidal radius  $r_t$ satisfies  \cite{vandenBosch:2017ynq,Lee:2020wfn}
\beq \label{eq:rtidal}
r_t = r_{\rm orbit} \left( \frac{M_{\rm clump}(r_t)}{M_{\rm host}(r_{\rm orbit})} \right)^{1/3}  \left(3- \left. \frac{d\log M_{\rm host}(r)}{d\log r} \right|_{r=r_{\rm orbit} } \right)^{-1/3} ~,
\eeq
where $M_{\rm clump}(r)$ is the mass of the clump inside a radius $r$, $M_{\rm host}(r)$ is the mass of the host (e.g. a compact halo, a halo from an adiabatic fluctuation, or  the Milky Way itself) inside the radius $r$, and $r_{\rm orbit}$ is the radius from the centre of the host at which the clump is orbiting.

To the accuracy that we require, the  factor in the second bracket of eq.~\eqref{eq:rtidal} can be set to $1$.  Then the tidal radius is the radius at which $\bar{\rho}(r_t) = \bar{\rho}_{\rm host} (r_{\rm orbit})$ where $\bar{\rho}_{\rm host}(r_{\rm orbit})$ is the mean dark matter density of the host inside the clump's orbit, $\bar{\rho}_{\rm host}(r_{\rm orbit}) = M(r_{\rm orbit})/\left(\frac{4}{3} \pi r_{\rm orbit}^3 \right)$, and $\bar{\rho}(r)=M_{\rm clump}(r)/\left(\frac{4}{3} \pi r^3\right)$  is the mean density of the clump within the radius $r$. 

For an orbit in the Milky Way, roughly the same distance from the centre of the galaxy as the Sun is, we have $\bar{\rho}_{\rm host}(8{\rm kpc}) \simeq 10^{11} M_\odot / (\frac{4}{3} \pi (8 \kpc)^3) \simeq  10^{-5} \eV^4$ (somewhat larger than the local dark matter density $\rho_{\rm local} \simeq 0.5 \GeV/\cm^3 \simeq 4\cdot 10^{-6} \eV^4$).

\subsubsection*{Dynamical friction}

If a clump is bound in a larger object, dynamical friction leads to the clump's orbit decaying and falling into the centre of the host object, where the clump will be destroyed by tidal forces. The orbit decays on a timescale \cite{2008gady.book.....B,Dai:2019lud} 
\beq \label{eq:tdecay}
t_{\rm decay} = t_{\rm orbit}  \frac{M_{\rm host}}{M_{\rm clump}} \left(  \log\left(\frac{M_{\rm host}}{M_{\rm clump}} \right) \right)^{-1} \simeq \frac{1}{\sqrt{G \bar{\rho}_{\rm host}}}  \frac{M_{\rm host}}{M_{\rm clump}} \left(  \log\left(\frac{M_{\rm host}}{M_{\rm clump}} \right) \right)^{-1} ~.
\eeq
where $t_{\rm orbit}$ is the time the clump takes to orbit the larger halo. 
Putting typical values of $M_{\rm clump}$ into eq.~\eqref{eq:tdecay}, dynamical friction is irrelevant for any of the objects of interest inside the Milky Way. However, it could be important for e.g solitons inside compact halos, or compact halos inside the lowest mass halos from the adiabatic perturbations, for which the ratio $M_{\rm host}/M_{\rm clump}$ is not too large.

\subsubsection*{Collisions with stars}

If it is in  a galaxy,  a DM clump can be disrupted by the energy transferred from a passing star. Typically such an encounter happens fast compared to the dynamical timescale of the clump. In this case, known as the impulse approximation, the energy transferred can be straightforwardly calculated. More precisely, the impulse approximation is valid if $r_{\rm clump}/b \ll v_{\rm rel}/\sigma$, where $r_{\rm clump}$ is the radius of the object, $b$ is the impact parameter of the collision, $v_{\rm rel}$ is the relative velocity of the collision, and $\sigma$ is the velocity dispersion of the clump (i.e. the variance of the magnitude of the DM velocity in the clump is $\sigma^2$)  \cite{Goerdt:2006hp}. 

The fuzzy halos around solitons and the compact halos are extended objects with densities that vary by five orders of magnitude between their centres and their outer edges. It is therefore plausible that the outer edges might be disrupted by a collision with a star while the centre remains. To account for this possibility we make the approximation that energy transferred to the mass in a shell at some distance from the centre of the clump can lead to this shell being lost from the object,  independently of what happens to the mass closer to the centre. Compared to the alternative estimate of requiring that the entire object is destroyed by an encounter, this means that the outer edges of objects are more easily destroyed. We discuss the disruption of solitons in Section~\ref{ss:soliton_dest}.

We consider only the possibility that a shell is removed by a single close encounter rather than a series of encounters. In the second case the energy transferred to a shell might be redistributed throughout the object. Ignoring this effect, and assuming the object and perturbing stars have a Maxwellian distribution,  the timescale for destruction by sequential heating and by a single catastrophic encounter happen to coincide \cite{Kavanagh:2020gcy}. Therefore our neglect of destruction by sequential heating will not have a major effect on our results, although further analysis would be worthwhile (see e.g. \cite{Goerdt:2006hp} for related studies in the context of WIMP dark matter).

We define $b_{\rm crit}$ to be the critical impact parameter, so that the object is destroyed if the impact parameter $b<b_{\rm crit}$ and is otherwise unaffected. Given that the probability of an encounter with impact parameter $b$ grows linearly with $b$ the majority of destructive encounters have $b$ not too much smaller than $b_{\rm crit}$ (the average destructive encounter has $b=2 b_{\rm crit}/3$). For all objects of interest  $b_{\rm crit} \gg  r_{\rm clump}$, and also the conditions for the impulse approximation to hold are satisfied. The energy transferred to a shell of dark matter, of mass $dm$, at radius $r<r_{\rm clump}$ from the centre of the clump by an encounter with a star of mass $M_s$ is\footnote{In the limit $b \ll R$ a similar expression can be obtained, and for $b \simeq R$ an approximate expression that interpolates between the two limiting behaviours can be used, see e.g.  \cite{Green:2006hh}. Note that the expression for the energy change of a particular particle gets further contributions, but these average to zero and are approximately the same magnitude as the term that remains anyway.}
\beq
\Delta E= \frac{4}{3} \frac{G^2 M_s^2 r^2}{v_{\rm rel}^2 b^4} dm ~.
\eeq
We approximate that the shell will be removed from the object if the transferred energy is greater than the gravitational binding energy of the shell, $GM(r)dm/r$, where $M(r)=4\pi\int_0^r r'^2 dr' \bar{\rho}(r')$. Consequently, the critical impact parameter is given by 
\beq
b_{\rm crit}^2= \frac{M_s}{v_{\rm rel}} \left( \frac{G}{\pi \bar{\rho}(r) } \right)^{1/2} ~,
\eeq
where, as before, $\bar{\rho}(r)$ is the mean density of the clump within the radius $r$.

We are particularly interested in objects that pass the neighbourhood of Earth on their orbits. Most such objects are on trajectories that do not entirely reside inside the galactic disk.  The cross section for a destructive encounter, $\pi b_{\rm crit}^2$, is proportional to the mass of the disrupting star. Therefore, the probability that a shell at radius $r$ is removed from the clump after the clump crosses the disk $n$ times is
\beq
p_{\rm dest} = \pi n \frac{S }{v_{\rm rel} }  \left( \frac{G}{\pi \bar{\rho}(r) } \right)^{1/2}  ~,
\eeq
where $S$ is the mass density encountered per unit transverse area of the disk, provided $p_{\rm dest} \ll 1$ (this analysis mirrors that applied to axion miniclusters in \cite{Tinyakov:2015cgg}). Consequently,
\beq
\begin{aligned} \label{eq:pdesthalo}
	p_{\rm dest} 	& = 0.4  \left( \frac{n}{100} \right)  \left( \frac{0.05 \eV^4}{\bar{\rho}\left(r\right)}\right)^{1/2} \left( \frac{S}{140 M_\odot {\rm pc}^{-2}} \right) \left( \frac{10^{-3}}{v_{\rm rel}} \right)  ~,
\end{aligned}
\eeq
with the reference value  $S=140 M_\odot {\rm pc}^{-2}$ a factor of $4$ larger than the column density transverse to the Milky Way disk to account for the fact that on average trajectories are not perpendicular to the disk \cite{Tinyakov:2015cgg}. $n\simeq 100$ corresponds to roughly the number of disk crossings since the formation of Milky Way for a clump on a typical orbit \cite{Tinyakov:2015cgg}, however this can vary significantly for different forms of orbit.

\subsubsection*{Tidal shocking by the disk}

A dark matter clump in a disk galaxy, could also be destroyed by the overall gravitational field of the disk, rather than by encounters with individual stars. As a clump moves through the disk, the energy of DM particles that are away from the centre of the clump in the direction perpendicular to the disk increases compared to those at the centre. We consider the dark matter of total mass $dm$ in a small region that is a distance $\Delta z$ away from the centre of the clump in the perpendicular direction. The relative  energy increase of this DM is  \cite{ostriker1972evolution} (see also e.g. \cite{Berezinsky:2014wya,2017JETP..125..434D}) 
\beq \label{eq:destcross}
\Delta E= 2 (2\pi G \sigma_s(r))^2  \frac{ (\Delta z)^2 }{ v_{z}^2 } dm (1+a^2)^{-3/2} ~,
\eeq
where $\sigma_s(r)$ is the surface mass density of the disk, and $v_z$ is the velocity of the clump perpendicular to the disk. At $8 {\rm kpc}$ from the Milky Way centre, $\sigma_s \simeq 10^8 M_\odot/{\rm kpc}^2$. The final factor in eq.~\eqref{eq:destcross} is a suppression due to adiabaticity of the process \cite{Weinberg:1994uy}. Intuitively, the destruction is less efficient if the tidal force changes slowly compared to the time a DM particle takes to orbit the clump. In particular,  $a= t_{\rm crossing}/ t_{\rm internal}$ where $t_{\rm internal}$ is the time the particle takes to orbit the clump, and  $t_{\rm crossing} \sim 5\cdot 10^{13} {\rm s}$  is the time the clump takes to cross the disk 	\cite{Berezinsky:2014wya}. In the Milky Way, $a\simeq 6 \left(\bar{\rho}/\eV^4 \right)^{1/2}$, which is relevant for the solitons, but only marginally important for the halos surrounding them.  

Similarly to before, we compare the energy increase in eq.~\eqref{eq:destcross} to the binding energy of the dark matter particles in the clump. A single crossing disrupts regions of the clump for which
\beq
\bar{\rho}(r) \lesssim 10^{-6} \eV^4 \left( \frac{\sigma_s}{10^8 M_\odot/{\rm kpc}^2} \right)^2 \left(\frac{10^{-3}}{v_z} \right)^2 ~.
\eeq
Even if the energy transferred adds up every disk crossing, the disruption through this process is negligible compared to that caused by collisions with stars.

\subsubsection*{Tidal shocking}

During hierarchical structure formation, as a DM clump becomes bound in a larger object it experiences tidal forces on timescales shorter than the dynamical time of the clump. These are known as tidal shocks  \cite{spitzer2014dynamical,gnedin1999tidal}, and are distinct from the steady state tidal forces analysed previously. During a typical tidal shock, a shell of dark matter of mass $dm$ at radius $r$ from the centre of a clump is expected to get a relative energy increase compared to the rest of the clump of
\beq \label{eq:tidal_shock}
\Delta E \simeq \frac{4\pi}{3} \gamma_1 G \bar{\rho}_{\rm host}  r^2  dm~,
\eeq
where $\gamma_1$ is expected to be of order 1 \cite{Berezinsky:2005py} and, as before, $\bar{\rho}_{\rm host}$ is the mean density of the host within the orbit of the clump.

We make the crude approximation that the energy transferred to the DM shell is not redistributed to DM in other parts of the clump. Then the rate of energy increase of the shell is $\dot{E}= \gamma_2 \Delta E/t_{\rm dynam} $, where $t_{\rm dynam} \simeq (G \bar{\rho}_{\rm host})^{-1/2}$ is the dynamical timescale of the larger object (parameterically the same as the time the clump takes to orbit the host $t_{\rm orbit}$ once it is bound, defined previously), and $\gamma_2$ parameterises the number of tidal shocks per dynamical time, which is expected to be roughly order 1. 
Hence the shell is likely to be disrupted on a timescale
\beq
\begin{aligned} \label{eq:tidal_shockt}
	t_{\rm shock} & \simeq \frac{\bar{\rho}(r)}{\sqrt{G} \gamma_1 \gamma_2 \bar{\rho}_{\rm host}^{3/2}} \\
	&\simeq  10^{9}{\rm year} \left( \frac{\bar{\rho}}{0.01 \eV^4} \right) \left(\frac{1}{\gamma_1 \gamma_2} \right) \left( \frac{10^{-4} \eV^4}{\bar{\rho}_{\rm host}} \right)^{3/2} ~.
\end{aligned}
\eeq
However, we stress that this estimate is very rough and a detailed analysis would be valuable in the future. We have little control of the parameters  $\gamma_1$ $\gamma_2$, and we do not know how long tidal shocking continues until the clump reaches a steady state orbit. Also it is not even certain if it is accurate to compare the energy increase per tidal shock, eq.~\eqref{eq:tidal_shock}, to the binding energy. For instance \cite{Bosch:2017} finds that disruption is not inevitable even if the energy increase is much larger than the binding energy. Moreover, as discussed, in a realistic history of structure formation, a DM clump is likely to be bound in a series of objects of increasingly large mass. Consequently, the clump might not need to survive too long inside a slightly higher mass clump until that itself is quickly destroyed when it is bound inside a larger object.

\subsection{Solitons} \label{ss:soliton_dest}

It is not clear if our approach of allowing the disruption of shells is an accurate approximation for solitons, because the de Broglie wavelength is comparable to the size of a soliton. Also, if the outer layer of a soliton were to be lost the profile of the remainder might change, affecting the subsequent destruction probability. In particular, the analysis of \cite{Du:2018qor} suggests that the remainder of the soliton might spread out, making later destruction more likely. For definiteness we make the approximation that the soliton will be disrupted if sufficient energy is transferred to remove a shell at radius $r\simeq 0.7 \lambda_J(\rho_s)$ (where the density profile switched from soliton like to NFW-like). If we instead required that the entirety of the soliton inside $ 0.7 \lambda_J(\rho_s)$ was destroyed, the destruction probability would decrease.

We have seen above that the probability that a shell at distance $r$ from the centre of a clump is destroyed depends on the mean DM density inside the shell, $\bar{\rho}(r)$, rather than the density at the distance $r$, which is $\rho(r)$.  At $ 2 \lambda_J(\rho_s)/3$, we have  $\bar{\rho} \simeq 0.2 \rho_s$. 
Solitons with $\bar{\rho} \gtrsim 0.05\eV^4$ are likely to survive destruction by the Milky Way central potential, collisions with stars, and dynamical friction in the Milky Way. Therefore, we conclude that these processes are likely to leave the majority of solitons produced around MRE undisrupted, since they have central densities $\rho_s \simeq \left( 0.1 \div 100 \right) \eV^4$ (see Figure~\ref{fig:HMF_soliton}, corresponding to $\bar{\rho} \simeq \left( 0.03 \div 30 \right) \eV^4$).

The possible disruption of solitons during hierarchical structure formation is less certain. Most solitons become bound in compact halos before being subsumed in an adiabatic halo. The probability that the soliton inside a compact halo is destroyed depends on the mean density of the compact halo inside the soliton's orbit $\bar{\rho}_{\rm host}(r_{\rm orbit})$. The compact halos have mean densities in the range $\bar{\rho}(r)_{\rm host} \sim (10^{-6} \div 1) \eV^4$, see Figure~\ref{fig:destruction_profile_2} (right) and masses between $M_{\rm halo} \simeq (10^2 \div 10^4) M_J(a_{\rm eq})$.

Because that fraction of dark matter bound in compact halos rapidly increases at late times, the majority of the solitons will be bound in the most massive and least dense compact halos. Given the low density of the these compact halos, tidal stripping will typically not destroy any solitons they contain. Additionally, the timescale for dynamical friction in eq.~\eqref{eq:tdecay} corresponds to
\beq \label{eq:tdynam_friction}
t_{\rm dyn} \simeq 3\cdot 10^{11} {\rm yr} \left(\frac{10^{-4}\eV}{\bar{\rho}(r)_{\rm host}} \right)^{1/2} \frac{M_{\rm host}/ M_{s}}{10^5} \frac{\log\left(10^5 \right)}{\log\left(M_{\rm host}/ M_{s} \right)} ~,
\eeq
where $M_s$ is the mass of the soliton and $M_{\rm host}$ is the mass of the compact halo that it is inside. Consequently dynamical friction will also not destroy a typical soliton in a fairly massive compact halo. 

Meanwhile, the timescale for destruction by tidal shocking, eq.~\eqref{eq:tidal_shockt} might be relevant for the lower mass solitons if they are contained in a relatively low mass (i.e. high density) compact halo. However, many of the compact halos will be destroyed as adiabatic structures form, in which case a soliton need only survive inside the compact halo until that happens. Similarly, a typical soliton is unlikely to be destroyed if it is bound into a relatively low mass adiabatic halo prior to ending up in the Milky Way. A detailed analysis will be needed to draw definite conclusions, and this would be worthwhile in the future.

So far we have neglected the wave-like nature of the dark photon inside a soliton. Although we do not attempt a full analysis of the effect of this on the solitons' survival, we note that this does allow the soliton to lose mass by tidal stripping even from the region inside $r_t$ defined in eq.~\eqref{eq:rtidal}. This has been analysed in \cite{Hui:2016ltb}, and considered in more detail in \cite{Du:2018qor}, which accounts for the fact that the soliton becomes less dense after mass is removed, accelerating the subsequent destruction. Using numerical simulations the latter reference finds that a soliton of mass $M$ orbiting a distance $r_{\rm orbit}$ will be unaffected by tidal stripping by a central potential from a halo with mean density $\bar{\rho}_{\rm host}(r)$ if the central density of the soliton $\rho_s$ satisfies
\beq
\rho_s/\bar{\rho}_{\rm host}(r_{\rm orbit}) \gtrsim 100 .
\eeq
For a soliton in the Milky Way $\bar{\rho}(r_{\rm orbit}) \simeq 5 \cdot 10^{-5} \eV^4$, so given that the typical soliton central densities are $0.1 \div 100 \eV$ they are unaffected by this effect. Likewise this will not affect a typical soliton that is bound in a compact halo.

Overall, although there are many uncertainties in our analysis, we conclude that it is plausible that the majority of the solitons survive to the present day.

\subsection{Fuzzy halos}

\begin{figure}[t!]
	\begin{center}
		\includegraphics[width=0.48\textwidth]{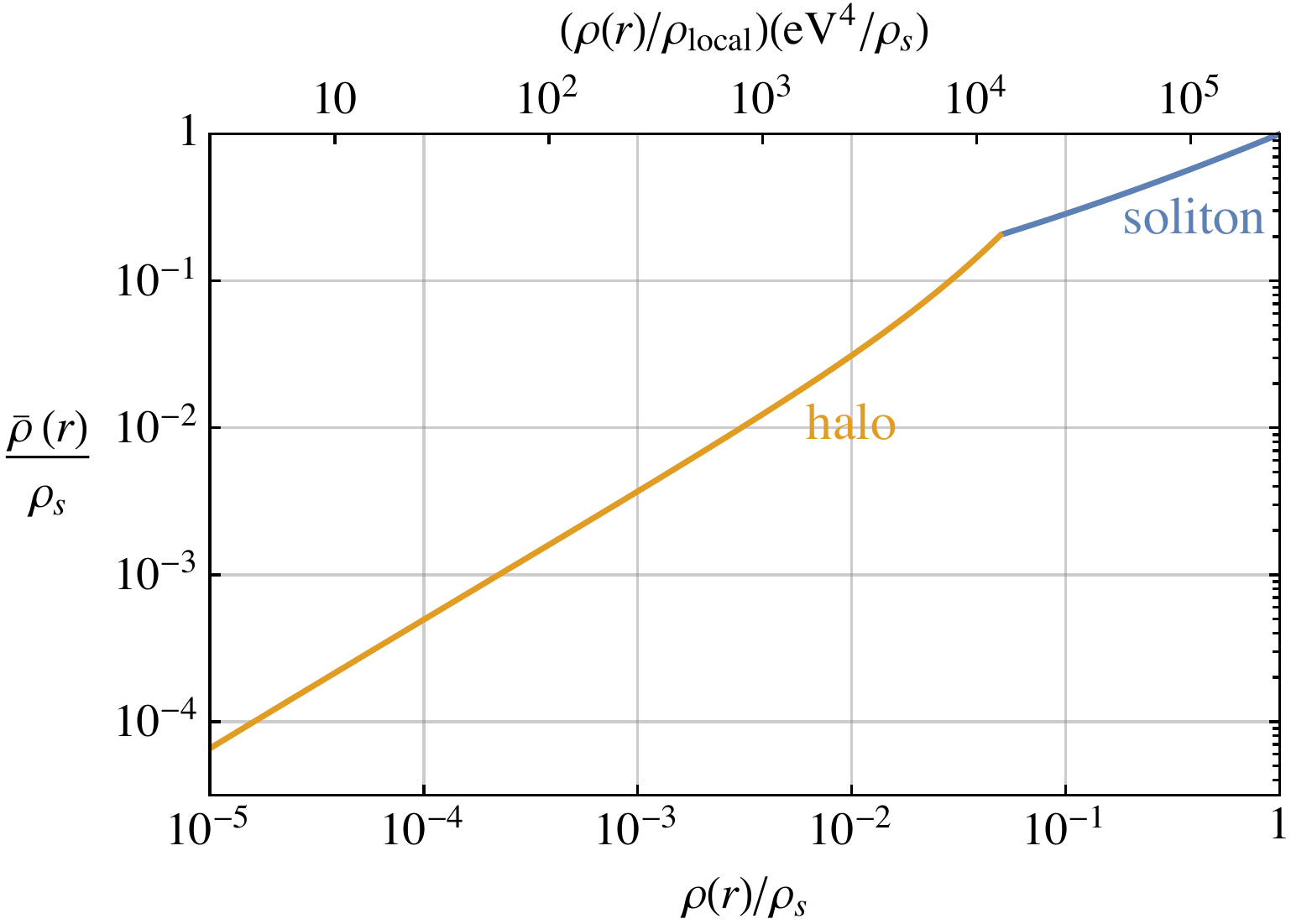}
		\includegraphics[width=0.48\textwidth]{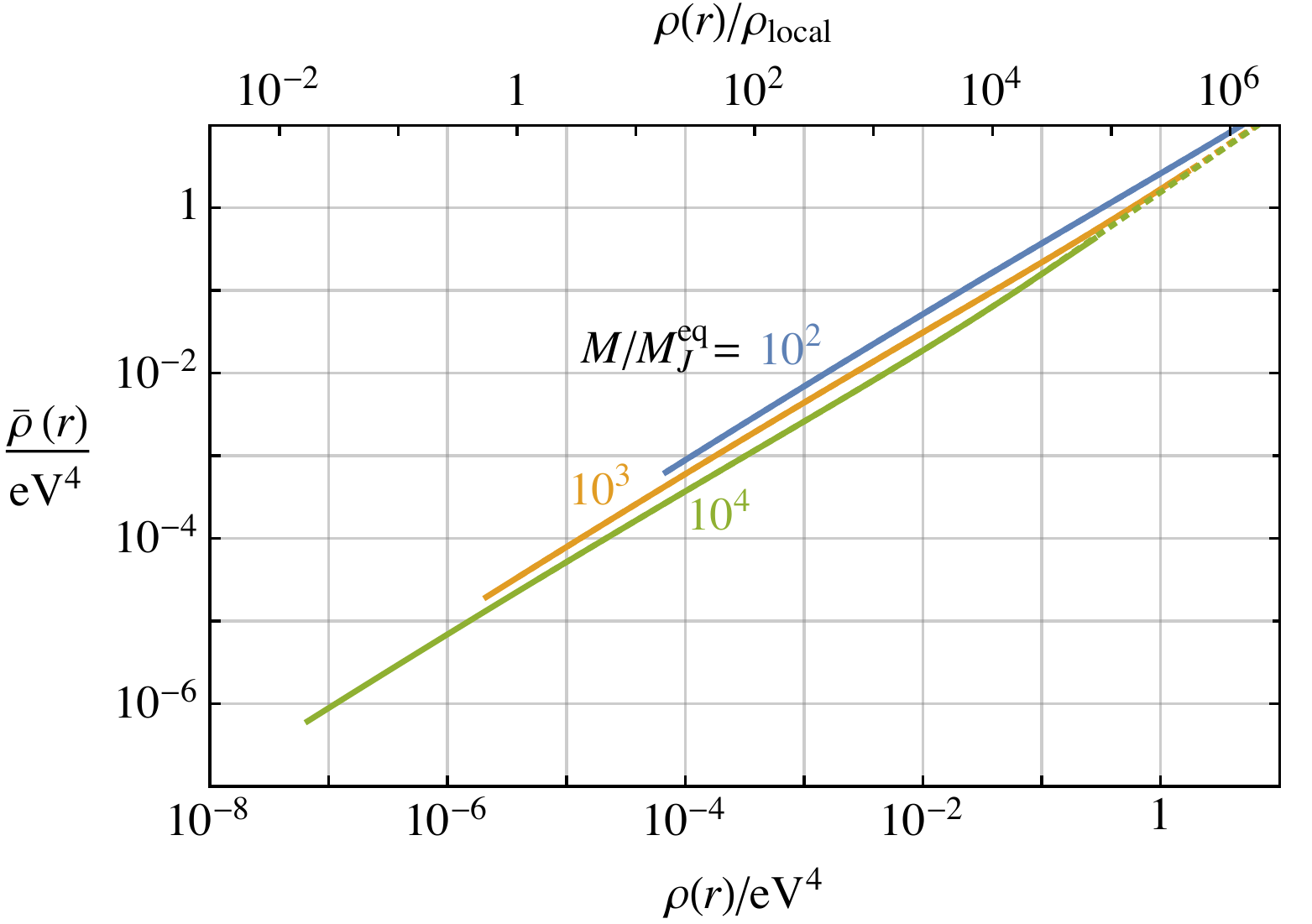}
	\end{center}
	\caption{ {\bf \emph{Left:}} The mean density inside a radius $r$, $\bar{\rho}(r)$, for a soliton and its surrounding fuzzy halo as a function of the local density at $r$, $\rho(r)$ (normalised to the density at the soliton core  $\rho_s$). These results are obtained by matching the soliton profile and NFW fit shown in Figure~\ref{fig:profile_log}. On the upper axis we indicate the corresponding enhancement over the local dark matter density, fixed to $0.5 \GeV/\cm^3$, which depends on $\rho_s$. Regions of the fuzzy halo with $\bar{\rho}(r)\gtrsim 0.05\eV^4$ are likely to survive undisrupted to the present day in the Milky Way.
	{\bf \emph{Right:}} The mean density inside a radius $r$, $\bar{\rho}(r)$, as a function of the local density $\rho(r)$ for compact halos of different masses. These results are obtained from the NFW fit described in Section~\ref{ss:fuzzy_halo}. The solid lines extend to densities where quantum pressure becomes relevant. The upper axis shows the corresponding enhancement over the local dark matter density. As with the fuzzy halos, regions with $\bar{\rho}(r)\gtrsim 0.05 \eV^4$ are likely to survive to the present day.
 		\label{fig:destruction_profile_2}} 
\end{figure}

Next we consider the fuzzy halos around the solitons. If such objects lie on trajectories that cross the neighbourhood of Earth, we can see from the analysis in Section~\ref{s:destroy} that destruction by stars is the most important of the effects of the Milky Way. From eq.~\eqref{eq:pdesthalo}, we expect that a halo is likely to survive out to a radius $r$ such that $\bar{\rho}(r) \simeq 0.05 \eV^4$. In Figure~\ref{fig:destruction_profile_2} (left) we plot $\bar{\rho}(r)$ as a function of the local density $\rho(r)$ for the solitons and their surrounding fuzzy halos (both normalised relative to the soliton's central density $\rho_s$, so that the plot applies all solitons and fuzzy halos, regardless of their mass). Values of $\bar{\rho}(r) \simeq 0.05 \eV^4$ in the range of interest corresponds to local densities of $\rho(r) \simeq  10^{-2} \eV^4$. Outside these $r$ the fuzzy halos are likely to be destroyed.

The halos could also be disrupted if it becomes bound in a compact halo prior to adiabatic structure formation (or in a low mass adiabatic halo prior to being bound in the Milky Way). Considering only the region with $\bar{\rho}(r)\gtrsim 0.05 \eV^4$, which has a chance of surviving interactions with stars, this part of the fuzzy halo is unlikely to be outside the tidal radius given by eq.~\eqref{eq:rtidal} when it  is bound in a compact halo. 

Meanwhile, the typical mass contained in a fuzzy halo inside $\bar{\rho}(r)\simeq  0.05 \eV^4$ is  roughly a factor of $10 \div 100$ greater than the soliton mass, so typically a factor of $10^4 \div 10^3$ smaller than the relatively high mass compact halos. Therefore, from eq.~\eqref{eq:tdynam_friction} dynamical friction might potentially be relevant. Likewise, tidal shocking, see eq.~\eqref{eq:tidal_shockt}, could be important.  However, as previously discussed the compact halos are themselves likely to be quickly destroyed. Finally we note that, as with the solitons, wave-like effects could be relevant for the fuzzy halos, and a detailed analysis of such effects would be worthwhile in the future.

\subsection{Compact halos}

The values of $\bar{\rho}(r)$ as a function of $\rho(r)$ for typical compact halos are plotted in Figure~\ref{fig:destruction_profile_2} (right). Like the halos around solitons, for a compact halo in the Milky Way the regions outside the radius $r$ such that $\bar{\rho}(r)=0.05 \eV^4$ are likely to be destroyed by collisions with stars. As a result, the majority of a compact halo with mass $\gtrsim 10^3 M_J(a_{\rm eq})$ is likely to be disrupted, possibly leaving a dense core behind. This means that the compact halos that contain the majority of the dark matter mass when adiabatic structure formation starts are likely to be mostly destroyed. Compact halos with masses  $\sim 10^2 M_J(a_{\rm eq})$ could survive to the present day, and these contain $\sim 10\%$ of the dark matter when adiabatic structure formation starts (see Figure~\ref{fig:HMF_white_noise_halo}). A small mass compact halo might also survive as substructure once bound inside a larger mass compact halo and survive to the present day.

\bibliography{darkphoton}
\bibliographystyle{JHEP}

\end{document}